\documentclass[numberedappendix, revtex4]{emulateapj}

\usepackage{graphicx}
\usepackage{epstopdf}
\usepackage{rotating}
\usepackage{pdflscape}
\usepackage{afterpage}
\usepackage{mathtools}
\usepackage{bm}
\usepackage{units}
\usepackage{enumitem}
\usepackage{appendix}
\usepackage{natbib}
\usepackage{xcolor}
\usepackage{hyperref}

\newcommand{\cholla}{{\it Cholla}~}		
\newcommand{\chollans}{{\it Cholla}}	
\newcommand{\nvidia}{NVIDIA~}
\newcommand{\nvidians}{NVIDIA}
\newcommand{\nvcc}{{\tt nvcc}~}
\newcommand{\Msun}{\mathrm{M}_{\odot}}

\begin{document}

\title{\cholla: A New Massively-Parallel Hydrodynamics Code for Astrophysical Simulation}

\author{Evan E. Schneider and Brant E. Robertson}
\affil{Steward Observatory, University of Arizona, 933 North Cherry Avenue,
    Tucson, AZ 85721, USA}

\begin{abstract}
We present \cholla (Computational Hydrodynamics On ParaLLel Architectures), a new three-dimensional hydrodynamics code that harnesses the power of graphics processing units (GPUs) to accelerate astrophysical simulations. \cholla models the Euler equations on a static mesh using state-of-the-art techniques, including the unsplit Corner Transport Upwind (CTU) algorithm, a variety of exact and approximate Riemann solvers, and multiple spatial reconstruction techniques including the piecewise parabolic method (PPM). Using GPUs, \cholla evolves the fluid properties of thousands of cells simultaneously and can update over ten million cells per GPU-second while using an exact Riemann solver and PPM reconstruction. Owing to the massively-parallel architecture of GPUs and the design of the \cholla code, astrophysical simulations with physically interesting grid resolutions ($\gtrsim256^3$) can easily be computed on a single device. We use the Message Passing Interface library to extend calculations onto multiple devices and demonstrate nearly ideal scaling beyond 64 GPUs. A suite of test problems highlights the physical accuracy of our modeling and provides a useful comparison to other codes. We then use \cholla to simulate the interaction of a shock wave with a gas cloud in the interstellar medium, showing that the evolution of the cloud is highly dependent on its density structure. We reconcile the computed mixing time of a turbulent cloud with a realistic density distribution destroyed by a strong shock with the existing analytic theory for spherical cloud destruction by describing the system in terms of its median gas density.
\end{abstract}


\section{Introduction}

Over the past fifty years, the field of computational hydrodynamics has grown to incorporate a wide array of numerical schemes that attempt to model a large range of astrophysical phenomena. From the pioneering work of \cite{Godunov59} and \cite{Courant67}, the sophistication of hydrodynamics solvers has steadily improved. Many astrophysical simulation codes now use high order reconstruction methods, implement very accurate or exact Riemann solvers, and model additional physics including gravity, cooling, magnetohydrodynamics, radiative transfer, and more \citep[e.g.][]{Kravtsov99, Knebe01, Fryxell00, Teyssier02, Hayes06, Stone08, Bryan14}. While these advanced techniques result in simulations of unprecedented physical accuracy, they can also be extremely computationally expensive. Given the detailed and expensive physical processes currently being modeled, new numerical approaches to modeling Eulerian hydrodynamics should be considered. This work presents a new, massively-parallel hydrodynamics code \cholla (Computational Hydrodynamics On ParaLLel Architectures) that leverages Graphics Processing Units (GPUs) to accelerate astrophysical simulations.

Historically, our ability to numerically model larger and more complex systems has benefitted from improvements in technology, especially increased storage and faster clock speeds for central processing units (CPUs). Algorithmic improvements such as adaptive mesh refinement \citep[e.g.,][]{Berger1984, Berger1989} have had a major impact on the ability of codes to achieve higher resolution, but much of the basic structure of static mesh grid codes has remained. In the last decade, computer speed has improved significantly as a result of increased parallelization, and the fastest supercomputers\footnote{{\tt http://www.top500.org}} now rely on hardware accelerators like GPUs or Intel Xeon Phi coprocessors to provide the bulk of their computational power. To leverage the full capabilities of these systems, multi-core CPU chips and accelerators must be used simultaneously in the context of a single hydrodynamics code. While similar parallelization and vectorization techniques apply to a variety of hardware accelerators, \cholla utilizes GPUs to perform all its hydrodynamical calculations. Engineering \cholla to run natively on GPUs allows us to take advantage of the inherently parallel structure of grid-based Eulerian hydrodynamics schemes, and enables the substantial computational performance gain demonstrated in this paper.

Accelerators and other special purpose hardware have been used in astrophysical simulations for many years
\citep[e.g.,][]{Sugimoto1990, Aarseth1999, Spurzem1999, Portegies2004, Harfst2007, Portegies2014}. Early work adapting Eulerian hydrodynamics solvers to the GPU indicated a promising avenue to accelerate simulations, with  developers reporting speedups of $50\times$ or more as compared to CPU-only implementations \citep[e.g.,][]{Brandvik07, Pang10, Bard10, Kestener10}. These preliminary efforts clearly illustrated the substantial performance gains that could be achieved using a single GPU. More recently, a number of multi-device GPU-based hydrodynamical simulation methods have been presented, including AMR techniques \citep{Schive10, Wang10}, two dimensional Galerkin approaches \citep{Chan2012}, Smoothed Particle Hydrodynamics (SPH) codes \citep{Sandalski12, Dominguez13}, and hybrid schemes \citep{Kulikov14}.

While these codes represent substantial advancement, the field of massively-parallel astrophysical hydrodynamics is still relatively new.  All of the aforementioned methods have been restricted to second-order spatial reconstruction, and many would require considerable modification to run on a cluster. In contrast, the hydrodynamics solver implemented in \cholla is among the most complex and physically accurate of those that have been adapted to GPU hardware. Successfully implementing such a complex solver in a hybrid environment on cluster scales displays our ability to merge the state-of-the-art in CPU hydrodynamics with a new generation of computer hardware.

As is evidenced by the number of CPU codes presented in the literature, room exists for many different approaches optimized for different purposes. With \chollans, we have built a fast, GPU-accelerated static mesh hydrodynamics module that can be used efficiently on its own or in conjunction with a variety of additional physics. Beyond accelerating the hydrodynamics calculation, offloading the work onto the GPU frees the CPU to perform other tasks. This excess computational capacity makes \cholla an excellent bedrock for developing complex physical models that require hydrodynamics.

The large dynamic range of spatial scales in many astrophysical problems requires simulations with both high resolution and a high level of physical accuracy. The interaction of a high mach number shock with a gas cloud falls into this category \citep{Klein94a}. Using the power of \chollans, we can efficiently run high resolution simulations of the cloud-shock problem to investigate how cloud density structure affects the destruction of high-density gas. Given the inhomogeneous nature of gas in galaxies, our results have wide-ranging implications, from the impact of supernovae on the gas in their immediate environment to the evolution of dense gas in galactic outflows.

In the following sections, we fully describe \chollans. The code models solutions to the equations of hydrodynamics using the Corner Transport Upwind (CTU) algorithm \citep{Colella90, GS08}, and includes multiple choices for both interface reconstruction and Riemann solvers. The CTU algorithm is presented in Section~\ref{sec:hydrodynamics} along with brief descriptions of the reconstruction methods and Riemann solvers, which are fully documented in the Appendices. The code structure, including the simulation setup, CUDA functions, optimization strategies necessary to take advantage of the GPU architecture, and Message Passing Interface \citep[MPI;][]{MPIForum94} implementation and scalability, is described in Section~\ref{sec:code_architecture}. We then demonstrate the excellent performance of \cholla on a suite of canonical hydrodynamics tests in Section \ref{sec:tests}. In Section \ref{sec:cloud}, we derive new results describing the interaction of a high mach number shock with a turbulent gas cloud. We conclude in Section \ref{sec:conclusion}.


\section{Hydrodynamics}\label{sec:hydrodynamics}

Hydrodynamics is relevant to many astrophysical processes and represents one of
the most computationally demanding parts of numerical simulations. Creating a fast 
hydrodynamics solver is therefore an important step in increasing the resolution 
and speed with which astrophysical calculations can be performed. In this section, we present the equations modeled by \chollans, and then describe the numerical algorithms used to model them. 
\cholla includes a variety of reconstruction techniques and Riemann solvers, each of which is described below.

In differential conservation law form \citep[see, e.g.,][]{Toro09}, 
the multi-dimensional Euler equations can be written:
\begin{equation}
\frac{\delta\rho}{\delta t} + \frac{\delta(\rho u)}{\delta x} + \frac{\delta(\rho v)}{\delta y} + \frac{\delta(\rho w)}{\delta z} = 0,
\label{eqn:mass_cons}
\end{equation}
\begin{equation}
\frac{\delta(\rho u)}{\delta t} + \frac{\delta(\rho u^2 + p)}{\delta x} + \frac{\delta(\rho uv)}{\delta y} + \frac{\delta(\rho uw)}{\delta z} = 0,
\label{eqn:momentum_cons_x}
\end{equation}
\begin{equation}
\frac{\delta(\rho v)}{\delta t} + \frac{\delta(\rho uv)}{\delta x} + \frac{\delta(\rho v^2 + p)}{\delta y} + \frac{\delta(\rho vw)}{\delta z} = 0,
\label{eqn:momentum_cons_y}
\end{equation}
\begin{equation}
\frac{\delta(\rho w)}{\delta t} + \frac{\delta(\rho uw)}{\delta x} + \frac{\delta(\rho vw)}{\delta y} + \frac{\delta(\rho w^2 + p)}{\delta z} = 0,
\label{eqn:momentum_cons_z}
\end{equation}
\begin{equation}
\frac{\delta E}{\delta t} + \frac{\delta[u(E + p)]}{\delta x} + \frac{\delta[v(E + p)]}{\delta y} + \frac{\delta[w(E + p)]}{\delta z} = 0.
\label{eqn:energy_cons}
\end{equation}
Here $\rho$ is the mass density, $u$, $v$, and $w$ are the $x$-, $y$-, and $z$-components of velocity, $p$ is the pressure, and $E$ is the total energy per unit volume,
\begin{equation}
E = \rho\left(\frac{1}{2}\bm{\mathrm{V}}^2 + e\right),
\end{equation}
where $\bm{\mathrm{V}} = [u, v, w]^{T}$ is the three-component velocity vector.
The total energy includes the specific internal energy, $e$, and the specific kinetic energy,
\begin{equation}
\frac{1}{2}\bm{\mathrm{V}}^2 = \frac{1}{2}\bm{\mathrm{V}} \cdot \bm{\mathrm{V}} = \frac{1}{2}\left(u^2 + v^2 + w^2\right).
\end{equation}
Equation~\ref{eqn:mass_cons} describes the conservation of mass, Equations~\ref{eqn:momentum_cons_x}-\ref{eqn:momentum_cons_z} the conservation of momentum, and Equation \ref{eqn:energy_cons} the conservation of energy. To model solutions to this system of conservation laws, an equation of state is also necessary. We use the equation of state for an ideal gas,
\begin{equation}
p = (\gamma - 1) \rho e,
\end{equation}
where $\gamma$ is the ratio of specific heats. Incorporating a real gas equation-of-state model \citep{Colella85} would not be incompatible with the structure of \chollans, though it is beyond the scope of our current work.

The Euler equations can also be written in vector notation. We define the vector of conserved quantities with components in three Cartesian dimensions,
\begin{equation}
\bm{u} = [\rho, \rho u, \rho v, \rho w, E]^\mathrm{T} 
\label{eqn:conserved_variables}
\end{equation}
including density, the three components of momentum, and total energy. We will also refer to the vector of primitive variables,
\begin{equation}
\bm{w} = [\rho, u, v, w, p]^\mathrm{T}
\label{eqn:primitive_variables}
\end{equation} 
that includes density, the three components of velocity, and pressure. We define three flux vectors
\begin{equation}
\bm{f} = 
\begin{bmatrix}
		\rho u \\
		\rho u^{2} + p \\
		\rho u v \\
		\rho u w \\
		(E + p) u
\end{bmatrix},
\label{eqn:x_flux}
\end{equation}
\begin{equation}
\bm{g} = 
\begin{bmatrix}
		\rho v \\
		\rho u v \\
		\rho v^{2} + p \\		
		\rho v w \\
		(E + p) v
\end{bmatrix},
\label{eqn:y_flux}
\end{equation}
and
\begin{equation}
\bm{h} = 
\begin{bmatrix}
		\rho w \\
		\rho u w \\
		\rho v w \\
		\rho w^{2} + p \\		
		(E + p) w
\end{bmatrix}.
\label{eqn:z_flux}
\end{equation}
Using these definitions, we can compactly write the three dimensional Euler equations in the conservative form:
\begin{equation}
\frac{\delta \bm{u}}{\delta t} + \frac{\delta \bm{f}}{\delta x} + \frac{\delta \bm{g}}{\delta y} + \frac{\delta \bm{h}}{\delta z} = 0.
\label{eqn:euler_compact}	
\end{equation}

The Euler equations can also be written using the primitive variables. In one dimension, the equations can be written as a set of linear hyperbolic equations of the form
\begin{equation}
\frac{\delta \bm{w}}{\delta t} + \bm{\mathrm{A}}(\bm{w}) \frac{\delta \bm{w}}{\delta x} = 0.
\end{equation}
The matrix $\bm{\mathrm{A}}(\bm{w})$ is diagonalizable and can be written
\begin{equation}
\bm{\mathrm{A}} = \bm{\mathrm{R}} \bm{\Lambda} \bm{\mathrm{L}},
\end{equation}
where $\bm{\mathrm{R}}$ is a matrix of right eigenvectors, $\bm{\Lambda}$ is a diagonal matrix of eigenvalues, and $\bm{\mathrm{L}} = \bm{\mathrm{R}}^{-1}$ is a matrix of left eigenvectors. The eigenvalues of $\bm{\mathrm{A}}$ are real and correspond to the speeds at which information propagates for the fluid equations. If we further define the characteristic variables, $\bm{\xi}$, according to
\begin{equation}
d\bm{\xi} = \bm{\mathrm{L}} d\bm{w},
\end{equation}
we can write the system a third way:
\begin{equation}
\frac{\delta \bm{\xi}}{\delta t} + \bm{\Lambda} \frac{\delta \bm{\xi}}{\delta x} = 0.
\end{equation}
This description is called the characteristic form of the Euler equations. The characteristic 
variables are sometimes called wave strengths, because they describe the magnitude of the jump in 
the primitive variables across an associated wave.
For an extensive treatment of the Euler equations and related subjects see, e.g., \citet{Laney98}, \citet{LeVeque02},
and \citet{Toro09}.

\subsection{The CTU Algorithm}\label{sec:CTU}

\cholla models the Euler equations using a three-dimensional implementation of the Corner Transport 
Upwind (CTU) algorithm optimized for magnetohydrodynamics \citep[MHD;][]{Colella90, Saltzman94, GS08}. 
The six-solve version of CTU used in \cholla is fully adapted for MHD and documented in both 
\cite{GS08} and \cite{Stone08}. We will describe here an abbreviated version including only the steps relevant for hydrodynamics calculations. 

The CTU algorithm is a Godunov-based method. A Godunov scheme uses a finite-volume 
approximation to model the Euler equations, evolving average values of the conserved quantities 
$\bm{u}$ in each cell using fluxes calculated at cell interfaces \citep{Godunov59}. In three dimensions, the calculation to update the conserved quantities $\bm{u}^{n}$ at timestep $n$ can be written as
\begin{equation}
\begin{aligned}
\bm{u}_{(i, j, k)}^{n+1} = \bm{u}_{(i, j, k)}^n &+ \frac{\Delta t}{\Delta x} [ \bm{F}_{(i - \frac{1}{2}, j, k)}^{n + \frac{1}{2}} - \bm{F}_{(i + \frac{1}{2}, j, k)}^{n + \frac{1}{2}} ] \\
&+ \frac{\Delta t}{\Delta y} [ \bm{G}_{(i, j - \frac{1}{2}, k)}^{n + \frac{1}{2}} - \bm{G}_{(i, j + \frac{1}{2}, k)}^{n + \frac{1}{2}} ] \\
&+ \frac{\Delta t}{\Delta z} [ \bm{H}_{(i, j, k - \frac{1}{2})}^{n + \frac{1}{2}} - \bm{H}_{(i, j, k + \frac{1}{2})}^{n + \frac{1}{2}} ] .
\end{aligned}
\label{eqn:simple_conserved_update}
\end{equation}
Here the superscript $n+1$ refers to the next time step, 
and $\bm{u}_{i, j, k}^{n+1}$ are the updated values of the conserved variables. The 
subscript $(i, j, k)$ refers to the three-dimensional Cartesian index of the cell. Indices 
that are displaced by half an integer refer to interfaces. For example, $(i - \frac{1}{2}, j, k)$ is 
the interface between cell $(i-1, j, k)$ and cell $(i, j, k)$.  The simulation time step is 
$\Delta t$, and $\Delta x$, $\Delta y$, and $\Delta z$ refer to the cell widths in each dimension. 
We use lowercase versions of $\bm{u}$ and $\bm{w}$ when referring to a cell-averaged quantity, and uppercase versions when referring to an estimated value at a cell edge (e.g., $\bm{U}$ and $\bm{W}$). The fluxes in Equation~\ref{eqn:simple_conserved_update} are averages in both space and time. Table~\ref{tab:notation} summarizes our notation.

\begin{deluxetable*}{c l}
\tablecolumns{2}
\tablewidth{0pt}
\tablecaption{Notation}
\tablehead{
\colhead{Symbol} & \colhead{Definition}}
\startdata
$\bm{u}^n$ & Conserved variable averages at time $n$ \\
$\bm{U}_L$, $\bm{U}_R$ & Reconstructed boundary values at the left and right of an interface \\
$\bm{U}^*_L$, $\bm{U}^*_R$ & Initial time-evolved boundary values \\
$\bm{F}^*$, $\bm{G}^*$, $\bm{H}^*$ & Initial one-dimensional fluxes \\
$\bm{U}^{n+\nicefrac{1}{2}}_L$, $\bm{U}^{n+\nicefrac{1}{2}}_R$ & Transverse-flux-evolved boundary values \\
$\bm{F}^{n+\nicefrac{1}{2}}$, $\bm{G}^{n+\nicefrac{1}{2}}$, $\bm{H}^{n+\nicefrac{1}{2}}$ & CTU fluxes \\
$\bm{u}^{n+1}$ & Updated conserved variable averages at time $n+1$ \\
\enddata
\label{tab:notation}
\end{deluxetable*}

When applied to every cell in the grid, Equation~\ref{eqn:simple_conserved_update} conserves each of the quantities in $\bm{u}$. However, the physical accuracy of a method based on Equation~\ref{eqn:simple_conserved_update} depends strongly on how the flux averages are calculated. Many hydrodynamics codes calculate the fluxes using only one-dimensional information as outlined in the method known as \citet{Strang68} splitting. While an elegant technique, Strang splitting may lead to asymmetries in hydrodynamics calculations and is not advantageous for some formulations of MHD \citep[see, e.g., the discussion in ][]{Balsara2004}. Therefore, we employ instead the unsplit CTU algorithm to improve on the one-dimensional calculation of fluxes by taking into account transverse fluxes that can cross cell interfaces in multi-dimensional simulations.

Before beginning a simulation the computational domain and boundaries must be initialized, as
described in Section~\ref{sec:code_architecture}. Once the fluid properties on the grid have 
been initialized, the first simulation time step, $\Delta t$, is calculated using the equation
\begin{equation}
\Delta t = C_0  \frac{\Delta x}{|u| + a},
\label{eqn:calc_dt_1D}
\end{equation}
where  $C_0$ is the Courant-Friedrichs-Lewy (CFL) number and $a$ is the average sound speed in the cell. In adiabatic hydrodynamics, the sound speed is a function of pressure and density
\begin{equation}
a = \sqrt{\gamma p / \rho},
\end{equation}
and can be calculated for each cell using the average values of the primitive variables. Thus, $|u| + a$ is the maximum wave speed in a cell with respect to the grid.

The minimum value of $\Delta t$ across the entire grid is determined and used in the CTU calculation, 
constraining the time step for every cell to be equal. In two or three dimensions, Equation \ref{eqn:calc_dt_1D} is modified such that the minimum required timestep is computed for each direction in the cell. In three dimensions, this
minimization is computed as 
\begin{equation}
\Delta t = C_0 \mathrm{min}\left(\frac{\Delta x}{|u| + a} \ , \frac{\Delta y}{|v| + a} \ , \frac{\Delta z}{|w| + a}\right).
\label{eqn:calc_dt_3D}
\end{equation}
With a suitable choice of $C_0$, Equation~\ref{eqn:calc_dt_3D} ensures that the Courant condition is satisfied in all three dimensions. Note that for the six-solve CTU algorithm the CFL number must be below $0.5$ for the solution 
to remain stable \citep{GS08}.

\begin{figure}
\centering
\includegraphics[width=1.0\linewidth]{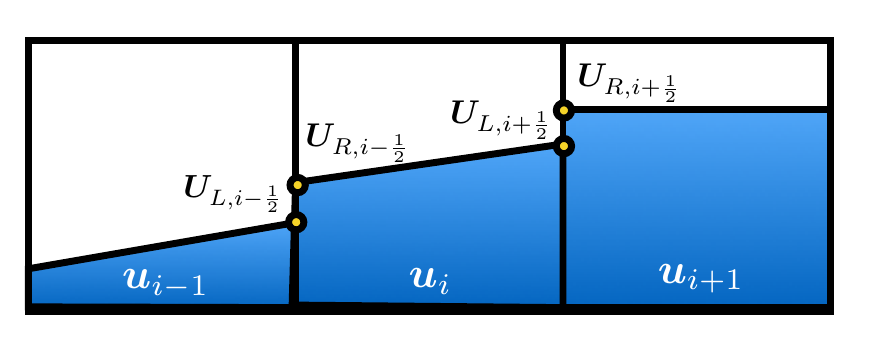}
\caption{Average values of the conserved variables, $\bm{u}$, are used to reconstruct boundary values on either side of each interface, $\bm{U}_L$ and $\bm{U}_R$, shown with circles. Shown is an example of piecewise linear reconstruction. The reconstructed boundary values are evolved in time to produce the initial time-evolved boundary values $\bm{U}^*_L$ and $\bm{U}^*_R$ (not shown). These $\bm{U}^*_L$ and $\bm{U}^*_R$ values are used as inputs to a Riemann problem whose
solution is used to compute fluxes across cell interfaces.}
\label{fig:reconstruction}
\end{figure}

Once we have calculated the timestep $\Delta t$ we carry out the following procedure:
\begin{enumerate}[leftmargin=*]
\item Reconstruct values of the conserved variables on both sides of every interface using the average value of the conserved quantities in adjacent cells. These reconstructed boundary values, denoted $\bm{U}_L$ and $\bm{U}_R$, represent an approximation to the true value of each conserved variable at the interface. For a multi-dimensional simulation, the reconstruction must be carried out in each dimension separately. We use additional subscripts to indicate which interface values we are calculating. For example, the left-hand reconstructed boundary value between cell $(i, j, k)$ and cell $(i+1, j, k)$ is denoted $\bm{U}_{L, (i+\frac{1}{2}, j, k)}$, while the right-hand value at that interface is $\bm{U}_{R, (i+\frac{1}{2}, j, k)}$. \cholla includes several different options for the interface reconstruction, which we describe in Section ~\ref{sec:reconstruction}. In order for the CTU algorithm to be second-order accurate in time, the reconstructed boundary values must be evolved half a time step before using them to calculate fluxes. The initial time evolution is considered part of the reconstruction, and is also inherently one-dimensional. We label the initial time-evolved boundary values $\bm{U}_{L}^{*}$ and $\bm{U}_{R}^{*}$. An example of a piecewise-linear reconstruction in the $x$-direction is shown in Figure~\ref{fig:reconstruction}.

\item Using the initial time-evolved boundary values as inputs, solve a Riemann problem at each cell interface in each 
direction. The solution to the Riemann problems yields a set of one-dimensional fluxes, $\bm{F}^{*}$, $\bm{G}^{*}$, and 
$\bm{H}^{*}$, corresponding to the $x$-, $y$-, and $z$-interfaces, respectively. Like the boundary value arrays, the
flux arrays contain five conserved value fluxes for each direction and interface. The Riemann solvers implemented 
in \cholla are described in Section~\ref{sec:riemannsolvers}.

\item Evolve the initial one-dimensional time-evolved boundary values half a time step using the transverse fluxes. For example, at the interface between cell $(i, j, k)$ and cell $(i+1, j, k)$ the transverse-flux-evolved boundary values are
\begin{equation}
\begin{aligned}
\bm{U}_{L, (i+\frac{1}{2}, j, k)}^{n+\frac{1}{2}} &= \bm{U}_{L, (i+\frac{1}{2}, j, k)}^{*} \\
  &+ \frac{1}{2}\frac{\Delta t}{\Delta y}[\bm{G}_{(i, j-\frac{1}{2}, k)}^{*} - \bm{G}_{(i, j+\frac{1}{2}, k)}^{*}] \\ 
  &+ \frac{1}{2}\frac{\Delta t}{\Delta z}[\bm{H}_{(i, j, k-\frac{1}{2})}^{*} - \bm{H}_{(i, j, k+\frac{1}{2})}^{*}], \\
\bm{U}_{R, (i+\frac{1}{2}, j, k)}^{n+\frac{1}{2}} &= \bm{U}_{R, (i+\frac{1}{2}, j, k)}^{*} \\
  &+ \frac{1}{2}\frac{\Delta t}{\Delta y}[\bm{G}_{(i+1, j-\frac{1}{2}, k)}^{*} - \bm{G}_{(i+1, j+\frac{1}{2}, k)}^{*}] \\ 
  &+ \frac{1}{2}\frac{\Delta t}{\Delta z}[\bm{H}_{(i+1, j, k-\frac{1}{2})}^{*} - \bm{H}_{(i+1, j, k+\frac{1}{2})}^{*}].
\end{aligned}
\label{eqn:transverse_evolution}
\end{equation}
For $y$ and $z$ interfaces, a cyclic permutation is applied. Thus, the $x$ interface states are evolved using the $y$ and $z$ fluxes, the $y$ interface states are evolved using the $x$ and $z$ fluxes, and the $z$ interface states are evolved using the $x$ and $y$ fluxes. This step is only relevant for multidimensional problems. In one dimension, the CTU algorithm reduces to just the initial reconstruction step, a flux calculation, and a final conserved quantity update.

\item Use the transverse-flux-evolved boundary values, $\bm{U}_L^{n+\frac{1}{2}}$ and $\bm{U}_R^{n+\frac{1}{2}}$, as inputs for a new set of Riemann problems. The solution to these Riemann problems yields the CTU fluxes $\bm{F}^{n+\frac{1}{2}}$, $\bm{G}^{n+\frac{1}{2}}$, and $\bm{H}^{n+\frac{1}{2}}$. Each flux is calculated using a one-dimensional Riemann problem at a single interface, but includes contributions from adjacent perpendicular interfaces as a result of the evolution in Equation~\ref{eqn:transverse_evolution}. The CTU fluxes are second-order accurate in time \citep{Colella90}.

\item Use the CTU fluxes to update the conserved quantities in each cell as in Equation~\ref{eqn:simple_conserved_update},
\begin{equation}
\begin{split}
\bm{u}_{(i, j, k)}^{n+1} &= \bm{u}_{(i, j, k)}^{n} \\
&+ \frac{\Delta t}{\Delta x} [\bm{F}_{(i-\frac{1}{2}, j, k)}^{n+\frac{1}{2}} - \bm{F}_{(i+\frac{1}{2}, j, k)}^{n+\frac{1}{2}}] \\
&+ \frac{\Delta t}{\Delta y} [\bm{G}_{(i, j-\frac{1}{2}, k)}^{n+\frac{1}{2}} - \bm{G}_{(i, j+\frac{1}{2}, k)}^{n+\frac{1}{2}}] \\
&+ \frac{\Delta t}{\Delta z} [\bm{H}_{(i, j, k-\frac{1}{2})}^{n+\frac{1}{2}} - \bm{H}_{(i, j, k+\frac{1}{2})}^{n+\frac{1}{2}}].
\end{split}
\label{eqn:conserved_update}
\end{equation}
The updated conserved quantities are used to calculate the next time step $\Delta t^{n+1}$.

\item Repeat the algorithm until the final simulation time is reached.

\end{enumerate}

\subsection{Interface Reconstruction}\label{sec:reconstruction}

\cholla currently implements five methods for cell interface reconstruction. These include the piecewise constant method (PCM), two versions of the piecewise linear method (PLM), and two versions of the piecewise parabolic method (PPM). Differences between the versions of piecewise linear and piecewise parabolic reconstruction are demonstrated in the tests presented in Section~\ref{sec:tests}. 
Access to multiple reconstruction options often proves useful, since lower-order methods are 
faster but higher-order methods are typically more accurate. Employing different versions of cell reconstruction also enables the impact of reconstruction on the evolution of a simulation to be quantified.
Here, we give a brief overview of each of the reconstruction techniques implemented in \cholla. Detailed descriptions of the piecewise linear and piecewise parabolic options can be found in Appendix~\ref{app:reconstruction}.

\subsubsection{Piecewise Constant Reconstruction}
The simplest reconstruction technique is the piecewise constant method \citep{Godunov59, Courant67}. In PCM, 
the initial time-evolved boundary values $\bm{U}_L^{*}$ and $\bm{U}_R^{*}$ are set equal to
the cell average quantities on either side of the interface, i.e.
\begin{equation}
\bm{U}_{R, (i-\frac{1}{2}, j, k)}^{*} = \bm{u}_{(i, j, k)}^{n},
\end{equation}
and
\begin{equation}
\bm{U}_{L, (i+\frac{1}{2}, j, k)}^{*} = \bm{u}_{(i, j, k)}^{n}.
\end{equation}
Note that in this notation, the boundary value at the \textit{right} of the interface is at the \textit{left} side of the cell, and vice versa. While the piecewise constant method is generally too diffusive for practical applications, it has merit for code testing, and is useful as a comparison to higher order reconstruction techniques.

\subsubsection{Piecewise Linear Reconstruction}

The second and third reconstruction techniques implemented in \cholla are both forms of the piecewise linear method, a scheme that is second-order accurate in space and time \citep[e.g.][]{Toro09}. The PLMP reconstruction method detailed below primarily involves the primitive variables, while the PLMC method subsequently explicated involves projecting the primitive variables onto wave characteristics.

PLMP follows the method outlined in Chapter 13.4 of \cite{Toro09}. First, the cell-average values of the primitive variables are used to calculate slopes of each variable across the left and right interface of each cell. We use the \citet{vanLeer79} limiter 
to monotonize the slopes, thereby reducing the likelihood of spurious oscillations in a numerical solution. 
The limited slopes are used to calculate reconstructed values of the primitive variables at the cell interfaces, $\bm{W}_L$ and $\bm{W}_R$. To convert these reconstructed boundary values into input states for the Riemann problem, we need to evolve them by half a time step. For PLMP, we do this by converting primitive quantities back into conserved variables and calculating the associated fluxes using Equations~\ref{eqn:x_flux}-\ref{eqn:z_flux}. We use these fluxes to evolve the reconstructed boundary values half a time step, generating the initial time-evolved boundary states for the first set of Riemann problems, $\bm{U}_L^*$ and $\bm{U}_R^*$.

The second linear reconstruction technique, PLMC, is based on the method outlined in \cite{Stone08}. This reconstruction also uses a linear approximation to model the distribution of the conserved quantities in each cell, but limits the slopes 
of the characteristic variables (rather than the primitive quantities) and evolves the reconstructed boundary values differently. Rather than simply evolving the boundary values using the associated fluxes as in PLMP, we employ the more sophisticated approach first described and termed ``characteristic tracing" in \cite{CW84}. We calculate a first approximation to the time-evolved boundary values by integrating under the linear interpolation function used to calculate $\bm{W}_L$ and $\bm{W}_R$. The domain of dependence of the reconstructed boundary value integral is defined by the minimum (for the left-hand interface) or maximum (for the right-hand interface) wave speed. This integration is then corrected by including the contribution from each of the other characteristics approaching the interface. Once the corrections have been made, the calculation provides the initial time-evolved boundary values $\bm{U}_L^*$ and $\bm{U}_R^*$ that act as input states for the first set of Riemann problems. This process is more fully described in Appendix~\ref{app:reconstruction}.

\subsubsection{Piecewise Parabolic Reconstruction}

The remaining two reconstruction techniques implemented in \cholla are both versions of the piecewise parabolic method (PPM) originally described in \cite{CW84}. We call the first method PPMP as it performs the reconstruction using primitive variables. Our PPMP implementation closely follows the FLASH code documentation \citep{Fryxell00}. 
The second method, abbreviated PPMC, uses an eigenvalue decomposition to project onto the characteristic variables 
and is based on the Athena code documentation \citep{Stone08}. 
Each PPM reconstruction method is described in detail in Appendix \ref{app:reconstruction}.

The approach to slope limiting differs slightly between the two parabolic reconstruction techniques. PPMP calculates slopes at each interface the same way as PLMP, using \citet{vanLeer79} limiting in the primitive variables. The slopes are limited in the characteristic variables for PPMC. In the parabolic methods, slopes are calculated using a stencil of five cells (two on either side of the cell for which we are calculating boundary values), which allows us to create a parabolic reconstruction of the primitive variables. This parabolic reconstruction makes PPM third-order accurate in space, though it remains only second-order accurate in time \citep{CW84}.

Several differences between the two parabolic reconstruction methods warrant further discussion. 
PPMP identifies and steepens the slopes near contact discontinuities, which results in a method 
that is less diffusive for contact waves. Downsides to contact steepening include the necessity of 
empirically determined criteria for selecting contact discontinuities and increased oscillatory behavior in the solution near shocks. In \cholla the ability to turn off contact steepening is retained, allowing an explicit comparison
of results obtained with and without the technique. PPMP also flattens shocks that have become too narrow to be treated accurately, which reduces the potential for severe post-shock oscillations. The more diffusive nature of PPMC renders a comparable correction unnecessary. Because the criteria for detecting a shock requires information from three cells on either side of an interface, the stencil for PPMP is larger than for PPMC. Both methods employ the characteristic tracing method of \cite{CW84} to translate from boundary extrapolated values based on the parabolic 
interpolation to input states for the Riemann problem, though the methods differ in detail 
(see Appendix~\ref{app:reconstruction}).

\subsection{Riemann Solvers}\label{sec:riemannsolvers}

\begin{figure}
\centering
\includegraphics[width=0.85\linewidth]{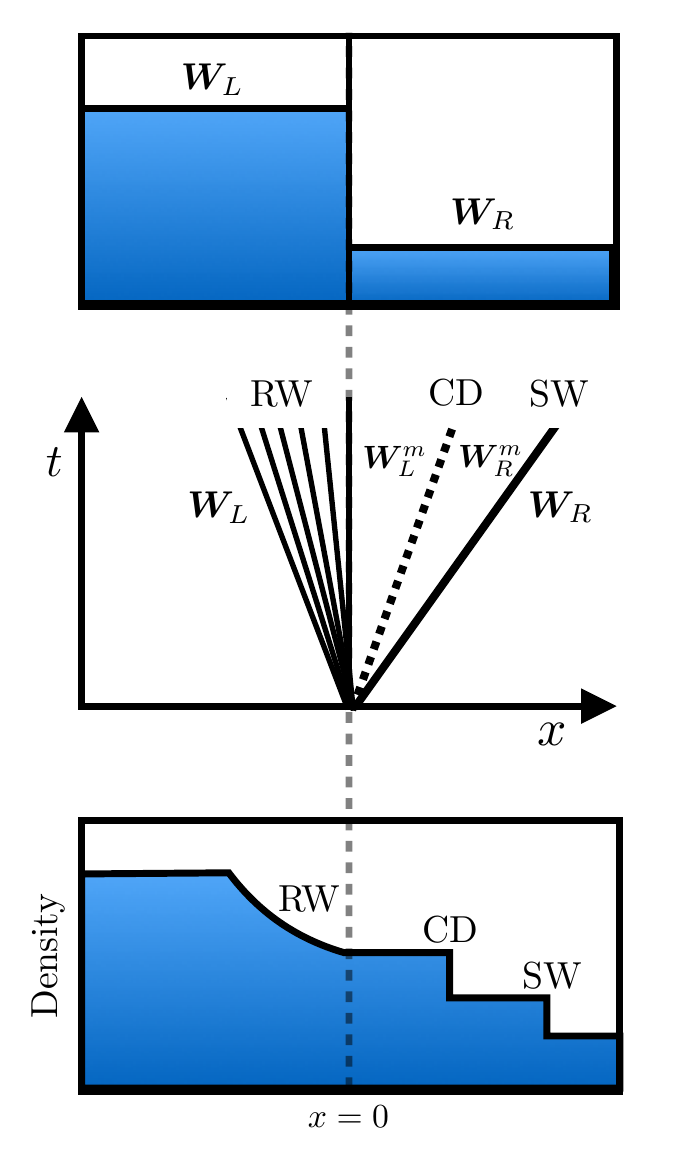}
\caption{An example Riemann problem. Top: Two initial states, $\bm{W}_L$ and $\bm{W}_R$ are separated by a 
discontinuity at $x = 0$. Middle: The solution to this Riemann problem displays three important features,
 consisting of a rarefaction wave (RW) expanding to the left, a contact discontinuity (CD) moving right, and a 
 shock wave (SW) moving right. Bottom: All three features can be seen in the solution for the density distribution.}
\label{fig:riemann_problem}
\end{figure}

Much effort has been devoted to finding efficient numerical algorithms to solve the Riemann problem \citep[e.g.,][]{Toro09},
an initial value problem consisting of two constant states separated by a jump discontinuity. As displayed in Figure~\ref{fig:riemann_problem}, the Riemann problem has an analytic solution that enables a numerical model for Eulerian hydrodynamics, as it allows for the calculation of the flux across an interface separating two initially discontinuous states. While Riemann solvers that calculate a numerical solution to the exact Riemann problem can be incorporated in hydrodynamics codes, the implicit nature of the Riemann solution requires an iterative step in the exact numerical solver. A large number of Riemann problems must be solved for every time step in a simulation, and the corresponding computational cost is substantial. As a result, a variety of approximate Riemann solvers have been engineered to quickly solve an approximation to the Euler equations at the expense of some physical accuracy.

\cholla computes numerical solutions to the Riemann problems on GPUs. Floating point operations are performed very efficiently on a GPU; additional factors like memory latency and data transfer contribute a larger share of the computational expense of the method. Thus, adding the extra operations needed for the iterative procedure in an exact solver versus an approximate one does not impact 
the performance speed of \cholla in the same way as for a CPU-based code. However, there are certain problems where the extra diffusion in an approximate solver is helpful, for example to deal with the well known carbuncle instability \citep{Quirk94} that affects grid-aligned shocks. For this reason, \cholla includes both an exact solver and the linearized solver first described by 
\citet{Roe81} that gives an exact solution to a linear approximation of the Euler equations. Detailed descriptions of our implementation of both solvers can be found in Appendix~\ref{app:riemann_solvers}.

\subsubsection{The Exact Solver}

\cholla implements the exact Riemann solver presented in \cite{Toro09}. The solver 
uses a Newton-Raphson iteration to calculate the pressure of the gas in the intermediate state $\bm{W}^m$ of
the Riemann solution that lies
between the initial states on the left and right of the interface, as
shown in Figure~\ref{fig:riemann_problem}. 
Once the pressure in the intermediate state has been found, the exact solution for the primitive variables between the left and right initial states can be calculated explicitly at any later point in time. The pressure and velocity are used to determine the solution at the cell interface, and the values of the primitive variables at that point are used to calculate the fluxes of conserved variables at the interface according to Equations~\ref{eqn:x_flux} - \ref{eqn:z_flux}. 
Transverse velocities are passively advected as scalar quantities.

The Toro Riemann solver gives a numerically exact solution to the Riemann problem in one dimension, and will never return negative densities or pressures if the input states are physically self-consistent. However, the input states on either side of the cell interface are estimated quantities, and because of the extrapolation involved in the reconstruction techniques they could be physically invalid.
In these situations, the solver may be presented with an initial value problem without a physically valid solution. 
To prevent artificial vacuum or negative pressure solutions owing to such a
circumstance, a pressure floor of $10^{-20}$ in the adopted unit scheme is enforced in \cholla. In practice, when using an exact Riemann solver the pressure floor has proved necessary only when performing 
the Noh test described in Section~\ref{sec:tests}.

\subsubsection{The Roe Solver}

One common alternative to calculating an exact solution to the Riemann problem is to linearize the non-linear conservations laws and solve the resulting approximate problem exactly. In one dimension, the non-linear Euler equations can be replaced with the following linearized equation
\begin{equation}
\frac{\delta \bm{u}}{\delta t} + \mathbf{A}(\bm{\tilde u})\frac{\delta \bm{u}}{\delta x} = 0,
\end{equation}
where $\textbf{A}$ is a constant Jacobian evaluated at 
some average state $\bm{\tilde u}$ that is a function of the initial states on either side of the cell interface. This method was employed by \cite{Roe81}, and \cholla includes a linearized solver very similar to the original Roe solver. 

The first step in the Roe solver is to calculate the average state $\bm{\tilde u}$. This average state, along with the eigenvalues, $\lambda^{\alpha}$, and left and right eigenvectors of the Jacobian $\mathbf{A}$, 
$\mathbf{L^{\alpha}}$ and $\mathbf{R^{\alpha}}$, can be used to calculate the Roe fluxes at the interface:
\begin{equation}
\bm{F}_{\mathrm{Roe}} = \frac{1}{2} \left( \bm{F_L} + \bm{F_R} + \sum\limits_{\alpha=1}^m \xi^{\alpha} | \lambda^{\alpha} | \mathbf{R^{\alpha}} \right).
\label{eqn:Roe_fluxes}
\end{equation}
Here, $\alpha = 1, m$ are the $m$ characteristics of the solution, and
\begin{equation}
\xi^{\alpha} = \mathbf{L^{\alpha}} \cdot \delta\bm{U}
\label{eqn:Roe_wave_strengths}
\end{equation}
are the characteristic variables, determined by projecting the differences in the initial left and 
right states, $\delta \bm{U} = \bm{U}_R - \bm{U}_L$, onto the left eigenvectors. $\bm{F}_L$ and $\bm{F}_R$ are fluxes calculated with the left and right input states using Equation~\ref{eqn:x_flux}. Expressions for the average state, $\bm{\tilde u}$, as well as the eigenvalues and eigenvectors are given in \cite{Roe81} and Appendix~\ref{app:riemann_solvers}. 
The matrix $\mathbf{A}$ is not actually needed in the calculation.

As pointed out by \cite{Einfeldt91}, there are certain Riemann problems that will cause any 
linearized solver to fail. In these cases, the linearized solution to the Riemann problem results in negative densities or pressures in the intermediate state calculated between the left and right input states. Because this intermediate state is used to calculate the fluxes returned by the solver, these unphysical solutions may
lead to numerical pathologies. A failsafe is needed to deal with the case where the Roe solver 
produces negative pressures or densities. Following the method of \cite{Stone08}, we check the intermediate densities and pressures before returning the fluxes calculated with the Roe solver. Should any of them be negative, we revert to using the simpler HLLE Riemann solver, described below.

\subsubsection{The HLLE Solver}

The HLLE solver is a modification of the HLL solver first described by \citet{Harten83} and later 
modified by \citet{Einfeldt88}. Although the method is extremely diffusive for contact discontinuities, as demonstrated by \citet{Einfeldt91} the HLLE solver is guaranteed to be positively conservative (that is, the density and internal energy remain positive). 
The HLLE solver calculates the interface flux using an average of the left and right state fluxes, together with bounding speeds comprising the largest and smallest physical signal velocities in the solution to the exact Riemann problem. If the Roe solver produces negative densities or pressures, we replace the Roe fluxes with a new numerical flux
\begin{equation}
\bm{F}_{\mathrm{HLLE}} = \frac{b^p \bm{F}_L - b^m \bm{F}_R}{b^p - b^m} + \frac{b^p b^m}{b^p - b^m}\delta\bm{U}.
\end{equation}
The fluxes $\bm{F}_L$ and $\bm{F}_R$, and the slopes $\delta\bm{U}$ are calculated as in the Roe solver. The signal velocities $b^p$ and $b^m$ are calculated using the largest and smallest eigenvalues of the Roe matrix as described in Appendix~\ref{app:riemann_solvers}. Because the HLLE solver quickly allows contact discontinuities to diffuse, we do not use it as a standalone Riemann solver in \chollans.


\section{Code Architecture}\label{sec:code_architecture}

\cholla is a grid-based hydrodynamics code that takes advantage of the massively 
parallel computing power of GPUs. In order to harness this power, \cholla was designed with the operation of the GPU in mind.  In this section, we describe the overall structure of \chollans, including optimization strategies necessary to benefit from the parallel architecture of GPUs. As is standard in GPU programming, we will use the term ``host" to refer to the CPU, and ``device" to refer to the GPU.

\cholla consists of a set of C/C++ routines that run on the host plus functions called \textit{kernels} that execute on a device. The device kernels and the host functions that call them are written in CUDA C, an extension to the C language introduced by \nvidians\footnote{\tt http://developer.nvidia.com}. All of the CUDA functions are contained in a separate hydro module so that they can be compiled independently with the \nvidia \nvcc compiler. In addition, we have written a C/C++ version of the hydro module that performs the same calculations as all of the GPU kernels, so it is possible to run \cholla without using graphics cards. We use this mode for testing, but it is not recommended for 
performance since the structure of the code is optimized for use with GPUs.

\subsection{Simulation Overview}

Before detailing each piece of the code, we give a general overview of the steps followed by \cholla when a simulation is run. Given the power of a single GPU, small problems can easily be run on a single host/device pair. For large problems, \cholla can be run using the MPI library, and we describe our MPI implementation in Section~\ref{sec:MPI}. If MPI is enabled, the simulation volume will be split into subvolumes according to the number of processes. Each subvolume will then be treated as a self-contained simulation volume for the duration of each simulation time step. The main difference between an MPI and non-MPI simulation is the method for applying boundary conditions at the end of each time step; we describe that method in Section~\ref{sec:MPI}.

\begin{figure}
\centering
\includegraphics[width=1.0\linewidth]{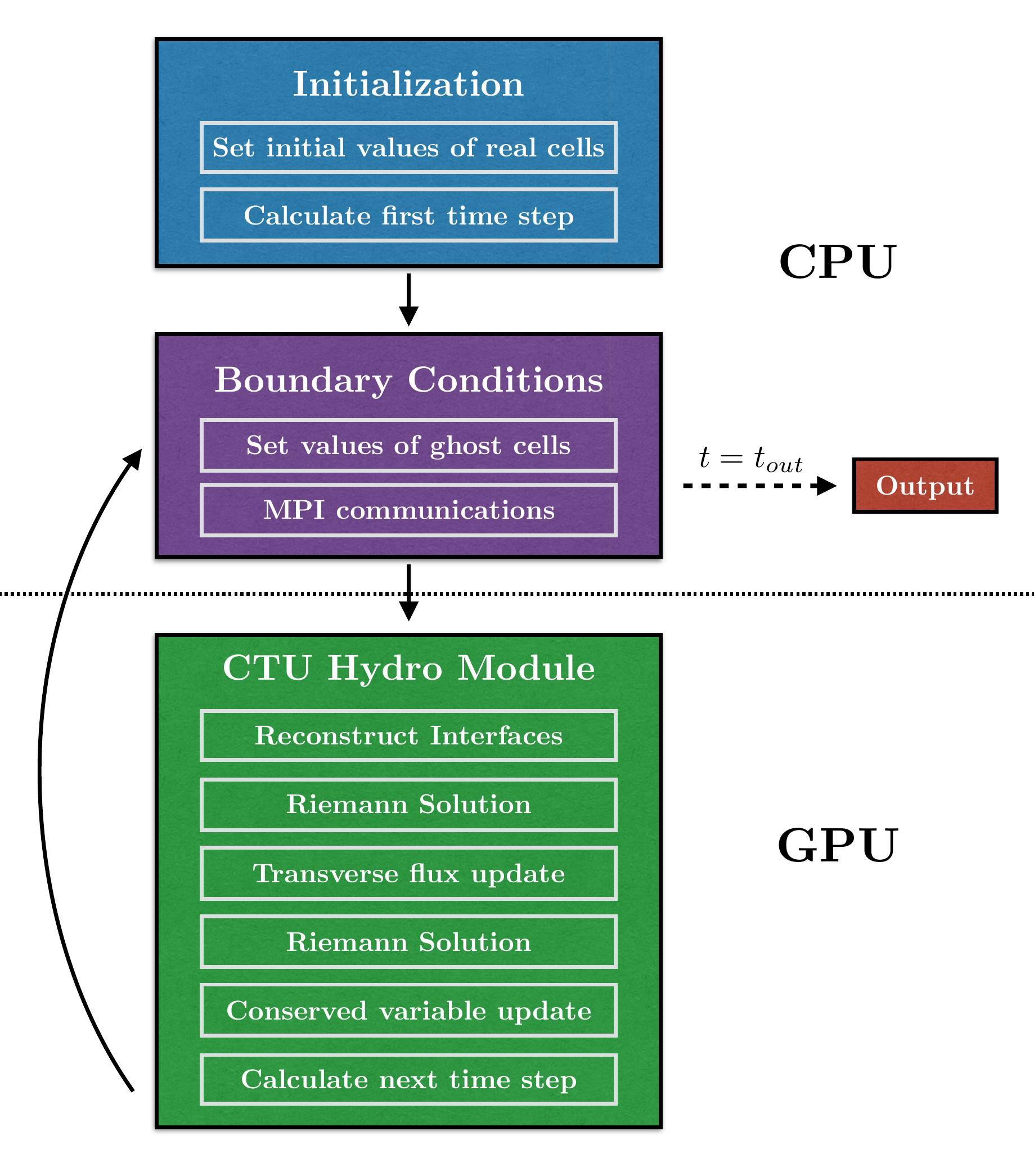}
\caption{Algorithmic procedure of a \cholla simulation. The initialization and application of boundary conditions are done on the CPU. The conserved variable array is passed to the GPU, where the hydro calculation is done. The updated conserved variables must then be passed back to the CPU after each time step so that boundary cell information can be exchanged and the data output written (if necessary).}
\label{fig:cholla_diagram}
\end{figure}

Portions of our algorithm that require information from potentially distant cells in the global simulation volume must be carried out on the host. The main host functions set initial conditions, apply boundary conditions, and perform any interprocess communications. 
Parts of the calculation that only require information from nearby cells can be carried out on the device. Because the bulk of the computational work resides in the CTU calculation that requires a stencil containing only local cells, essentially all of the hydrodynamical computations are performed on the GPU.

The steps in the \cholla algorithm are listed below and illustrated in Figure~\ref{fig:cholla_diagram}.

\begin{enumerate}[leftmargin=*]
\item Initialize the simulation by setting the values of the conserved fluid quantities for all cells in the simulation volume, and calculate the first time step.
\item Transfer the array $\bm{u}$ of conserved variables 
to the GPU. The conserved variable array contains the values of each conserved quantity for every cell in the simulation volume.
\item Perform the CTU calculation on the GPU, including 
updating the conserved variable array and computing the next time step.
\item Transfer the updated conserved variable array back to the CPU.
\item Apply the boundary conditions. When running an MPI simulation, this step may
require interprocess communication to exchange information for cells at the edges of subvolumes.
\item Output simulation data if desired.
\end{enumerate}

The initialization of the simulation is carried out on the host. The initialization includes setting the values of the conserved variables for both the real and the ghost cells according to the conditions specified in a text input file. Ghost cells are a buffer of cells added to the boundaries of a simulation volume to calculate fluxes for real cells near the edges. The number of ghost cells reflects the size of the local stencil used to perform fluid reconstruction. Because updating the ghost cells at each time step may require information from cells that are not local in memory, the values of the ghost cells are set on the host before transferring data to the GPU.

Once the simulation volume has been initialized on the CPU, the hydrodynamical calculation begins. The host copies the conserved variable array onto the device. Because the GPU has less memory than the CPU, the conserved variable array associated with a single CPU may be too large to fit into the GPU memory at once. If so, \cholla uses a series of splitting routines described in Section \ref{sec:subgrid_splitting} to copy smaller pieces of the simulation onto the GPU and carries out the hydrodynamics calculations on each subvolume. At the end of the hydro calculation the next time step is calculated on the device using a GPU-accelerated parallel reduction. The updated conserved variables and new time step are then transferred back to the host. The host updates the values of the ghost cells using the newly calculated values of the real cells, and Steps 2 - 5 repeat until the desired final simulation time is reached.

After each time step the values of the ghost cells are reset using the newly updated values of the conserved variables. \cholla includes three standard boundary conditions: periodic, reflective, and transmissive. These can be set in any combination on any of the borders of the simulation volume. For periodic and transmissive boundaries, the conserved variable values of each ghost cell are copied from the appropriate real cell. For reflective boundaries we follow the same process but reverse the sign of 
the perpendicular component of momentum. \cholla also includes the capability to define custom boundary conditions, such as the analytic boundaries specified in the Noh Strong Shock test (see Section~\ref{sec:noh3D}).  In a simulation performed using MPI communication, any necessary boundary regions are exchanged between relevant processes 
as described in Section~\ref{sec:MPI}.

\subsection{Memory Structure}\label{sec:memory_structure}

The data for a simulation in \cholla are contained in two structures. A header stores information about the size and shape of the domain, as well as global variables including the simulation time. A second structure contains the values of the conserved variables for each cell in the simulation. In an object oriented programming model, these values would often be stored in memory as an array of structures,
\begin{gather*}
\mathrm{Cell}[0].\{ \rho \ \rho u \ \rho v \ \rho w \ E \}, \\
\vdots \\
\mathrm{Cell}[N-1].\{ \rho \ \rho u \ \rho v \ \rho w \ E \},
\end{gather*}
where $N$ is the total number of cells in the grid. In a CPU-based simulation code, this configuration can improve the performance of memory accesses.

The object oriented model is intuitive, but the memory structure is not efficient when implemented on the GPU. On NVIDIA GPUs, calculations are performed simultaneously by thousands of individual computational elements called {\it cores}, analogous to but individually much less powerful than a typical CPU core. The set of instructions carried out on a single GPU core is called a \textit{thread}. The efficiency of the GPU comes in part from its ability to efficiently schedule the execution of millions of threads requested in a single kernel call. The scheduling is organized by {\it streaming multiprocessors} on the device that schedule threads for execution in groups called \textit{warps}. Each thread warp performs a given set of operations simultaneously in the execution model often referred to as Single Instruction Multiple Data (SIMD). Given that data operations across cores on the GPU are rapidly executed in a massively parallel manner via the SIMD approach, hardware timescales such as the GPU global memory access time can represent a considerable fraction of the total computational expense of a calculation. Techniques to reduce the expense of global memory accesses include the organization of data needed by each thread warp into adjacent regions in physical memory. To facilitate this advantageous data locality, \cholla organizes conserved variables into a structure of arrays: 
\begin{gather*}
\{ \rho_0 \ ... \ \rho_{N-1} \}, \\
\{ \rho u_{0} \ ... \ \rho u_{N-1} \}, \\
\{ \rho v_{0} \ ... \ \rho v_{N-1} \}, \\
\{ \rho w_{0} \ ... \ \rho w_{N-1} \}, \\
\{ E_0 \ ... \ E_{N-1} \}.
\end{gather*}
The thread warps can retrieve the conserved quantities within this structure of arrays in global memory with unit stride in memory accesses, reducing collisions in the access pattern.

The process of initiating a data transfer from the host to the device involves an associated computational overhead. Limiting the number of transfers required by the algorithm mitigates this overhead, as hardware latency may
cause many small transfers to
take longer than one large transfer. The allocation of a single structure containing the conserved variable arrays ensures a contiguous data layout in memory, and limits the required data transfers to two (single transfers to and from the device).

\subsection{The GPU Grid}\label{sec:gpu_grid}

\begin{figure}
\centering
\includegraphics[width=1.0\linewidth]{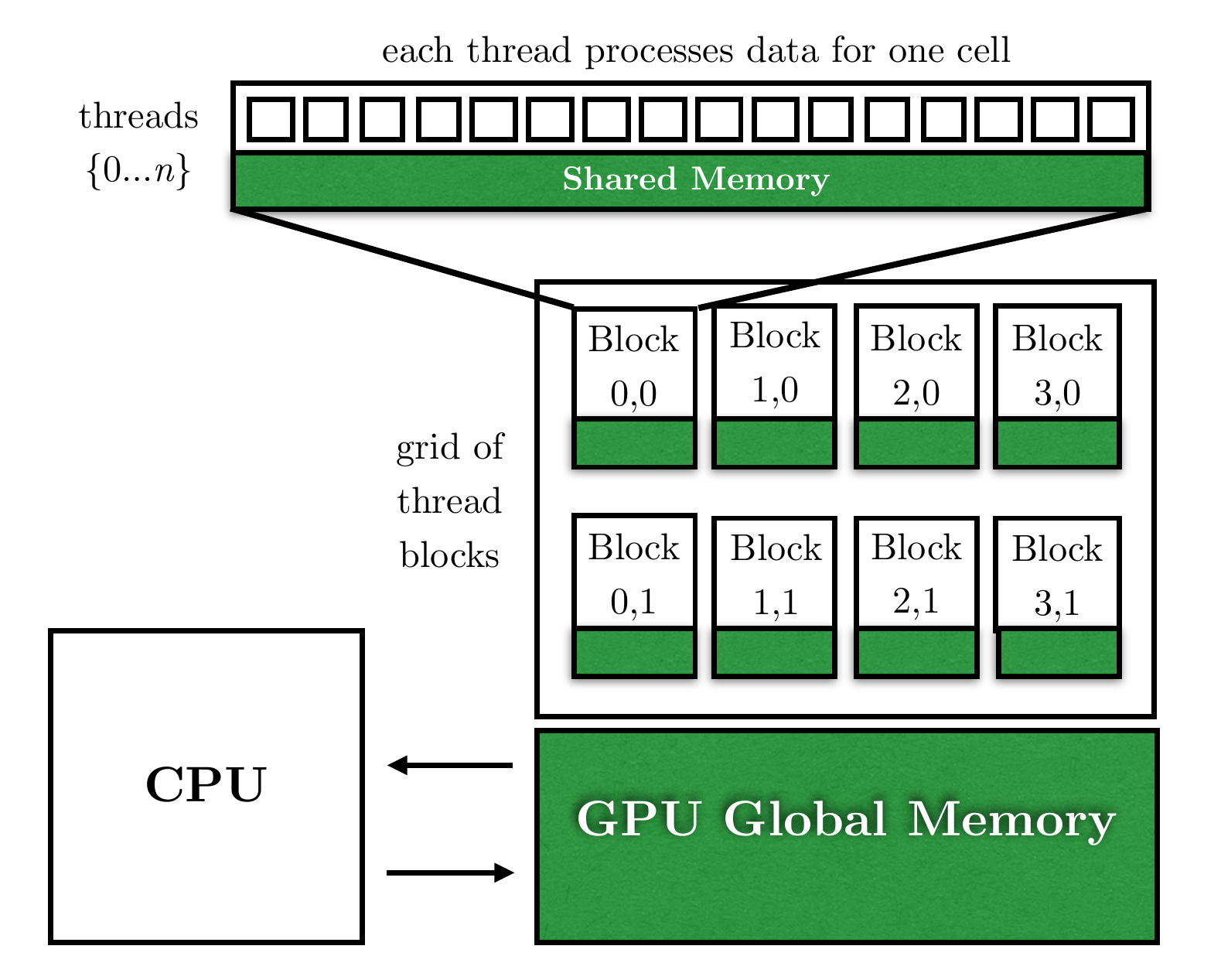}
\caption{\cholla memory structure. After the conserved variable array is copied from the CPU into global memory on the GPU,
the GPU initializes a grid of one dimensional thread blocks with set numbers of GPU threads. Each thread then calculates the information for a single grid cell. All threads can access global memory, but only threads in the same block have access to the much smaller amount of per-block shared memory.}
\label{fig:cpu_gpu}
\end{figure}

Once the simulation volume has been initialized on the host and the values of the ghost cells have been set, the array of conserved variables is transferred to global memory on the GPU. When kernels are then executed on the device, the GPU launches a {\it grid} of {\it thread blocks}. The GPU grid and thread block dimensions are set by the programmer and are application dependent. \cholla typically uses one or two dimensional grids of one dimensional thread blocks; the latter arrangement is illustrated in Figure~\ref{fig:cpu_gpu}. We emphasize that the dimensions of the GPU grid are not constrained to match the dimensions of the simulation, as the location of a cell in the simulation volume can always be mapped to a unique index within the GPU grid of thread blocks.

The dimensions of the GPU grid can affect the efficiency with which the device performs calculations and dictate the mapping from a real-space cell index to a thread index. To define the thread index, the CUDA programming model includes built-in data elements that return specific values for each thread, as shown in the following pseudo-code:
\begin{displaymath}
\texttt{tid = threadIdx.x + blockIdx.x * blockDim.x}.
\end{displaymath}
Here, \texttt{threadId} returns the ID of the thread within the block, \texttt{blockIdx} returns the ID of the thread block within the grid, and \texttt{blockDim} returns the dimensions of the block. By combining these pre-defined quantities, a unique global index can be calculated for each thread. This pseudo-code assumes a one dimensional grid of one dimensional blocks, but could easily be adjusted to create an equivalent mapping for two or three dimensional blocks or grids. We use the thread index to assign each thread the work of computing the conserved variable update for a single cell in the simulation. For \chollans, we choose a one dimensional block of threads because most of the kernels are one dimensional in nature. The PPM reconstruction, for example, requires only a one dimensional stencil and is carried out separately for the $x$, $y$, and $z$ interfaces. Because a new GPU grid with different dimensions can be initiated every time a device function is called, the thread index calculation and subsequent mapping to a real cell index must be performed within each GPU kernel. No data needs to be transferred back to the CPU between kernels, as all information needed between kernels is stored in the GPU global memory.

\subsection{The GPU Kernels}

\cholla leverages a modular design that enables an easy selection of the reconstruction method or Riemann solver, and
facilitates the incorporation of new features. Each reconstruction method and Riemann solver is performed through an associated kernel executed by the GPU. A routine implementing the CTU algorithm calls these kernels through a wrapper function that segregates CUDA calls from the rest of the code. This organization allows for a flexible compilation structure in which non-CUDA code (including MPI calls) can be compiled with standard C compilers. Within the CUDA wrapper for the CTU algorithm, the following steps are followed:
\begin{enumerate}[leftmargin=*]
\item Allocate arrays within GPU global memory to hold the conserved variables $\bm{u}$, the initial time-evolved boundary values $\bm{U}^*$, the initial one-dimensional fluxes $\bm{F}^*$, the transverse-flux-evolved boundary values $\bm{U}^{n+\frac{1}{2}}$, and the CTU fluxes $\bm{F}^{n+\frac{1}{2}}$.
\item Transfer the conserved variable data from the host to the device and store in the newly allocated arrays in GPU global memory.
\item Call the reconstruction kernel for each dimension.
\item Call the Riemann solver kernel for each dimension.
\item Call the kernel to perform the transverse flux update (Equation~\ref{eqn:transverse_evolution}).
\item Call the Riemann solver kernel for each dimension again. 
\item Call the kernel to update the conserved variables and calculate the next time step (Equation~\ref{eqn:conserved_update}).
\item Transfer the conserved variable arrays back to the CPU.
\end{enumerate}
Step 1 involves the allocation of memory on the GPU, and it should be noted that 
the global GPU memory available is typically small compared with the CPU memory. Depending on the device, the simulation size, the number of MPI processes, and the domain decomposition, each process's conserved variable data may exceed the available GPU memory. An excess may occur even if the local grid governed by each process is small (e.g., $128^3$). When necessary, \cholla uses a set of splitting routines to divide the simulation volume into more manageable subvolumes that are then treated according to the steps listed above. Section~\ref{sec:subgrid_splitting} describes these splitting routines in more detail.

The GPU kernel calls in Steps 3-7 resemble traditional C function calls, 
but kernels are implemented with additional variables that establish the dimensions of the grid of thread blocks launched by the GPU. For example, the syntax for calling \chollans's PPM reconstruction function is: 
\begin{gather*}
\texttt{PPM\_reconstruction<<<BlocksPerGrid,} \\
\texttt{ThreadsPerBlock>>>(<function\_parameters>)},
\end{gather*}
where the triple chevron syntax, \texttt{<<<,>>>}, informs the CUDA-enabled compiler that this function should be executed
on the GPU device. Since the amount of data processed at once by the GPU is limited by its available global memory, \texttt{BlocksPerGrid} can always be set large enough to assign a thread to each cell. 

Separate kernels carry out different parts of the CTU algorithm, but each kernel shares common elements. Every kernel must begin with a calculation of the index of each thread, as described in Section~\ref{sec:gpu_grid}. Using the appropriate mapping, the index of each thread of the kernel can be translated to a unique real-space cell index in the simulation volume. The threads within the kernel then retrieve necessary cell data
from the GPU global memory. For the reconstruction function, these data 
would include the values of the conserved variables for the cell assigned to that thread, as well as those of the nearby cells within the reconstruction stencil. Once the data have 
been retrieved, the threads carry out any relevant calculations, 
and load the result into the relevant GPU global memory array. Once all of the threads have finished 
their calculations the kernel returns, and the process continues through each of the steps listed above.

\subsection{Time Step Calculation}\label{sec:timestep}

The implementation of most kernels described in the previous section (reconstruction, Riemann solver, and transverse flux update) follows closely the descriptions of the CTU calculation given in Section~\ref{sec:hydrodynamics}. However, the final conserved variable update kernel in each iteration of the algorithm is extended to include the calculation of the next simulation time step $\Delta t^{n+1}$ via a parallel reduction operation performed on the GPU. Reductions on the GPU are a commonly used process, and examples of this operation can be found in e.g., the CUDA toolkit\footnote{{\tt https://developer.nvidia.com/cuda-toolkit}}. We include a brief explanation of our implementation of the parallel reduction operation here as a concrete example of the advantage of moving a given function from the CPU to the GPU.

Thread blocks on the GPU have a limited amount of ``shared memory" that each thread in the block can access rapidly, as illustrated in Figure~\ref{fig:cpu_gpu}. At the end of the conserved variable computation, the updated conserved variable data for each cell are stored in the private register memory assigned to each thread. The updated values of the conserved variables are used by each thread to calculate the minimum time step associated with its cell, according to Equation~\ref{eqn:calc_dt_3D}. The individually calculated time steps are then loaded into an array the size of the thread block in shared memory - note that each thread block has its own array. The threads in the block then perform a tree-based parallel reduction on the time step array, finding the minimum time step for the entire block. This minimum value is uploaded into an array in the GPU global memory, and is then passed back to the CPU, where the final reduction is performed.

Calculating the time step on the GPU achieves a performance gain relative to a CPU since executing a large number of floating point operations is extremely efficient on the GPU. The shared memory reduction on the GPU reduces the number of loops needed on the CPU by a factor of \texttt{ThreadsPerBlock} (typically set to $128$ for an \nvidia Kepler K20X GPU). For reference, we find the parallel GPU reduction time step calculation for a $1920\times1080$ simulation can achieve a $100\times$ performance gain relative to a single CPU core depending on the architecture ($\sim1\unit{ms}$ vs. $\sim100\unit{ms}$).

\subsection{Subgrid Splitting}\label{sec:subgrid_splitting}

As mentioned in previous sections, the total amount of memory on a single device may be quite limited when compared to the memory available on the host. The memory footprint on the GPU for each cell in the simulation volume is of order $0.5$ kilobytes, including conserved variables, interface values, and fluxes. At present, a typical GPU has only a few gigabytes of global memory, though this number has been increasing with each new generation of devices. Therefore, current devices can typically only hold the information for a 3D hydrodynamical simulation of size $\sim 228^3$. In \chollans, slightly more cells can fit for 1D and 2D simulations owing to the reduced number of transverse interface states and fluxes that must be stored. The limited memory resources often require that the simulation volume associated with a local process may need to be successively subdivided to fit on a GPU. We term this subdivision process ``subgrid splitting". Similar methods have been used in CPU-based codes such as HERACLES \citep{Gonzales07}.

In practice, subgrid splitting is typically only needed for multidimensional simulations or 1D simulations with millions of cells. A description of the 1D subgrid splitting is provided below as a straightforward example. First, the size of the simulation volume that will fit on the GPU at once given the global memory available is calculated and stored in a variable, e.g. \texttt{MaxVol}. The local volume governed by each local process is further split into subvolumes of size less than or equal to \texttt{MaxVol}. We refer to these subvolumes as ``subgrid blocks''. The CTU calculations for each subgrid block are performed on the GPU sequentially, \textit{including any necessary ghost cells from nearby subvolumes}. Memory buffers on the CPU are used to avoid overwriting grid
cells that act as ghost cells for neighboring subgrid blocks. The procedure of copying data to the GPU, calculating, and transferring data back to the CPU is repeated until the hydro step for the entire simulation volume has completed. Because the conserved variables for the simulation are contiguous in memory on the host, copying them into buffers via \texttt{memcpy} contributes a negligible amount to the total simulation run time.

\begin{figure}
\centering
\includegraphics[width=1.0\linewidth]{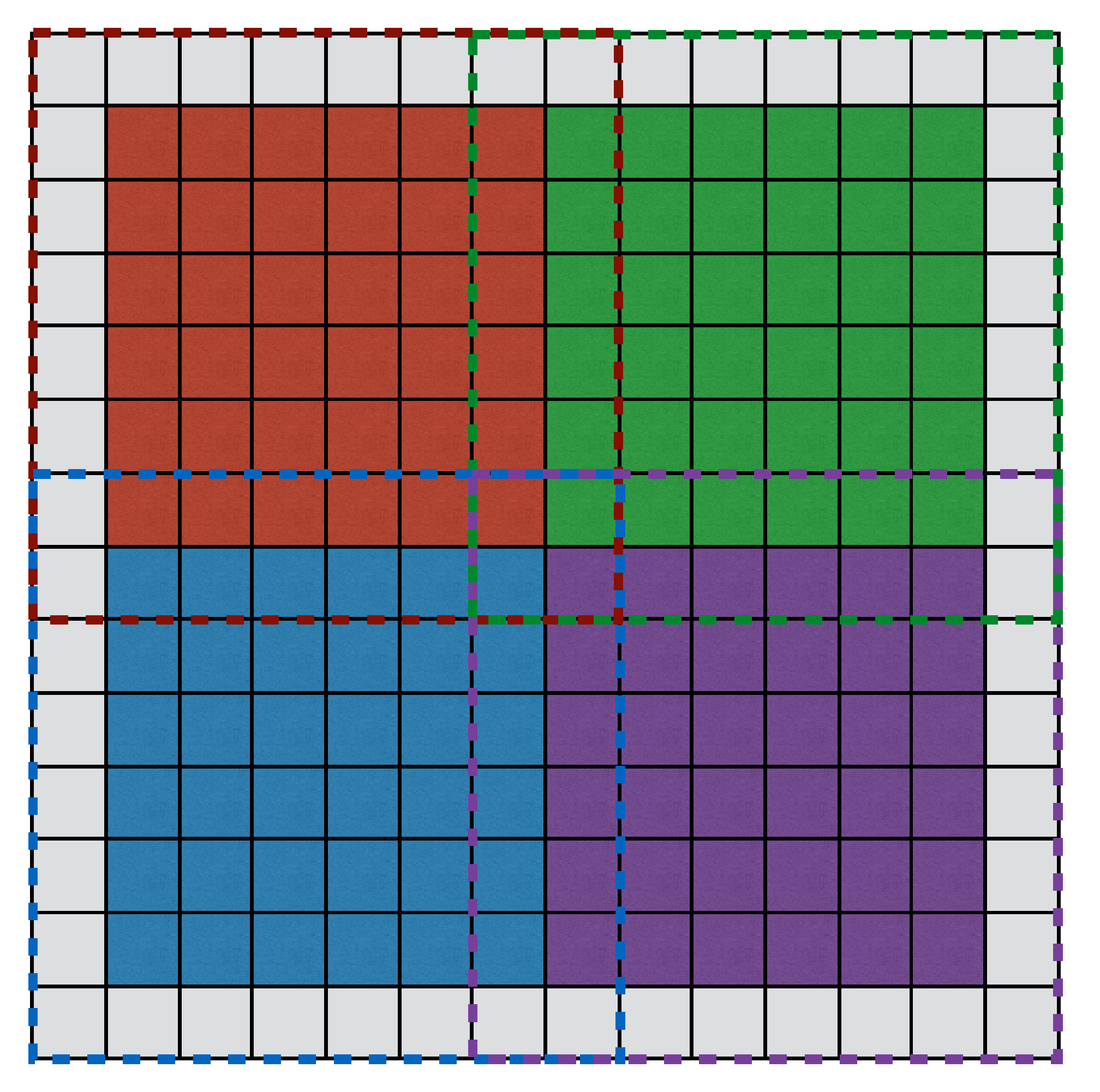}
\caption{The well-known ``ghost cell pattern'' as applied to the subgrid blocking algorithm in \cholla. When the total area of a 2D simulation is too large to fit in global memory on the GPU, the simulation volume must be split into smaller subgrid blocks for GPU computation of the hydrodynamical calculation. When copying a subgrid block of the simulation onto the GPU, memory buffers are utilized such that each subgrid block can be copied to the device and the conserved variables updated without overwriting real cell data that will be needed as ghost cells for neighboring subgrid blocks. For this illustration, ghost cells on the global outer boundary of the simulation are shown in gray. The dashed lines outline the cells needed to perform the CTU calculation for each colored subgrid block. See e.g., Figure 9 of \cite{Kjolstad10}.}
\label{fig:subgrid_splitting}
\end{figure}

To illustrate the subgrid blocking method, Figure~\ref{fig:subgrid_splitting} displays a two-dimensional grid with an example subgrid blocking by a factor of four. Each of the four subgrid blocks (red, green, blue, purple) require real cells from adjacent subgrid blocks to act as {\it subgrid ghost cells} to form the full computational stencil for the fluid reconstruction and CTU calculations (indicated
by colored dashed lines). Since these subgrid blocks also abut either local simulation boundaries between local 
processes or global boundaries at the edges of the illustrated region, they also require standard ghost cells (gray regions) to complete their computational stencils. The subgrid block regions outlined by the dashed lines are transferred sequentially to the GPU for the CTU calculation, taking care to preserve in memory the subgrid ghost cell boundary regions between subgrid blocks until the entire local volume has been processed. The standard ghost cells are updated after the CTU calculation, since they depend either on communication between MPI processes or the global boundary conditions of the simulation. Note that Figure~\ref{fig:subgrid_splitting} illustrates a very small grid for convenience, and the actual subgrid regions of a 2D simulation would be orders of magnitude larger.

We performed a variety of tests using subgrid blocks of different shapes in order to determine a method of division that helps minimize GPU communication overhead. For 2D simulations, splitting the volume into approximately square regions works well. For 3D simulations, we find that maintaining an aspect ratio that is approximately even in the $y$ and $z$ directions with a longer aspect ratio in the $x$-direction works well. In practice, we keep the aspect ratio even in $y$ and $z$ while maintaining an aspect ratio in $x$ that is roughly 5 times the geometric mean of the $y$- and $z$-aspect ratios. Memory access overheads can be reduced by first
copying multiple subgrid blocks into buffers on the CPU, and then transferring subarrays containing individual subgrid blocks to the GPU for computation. Even in simulations where local volumes must be subdivided into subgrid blocks dozens of times the overhead associated with copying the conserved variables into buffers on the CPU is insignificant, typically limited to $5\%$ of the total time taken for the CTU calculation.

Transferring data to and from the GPU at each time step is time consuming, often taking $\sim 30\%$ of the entire GPU computation time for the hydro module. Therefore, strategies that reduce the fraction of the simulation volume transferred at each time step are desirable. For example, simulations that do not require subgrid splitting might achieve a performance boost by only transferring ghost cells that need to be updated via MPI. Such a strategy is beyond the scope of the current work, but is certainly worth exploring in future versions of \chollans.

\subsection{MPI Implementation and Scaling}\label{sec:MPI}

The massively parallel algorithm implemented by \cholla can be adapted to execute on multiple GPUs simultaneously. \cholla can thereby gain a multiplex advantage beyond the significant computation power afforded by a single GPU. The parallelization is implemented using the MPI library. The global simulation volume is decomposed into subvolumes, and the subvolumes are each assigned a single MPI process. In \chollans, each MPI process runs on a single CPU that has a single associated GPU, such that the number of MPI processes, CPUs, and GPUs are always equal. When the simulation volume is initialized, each process is assigned its simulation subvolume and surrounding ghost cells. Since the hydrodynamical calculation for every cell is localized to a finite stencil, only the ghost cells on the boundary of the volume may require updating from other processes via MPI communication every time step. Compared with a simulation done on a single CPU/GPU pair, additional overheads for a multi-process simulation can therefore
include MPI communications needed to exchange information at boundaries and potential inefficiencies in the GPU computation 
introduced by the domain decomposition. While domain decomposition influences communications overheads in all MPI-parallelized codes by changing the surface area-to-volume ratio of computational subvolumes, domain decomposition additionally affects the performance of a GPU-accelerated code by changing the ratio of ghost to real cells in memory that must be transferred to the GPU.
Since memory transfers from the CPU to the GPU involve considerable overhead, domain decompositions that limit the fraction of ghost cells on a local process are favorable. \cholla therefore allows for two different domain decompositions, described below.

\subsubsection{Slab Decomposition}

Following the domain decomposition utilized by the {\it Fastest Fourier Transform in the West} discrete Fourier transform library \citep[FFTW;][]{Frigo05}, \cholla can use a slab-based decomposition in which the simulation volume is sliced only in one dimension.
In the slab decomposition a maximum of two boundaries may be shared between processes, and because there are limited communications the slab decomposition proves efficient for simulations run with a small number of processes. With the addition of more processes the slabs grow narrower, the ratio of boundary ghost cells to real cells for each subvolume increases rapidly, and the time required to exchange boundary cells between processes remains nearly constant. Although these features cause a computational inefficiency that continues to degrade with increasing numbers of processes, \cholla nonetheless includes an optional slab decomposition for use with limited processes and in conjunction with FFTW.

The division of the simulation volume for \chollans's slab decomposition is straightforward. When the FFTW library slab decomposition is used, the slab width on each process is optimized for accelerating discrete Fourier transform computations. Otherwise, the number of cells in the $x$-dimension spanning the total simulation volume is divided evenly across the number 
of processes, and any remaining cells are split as evenly among the processes as possible. Once the domains have been assigned, each process initializes the real cells associated with its volume and exchanges boundary cells. First, each process posts a receive request for each MPI boundary. If the process has a global simulation boundary along the $x$-direction, it posts either one receive request in the case of reflective, transmissive, or analytic global boundary conditions, or two in the case of global periodic boundary conditions. Processes that are surrounded by other processes will always have two MPI boundaries. The processes then send the initialized values of the real cells from their subvolume that are needed by other processes. While waiting for the cell exchange communications to complete, each process computes the cell values on its non-MPI boundaries (typically the $y$- and $z$ boundaries). This asynchronous ordering of communication and boundary computation minimizes the amount of time the CPU must sit idle while waiting to receive boundary cell information. Once all the receives have completed, each process proceeds through 
the CTU step as though it were an independent simulation volume. At the end of each time step, boundary cells must again be exchanged along with the information from each local subvolume required to determine the global simulation time step.

\subsubsection{Block Decomposition}

In addition to begin computationally inefficient, a slab decomposition limits the total number of processes that can be run as a result of the finite dimensions of the simulation volume. To improve upon both these factors, \cholla includes a block decomposition that seeks to minimize the aspect ratios of the simulation subvolume evolved by each process. For a block decomposition, up to six MPI communications for cell exchanges may be required per time step. Despite this increased number of communications, the reduction of the surface area-to-volume ratio of the block decomposition improves its efficiency beyond that achieved by a slab decomposition for large numbers of processes.

In the case of a block decomposition the $x$-, $y$-, and $z$-dimensions for each subvolume are kept as even as possible. Once a simulation volume is appropriately divided and local real cells initialized by each process, boundary cells between subvolumes must be exchanged. To keep the number of MPI communications to a maximum of six, the processes exchange boundary cells in a specific order. First, each process posts a receive request for any MPI boundaries on an $x$-face. While waiting for those data 
to arrive, the processes set $x$-ghost cells on any non-MPI $x$ faces. Once the $x$-boundaries arrive, the processes post receive requests for MPI boundaries on $y$-faces, including the corner regions just transferred in the exchange along the $x$-direction. While waiting for the $y$-boundaries to arrive, each process computes the ghost cell values along non-MPI $y$-boundaries. By first sending the $x$-boundaries, processes can receive information needed for the $y$-boundary exchange from diagonally-adjacent processes without directly exchanging information with those processes. The same procedure is followed for the $z$-boundaries.

\begin{figure}
\centering
\includegraphics[width=1.0\linewidth]{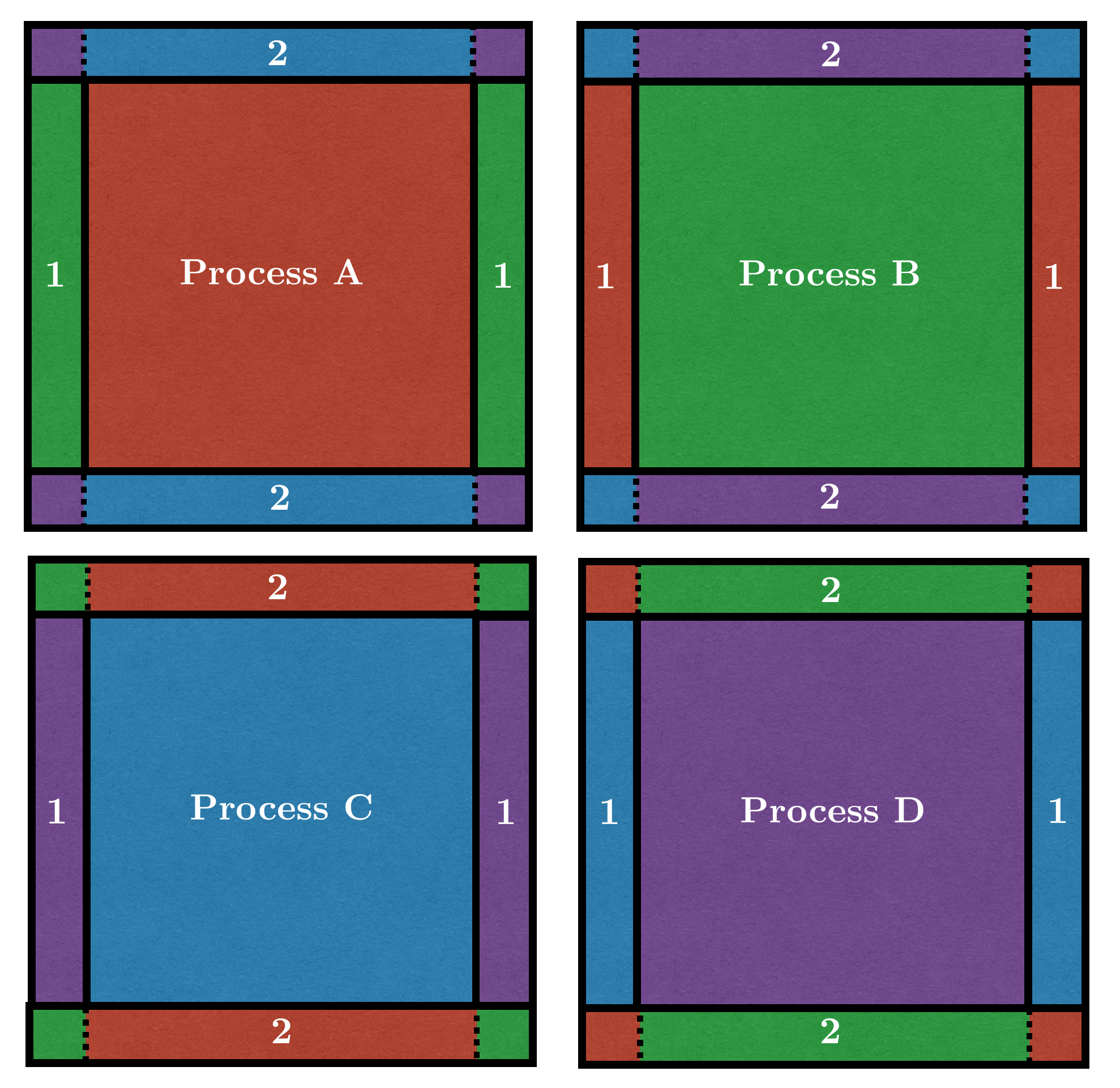}
\caption{Ghost cell information is exchanged by MPI processes for the case of a 2D simulation with periodic boundaries. The real-cell domain of each of the four processes is represented by a colored square (A, B, C, and D). Each process needs information from the other three processes in order to set all of its ghost cells. By first exchanging $x$ boundaries (represented by the outlined rectangles labeled step $1$), then exchanging $y$ boundaries (step $2$), the processes are able to access the needed information without explicitly communicating with every other process. For example, on step $2$, process D receives the red boundary cells from Process A that it needs to update its corner ghost cells, without ever communicating directly with process A.}
\label{fig:MPI_boundary_exchange}
\end{figure}

Figure \ref{fig:MPI_boundary_exchange} illustrates this process for a 2D simulation with periodic boundaries and a four-process decomposition. Each process first initializes its real cells, represented by the large colored squares. To perform the hydrodynamical simulation timestep in parallel across separate processes, each process must receive boundary ghost cells from the real cells hosted by surrounding processes. The boundary regions for each process are outlined in Figure~\ref{fig:MPI_boundary_exchange} and labelled 1 and 2. The processes first exchange $x$-boundary information (region 1) via two MPI communications. Once the $x$-boundary exchange is complete,  $y$-boundary information (region 2) is exchanged, including the corner regions received from other processes. The same procedure is repeated for $z$-boundaries in a 3D simulation. Following this pattern keeps the required number of MPI communications to a maximum of six, instead of the potential twenty-six communications that would be required
to separately exchange each face, edge, and corner boundary region for an interior subvolume in a 3D simulation. When using the block decomposition with a large number of GPUs, the MPI communications typically comprise only a few percent of the total computation time.

We note briefly that the block decomposition implemented in \cholla may also be adapted to enable the use of Fast Fourier Transform libraries that use a block decomposition, such as the Parallel FFT package written by Steve Plimpton\footnote{\tt http://www.sandia.gov/~sjplimp/docs/fft/README.html}.

\subsubsection{Scaling}

\begin{figure}
\centering
\includegraphics[width=1.0\linewidth]{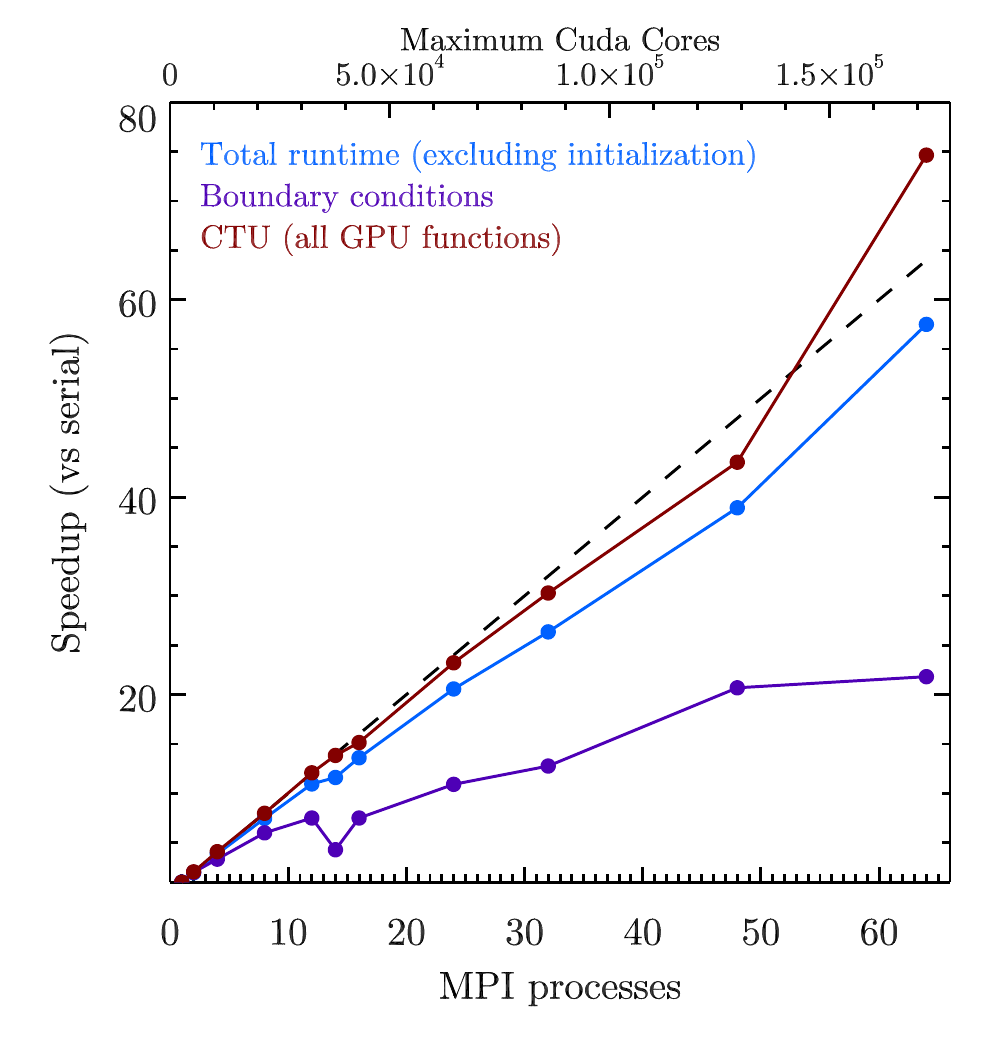}
\caption{Strong scaling on a $512^3$ double precision sound wave perturbation test with periodic boundaries measured relative to the calculation done on a single GPU (no MPI). Ideal strong scaling is shown by the dashed one-to-one line. The total runtime for the simulation remains close to ideal up to 64 processes, with non-ideal scaling coming primarily from the MPI communications needed to set boundary conditions. The portion of the code executing on the GPU is incorporated entirely within the CTU function, shown by the red points. We exclude the time taken to initialize the grid because the test was short and the initialization was a significant fraction of the total runtime, and therefore would heavily bias the total runtime results.}
\label{fig:strong_scaling}
\end{figure}

\begin{figure}
\centering
\includegraphics[width=1.0\linewidth]{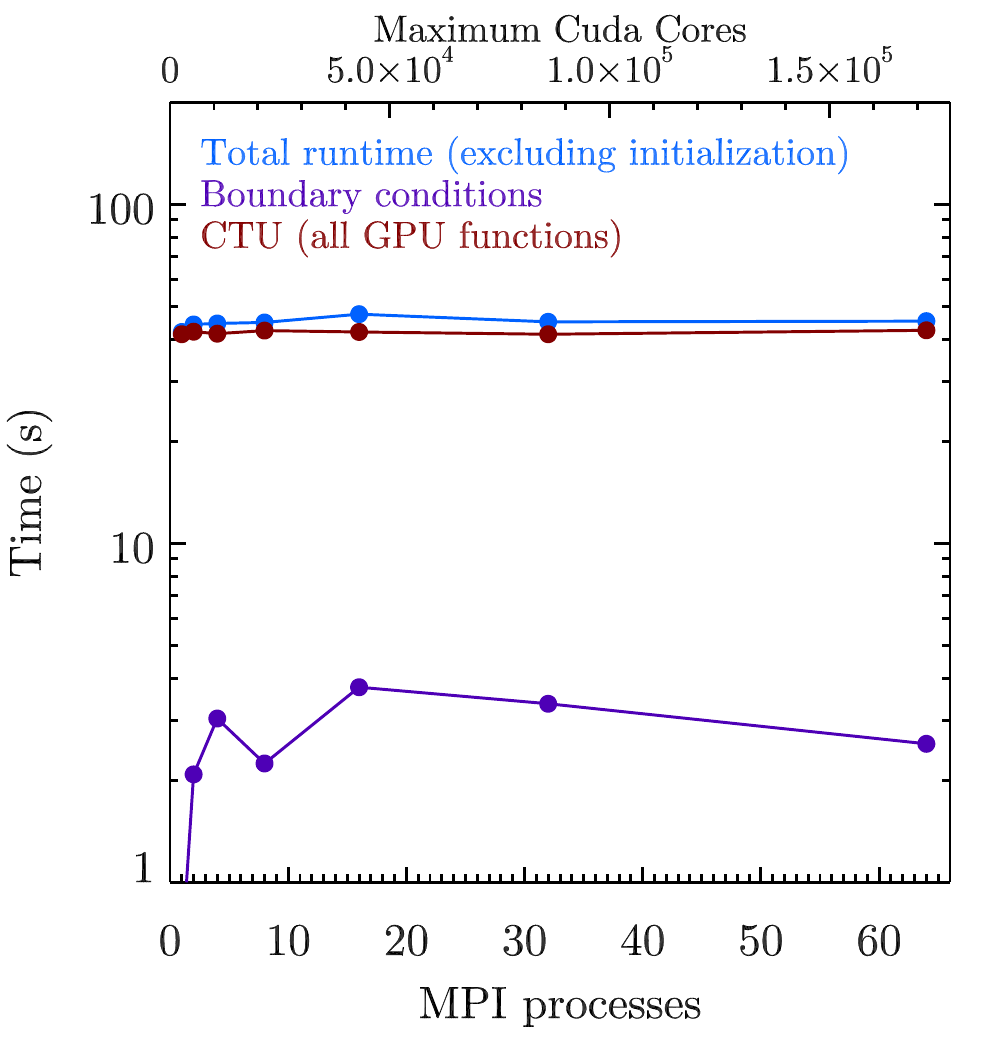}
\caption{Weak scaling performance of \cholla. A double precision sound wave perturbation test is used, with the total number of cells calculated by each process is scaled to maintain $\approx322^3$ cells for a single process. The test is run for 10 time steps. The total runtime (blue points) remains roughly constant up to 64 processes, as does the time taken for the GPU portion of the code (red points). The time taken for the boundary conditions (purple points) increases at low numbers of processes as the maximum number of MPI communications increases, but then remains flat as more processes are added.} 
\label{fig:weak_scaling}
\end{figure}

The scalability of the \cholla MPI implementation to more than one GPU warrants a discussion. To study the scaling of the code, the GPU-accelerated International Business Machines iDataplex cluster {\it El Gato} at the University of Arizona was used. Using {\it El Gato}, we have tested both the strong and weak scaling of \cholla using up to 64 GPUs. The results are shown in Figures~\ref{fig:strong_scaling} and \ref{fig:weak_scaling}. For both scaling tests, a three-dimensional sound wave perturbation problem with periodic boundary conditions and a block decomposition is used to maximize the number of MPI communications required per timestep. In both the strong and weak scaling tests, \cholla updates an average of $6.7\times10^6$ cells per second using the NVIDIA Kepler K20X GPUs available on {\it El Gato}. The 3D sound wave perturbation requires work to be done by every cell and uses third-order spatial reconstruction and an exact Riemann solver. The test is therefore relatively inefficient. By contrast, on a 2D sound wave test using second-order spatial reconstruction and a Roe solver, \cholla updates an average of $1.8\times10^7$ cells per second. All tests have been performed using double precision.

For the strong scaling test, a $512^3$ grid is evolved for $10$ time steps. The timing results for the total test runtime,
the CTU algorithm (performed on the GPU), and the boundary computation including ghost cell exchange communication are tracked separately. We exclude the simulation initialization from the runtime, as it comprises a significant fraction of the runtime for these short tests and obscures the results. As Figure~\ref{fig:strong_scaling} shows, the overall scaling of \cholla is close to ideal, with the CTU step scaling slightly better than ideal beyond 8 processes. The increased efficiency at 16 processes or more owes to the decomposition decreasing the cells per process below the number that necessitates subgrid splitting, thereby reducing the CPU-GPU communications overhead. All of the GPU calculations contained within the CTU step scale better than ideal at 64 processes. That the boundary condition computation does not scale as well primarily owes to the reduced number of MPI communications needed in runs with small numbers of processes compared with tests utilizing large numbers of processes where every subvolume
boundary requires an MPI communication per timestep. The scaling between an 8 process run and a 64 process run, both of which require MPI communications for all boundaries, is close to ideal. We achieve an effective bandwidth for the MPI ghost cell exchange of 2.4 gigabits per second in all runs.

The weak scaling performance of \cholla is shown in Figure~\ref{fig:weak_scaling}. A double precision sound wave perturbation with periodic boundaries is again used, but in this test the size of the total computational volume is rescaled with the number of processes 
to keep the number of cells per process approximately constant at $\approx322^3$ (note that each process uses its own distinct GPU during the simulation). The test is run for 10 time steps. The total runtime efficiency as a function of the number of processes remains roughly constant beyond a single process. The CTU algorithm comprises the majority of the computational cost of each timestep, and exhibits nearly perfect weak scaling. The boundary condition calculation for the serial case does not involve MPI communications and is correspondingly inexpensive when using a single process. With two or more processes, MPI communications induce an additional overhead beyond the single process case. However, the weak scaling of the boundary conditions is reasonably maintained to 64 processes.


\section{Tests}\label{sec:tests}

A large variety of hydrodynamics tests exist in the literature, some of which have been used for several decades \citep[e.g.][]{Sod78}. Their ubiquity makes these canonical tests an excellent way to compare the performance of \cholla with other codes. In addition, many tests have been designed to explicitly show the failings of hydrodynamics solvers in various environments or highlight the circumstances where they perform exceptionally well. In choosing the tests shown below, we attempt to demonstrate the breadth of problems \cholla can simulate. We also demonstrate the effects of changing reconstruction methods or Riemann solvers, and show differences in the outcomes of tests where they are relevant. If not otherwise specified, the following tests were performed using piecewise parabolic reconstruction with the characteristic variables (PPMC) and an exact Riemann solver. All tests were performed in double precision on GPUs.

Before delving into the specifics of each test, we make a note about the convergence rate of \cholla.  
Many shock-capturing methods revert to first order at shocks \citep[see, e.g.,][]{Laney98}, so to test the convergence rate 
of \cholla the smooth perturbation test described in \cite{Stone08} is employed.
Both the PPMP and PPMC implementations in \cholla
demonstrate second-order convergence in the L1 error norm out to grid
resolutions of $1024$ cells.


\subsection{1D Hydrodynamics}\label{sec:1Dhydro}

\subsubsection{Sod Shock Tube}

\begin{figure}
\centering
\includegraphics[width=1.0\linewidth]{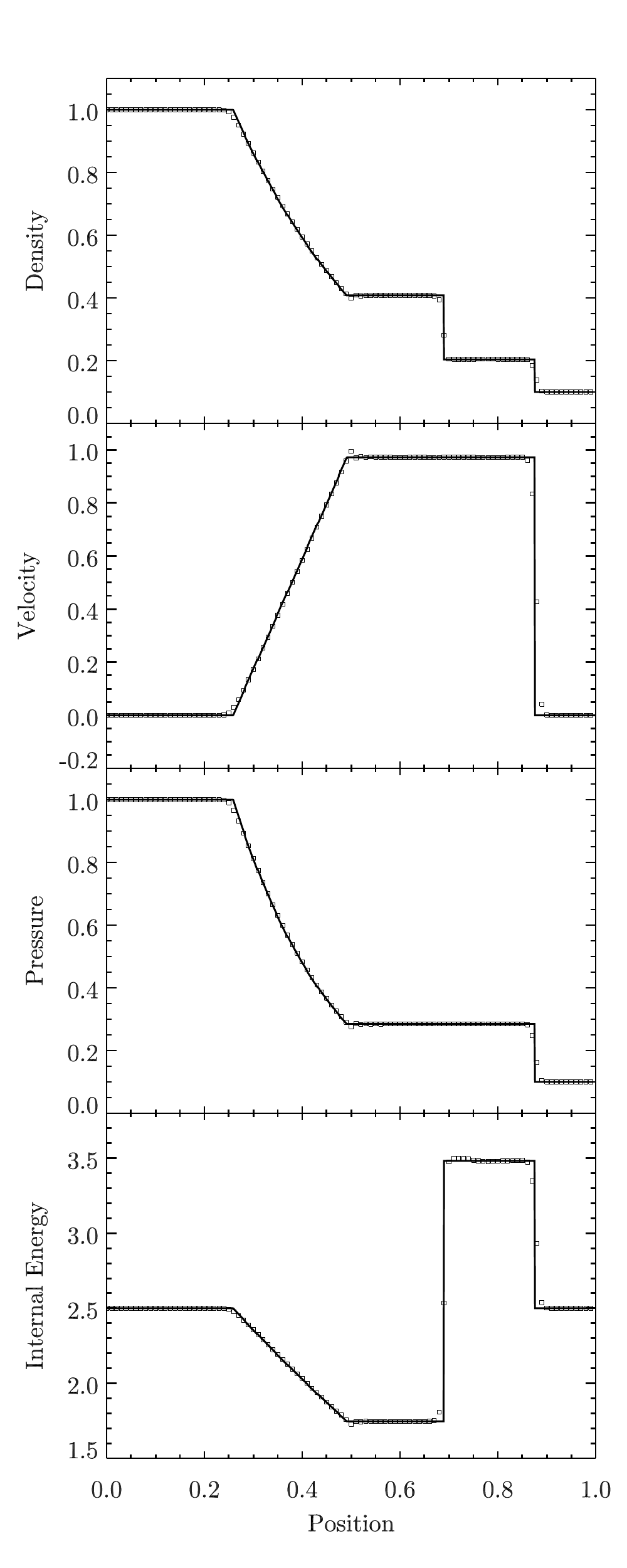}
\caption{The solution to the Sod shock tube test using PPMP with a resolution of 100 cells. The exact solution is shown as a line with points from the \cholla simulation over plotted. Features seen in the density plot include a rarefaction wave expanding from the initial discontinuity at $x = 0.5$, a rightward moving contact discontinuity at $x\approx0.7$, and a rightward moving shock at $x\approx0.85$.}
\label{fig:sod_flash}
\end{figure}

\begin{figure}
\centering
\includegraphics[width=1.0\linewidth]{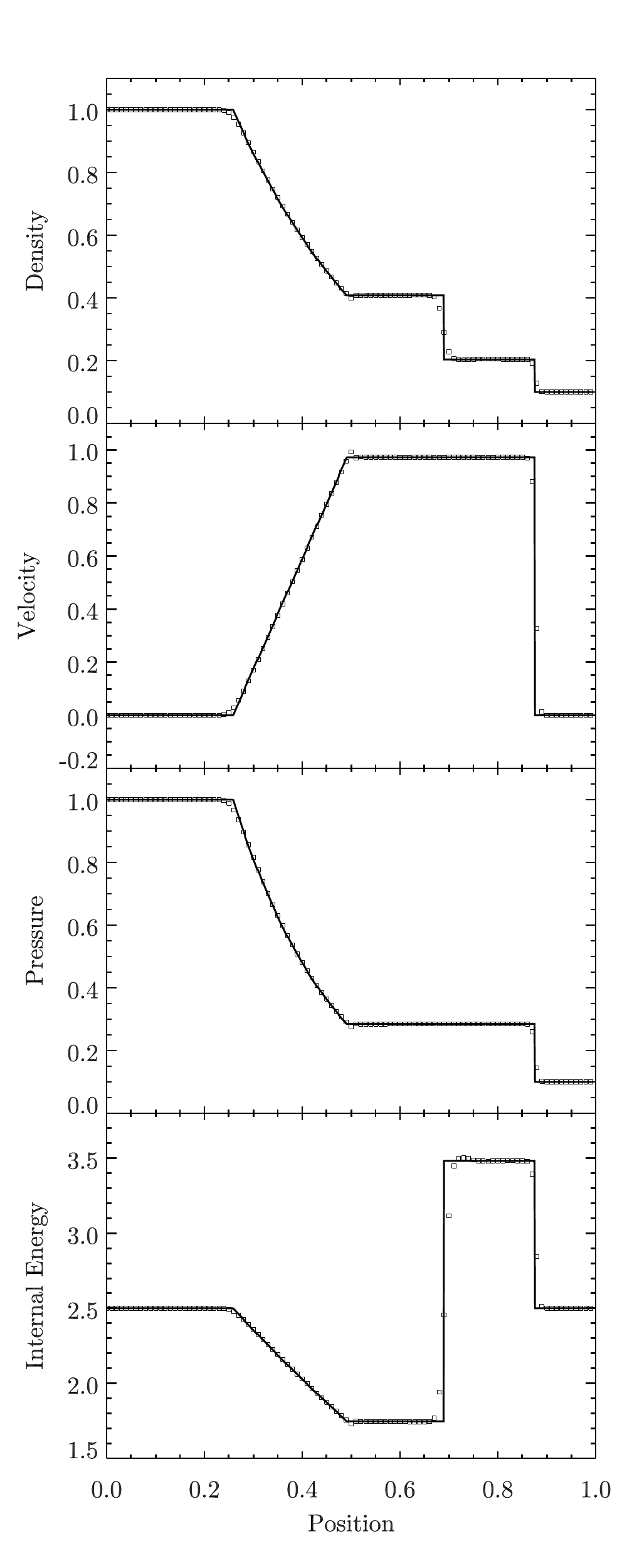}
\caption{The solution to the Sod shock tube test using PPMC with a resolution of 100 cells. The exact solution is shown as a line with points from the \cholla simulation over plotted. The same fluid features are seen as in Figure~\ref{fig:sod_flash}, with the contact discontinuity slightly less narrowly resolved.}
\label{fig:sod_athena}
\end{figure}

The Sod problem \citep{Sod78} is often the first test performed by hydrodynamics codes, 
and we do not diverge from precedent here. Though the Sod problem is not a difficult test to run, the solution contains several important fluid features. We present the test here as an example of the ability 
of \cholla to resolve both shocks and contact discontinuities within a narrow region 
of just a few zones. The initial conditions are simply a Riemann problem, with density and pressure
$\rho_L = P_L = 1.0$ on the left, $\rho_R = P_R = 0.1$ on the right, and an
initial velocity $u_L = u_R = 0.0$. The initial discontinuity is at position $x = 0.5$. 
For this and all of the following one dimensional tests, orthogonal velocities
are set to zero. We use an ideal gas equation of state with $\gamma = 1.4$ for all tests, unless otherwise noted.

The test is computed on a grid of 100 cells until a final time of $t = 0.2$. By that time a shock, a contact discontinuity, and a rarefaction fan have formed and spread enough to be clearly visible as seen in Figures~\ref{fig:sod_flash} and \ref{fig:sod_athena}. As described in Section \ref{sec:reconstruction}, \cholla has two versions of piecewise parabolic interface reconstruction. PPMP follows the FLASH code documentation \citep{Fryxell00} and includes contact discontinuity steepening and shock flattening, while PPMC is based on the Athena code documentation \citep{Stone08} and reconstructs the interface values using characteristics without explicit steeping or flattening. As can be seen in the density plot, the contact discontinuity is resolved over just two zones using PPMP, and over three to four zones using PPMC. Because of its the explicit treatment of contacts, the PPMP method is slightly better at resolving contact discontinuities, but is also more susceptible to nonphysical oscillations as demonstrated in later tests.

\subsubsection{Strong Shock Test}

\begin{figure}
\centering
\includegraphics[width=1.0\linewidth]{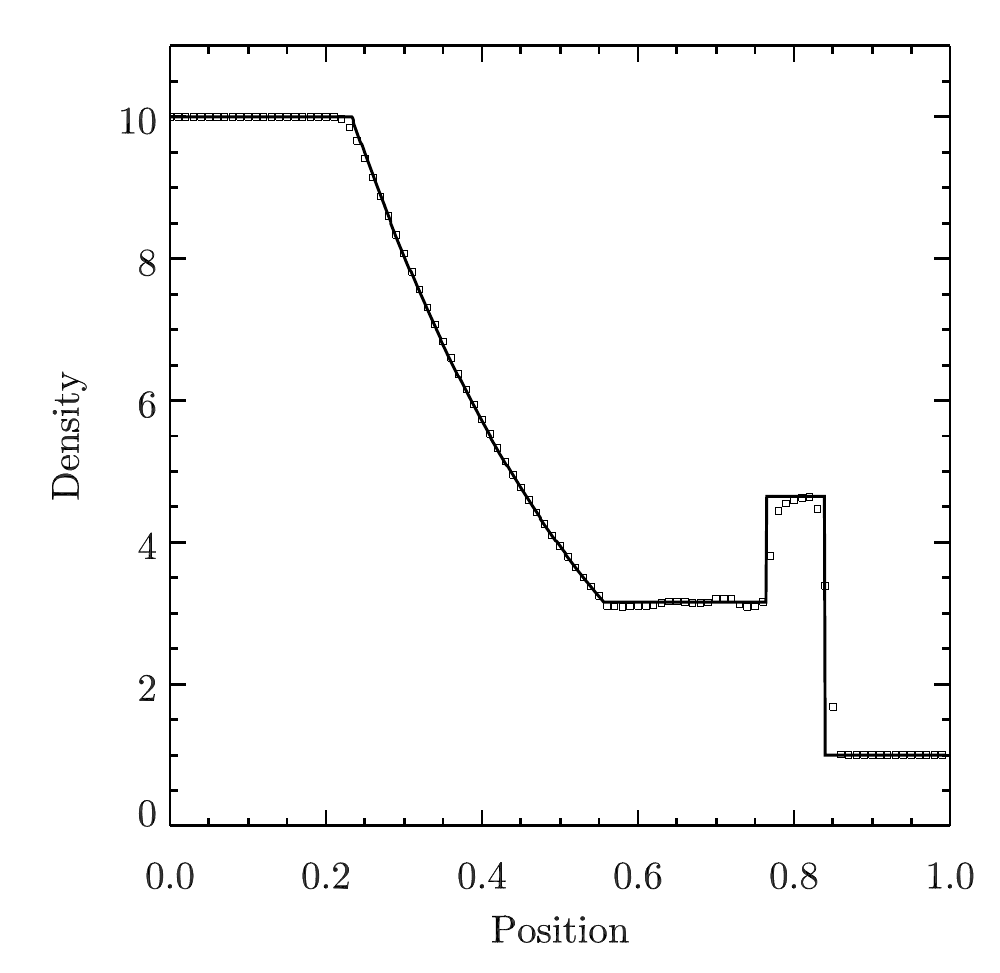}
\includegraphics[width=1.0\linewidth]{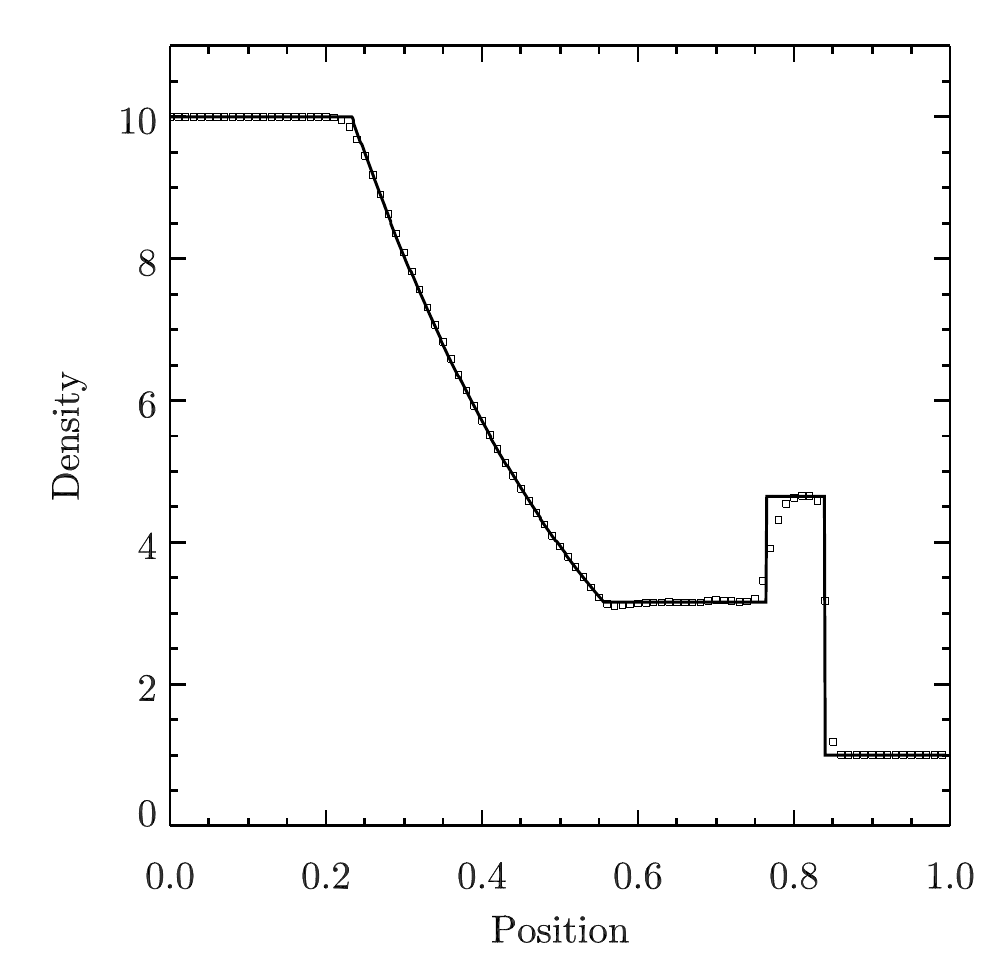}
\caption{Numerical solution to the strong shock test at time $t=0.2$
using PPMP (top) and PPMC (bottom) with a resolution of 100 cells, as compared to the exact solution shown by the solid 
line. The detailed initial conditions are described in the text. 
The contact discontinuity is better resolved with PPMP, but there fewer oscillations in the solution calculated with PPMC.}
\label{fig:strong_shock}
\end{figure}

The strong shock test \citep{Fryxell00} resembles the Sod shock tube, but is more 
discriminating owing to the much more severe differences between the left and right initial states. 
This test starts with an initial  discontinuity at $x=0.5$, with left and right densities $\rho_L = 10.0$ and $\rho_R = 1.0$. Initial pressures are $P_L = 100$ and $P_R = 1.0$. The initial velocities are set to zero, as in the Sod test. 
The problem is calculated on a grid of 100 cells, and the resulting 
density in the numerical solution using both PPMP and PPMC is shown at time $t = 0.07$ in Figure~\ref{fig:strong_shock}.

As can be seen in Figure~\ref{fig:strong_shock}, both PPMP and PPMC do a decent job reproducing the exact solution on this difficult problem. However, the differences between the two reconstruction methods have more discernible effects in this test. The contact discontinuity at $x = 0.75$ is better resolved with PPMP, but the solution is more oscillatory in the region between the contact discontinuity and the tail of the rarefaction fan. In constructing the linear slopes across interfaces, both PPMP and PPMC use limiters that are designed to be total variation diminishing (TVD). However, the third-order reconstruction leads to added complications \citep{CW84}. Despite attempts to preserve monotonicity (see Appendix~\ref{app:reconstruction}), problems with strong shocks are observed to cause oscillations in both methods. Due to its more diffusive nature, we find that PPMC is less susceptible to oscillations in regions with strong density and pressure contrasts. 
The inclusion of contact discontinuity steepening in PPMP keeps contacts sharp but tends to exacerbate the oscillations. 
The discontinuity detection relies on a number of heuristically 
determined constants, and the resulting slopes are not always TVD. The oscillations present in the upper panel of Figure~\ref{fig:strong_shock} can be significantly reduced by lowering the value of the constant that determines whether a zone contains a density discontinuity or a shock (see Equation~\ref{eqn:pressure_jump}). This constant is labeled ``$K_0$" in \citealt{CW84}, not to be confused with ``$K$", the coefficient used in their artificial dissipation scheme. When the discontinuity detection in PPMP is turned off entirely (equivalent to setting $K_0 = 0$) the two methods produce very similar results on the strong shock test.

\subsubsection{Strong Rarefaction Test}

\begin{figure}
\centering
\includegraphics[width=1.0\linewidth]{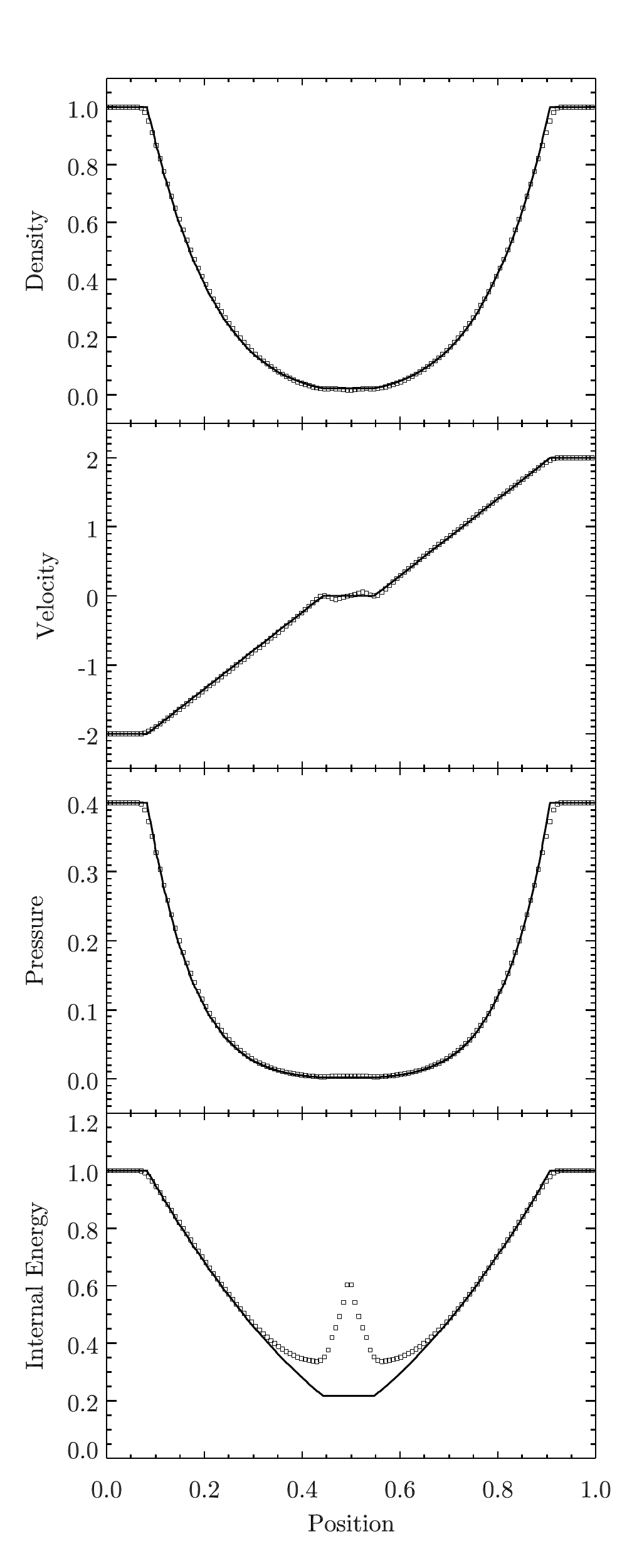}
\caption{Numerical solution to the Einfeldt strong rarefaction test at $t=0.07$
using PPMP and the exact Riemann solver with a resolution of 128 cells. The exact solution is shown as a line, with the solution from \cholla over plotted. This test will cause linearized solvers to fail without modification.}
\label{fig:123}
\end{figure}

The strong rarefaction test, or \textit{123 problem}, was originally used by \cite{Einfeldt91} to illustrate a scenario that causes a subset of approximate Riemann solvers to fail. Because the solution contains a region where the energy is largely kinetic and the pressure is close to vacuum, the Roe solver (or any other linearized solver) will produce negative densities or pressures in the numerical solution \citep{Einfeldt91}. The initial conditions consist of a fluid with constant density and pressure but opposite receding velocity at the center. Specifically, we set $\rho_L = \rho_R = 1.0$, $P_L = P_R = 0.4$, $u_L = 2.0$, and $u_R = 2.0$. In Figure \ref{fig:123}, we show the results of this test on a grid of 128 cells at time $t = 0.15$ using PPMP reconstruction and the exact solver. This test is not a challenge using the exact solver, but without modification the Roe solver would fail on this problem. For this reason, we test the density and pressure produced in the solution by the Roe solver and revert to the HLLE solver if necessary. With that fix we can run the problem with either solver, and in fact the HLLE fluxes are only needed on the first step of the simulation.

\subsubsection{Shu and Osher Shocktube}

\begin{figure}
\centering
\includegraphics[width=1.0\linewidth]{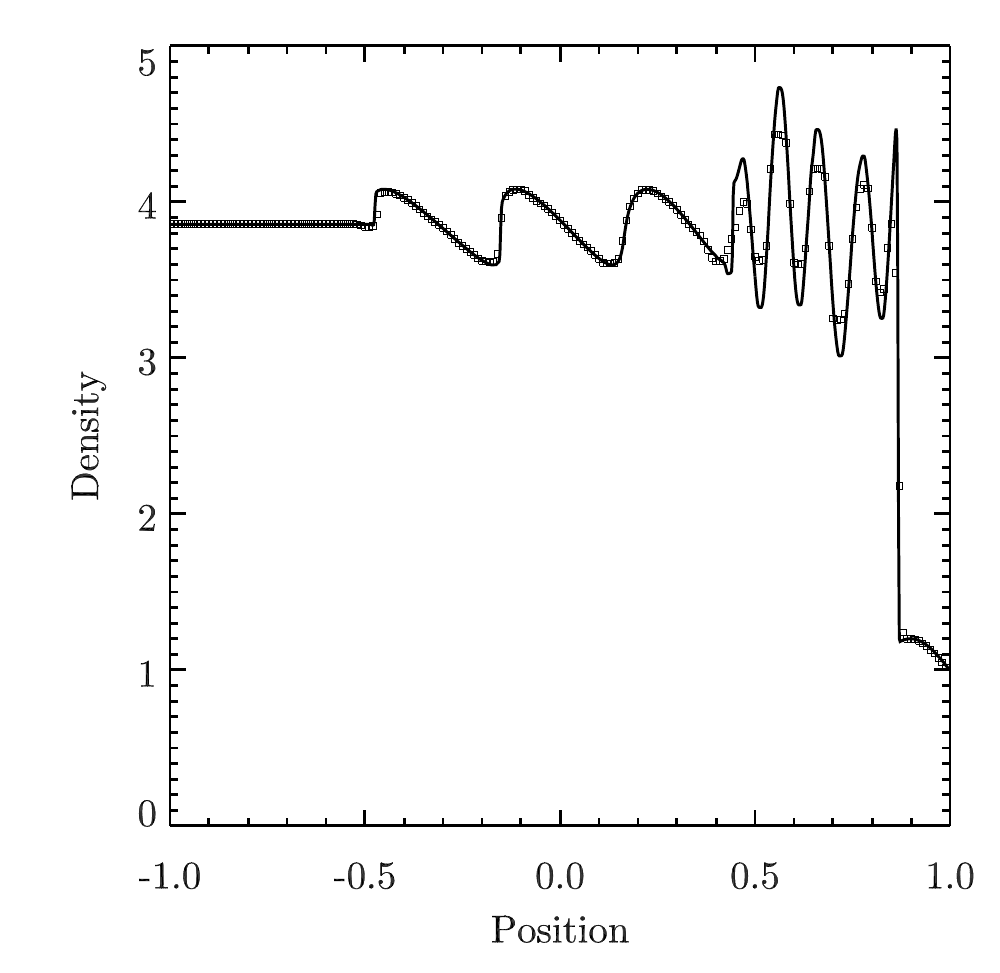}
\caption{Numerical solution to the Shu \& Osher shock tube problem. A low resolution solution with 200 cells (points)
is plotted over a high resolution solution with 800 cells (line). 
This test shows the result when PPM limiters cut off maxima in low resolution models of smoothly varying solutions.}
\label{fig:shu_osher}
\end{figure}

The {\it Shu-Osher shocktube test} shows the tendency of PPM to cut off maxima in smoothly varying problems as a result of the slope limiters imposed in the reconstruction method \citep{Shu89}. The test consists of a strong shockwave propagating into a region with a sinusoidally varying density. The initial conditions are $\rho_L = 3.857143$, $u_L = 2.629369$, and $p_L = 10.3333$; $\rho_R = 1 + 0.2\mathrm{sin}(5\pi x)$, $u_R = 0$, and $p_R = 1.0$. We run the problem on the domain $x = [-1, 1]$ with the initial discontinuity at $x = -0.8$. The results of the test using both 200 cells and 800 cells are shown in Figure~\ref{fig:shu_osher}. As can be seen, the low resolution solution does lose some of the amplitude of the peaks. Using newer versions of the limiting functions can help alleviate this problem \citep{Colella08, Stone08}, although we have not yet implemented these limiters in \chollans.

\subsubsection{Interacting Blast Waves}

Originally described in \cite{CW84}, the {\it interacting blast wave test}
helps quantify the behavior of a code near strong shocks and contact discontinuities.
The test consists of a medium with initially constant density $\rho = 1.0$, with $\gamma = 1.4$ on the domain $x = [0, 1]$. Reflecting boundary conditions are used. Two shocks are initialized on 
either side of the domain, with $p = 1000$ for $x < 0.1$, $p = 100$ for $x > 0.9$, and $p = 0.01$ in between. 
The problem is run until time $t = 0.038$, at which point the shocks and rarefactions in the initial solution have interacted multiple times.

\begin{figure}
\centering
\includegraphics[width=1.0\linewidth]{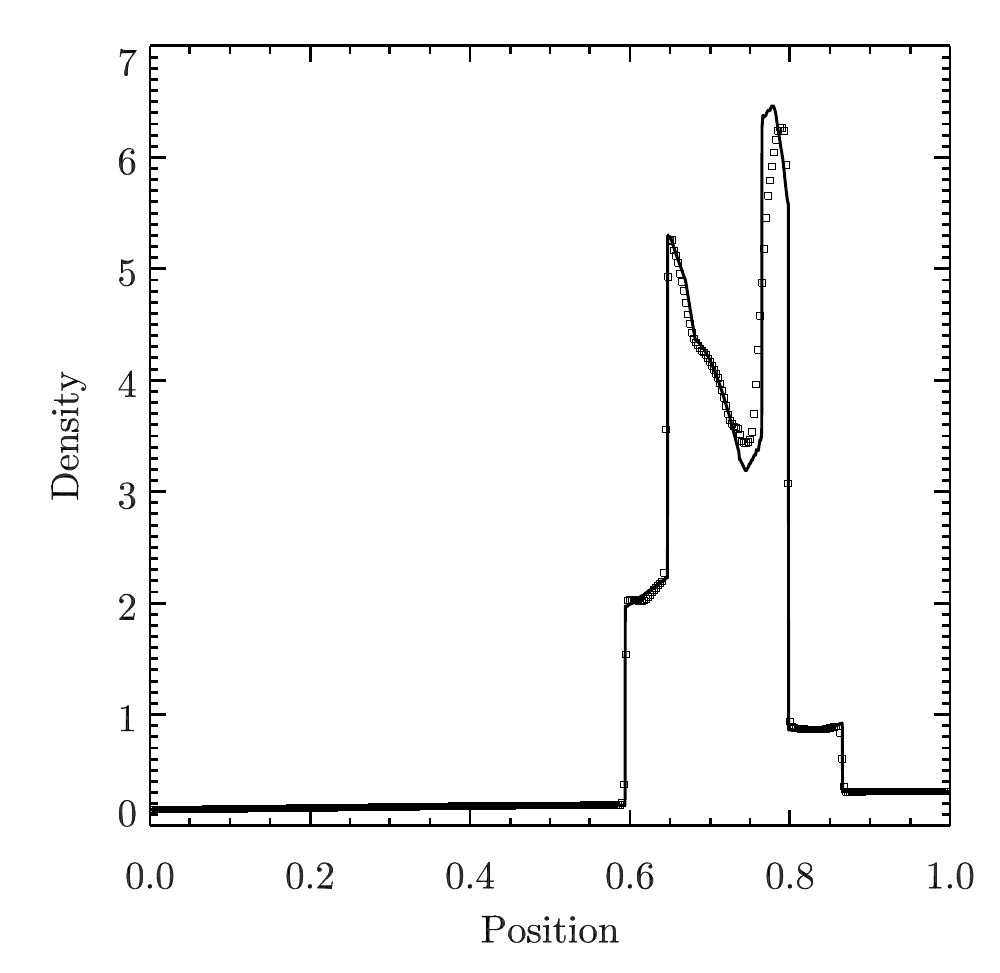}
\includegraphics[width=1.0\linewidth]{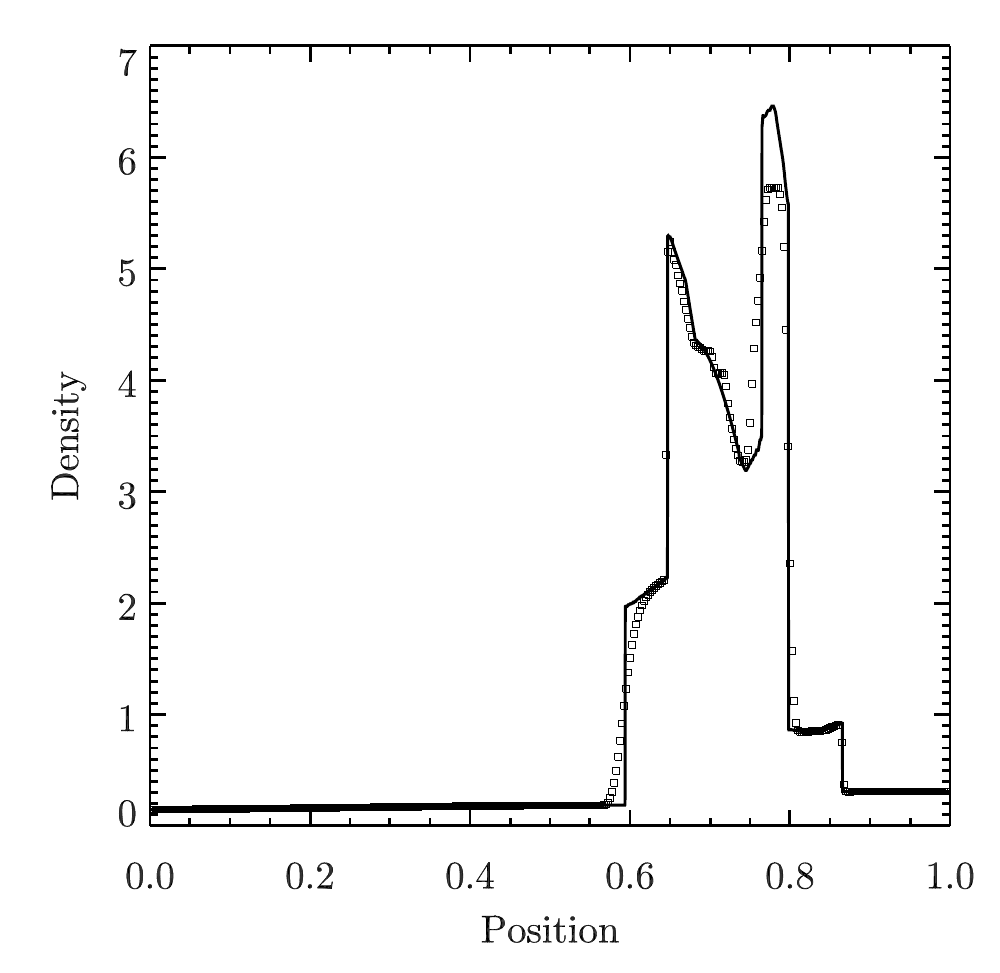}
\caption{Numerical solution for the interacting blast wave test using PPMP (top) and PPMC (bottom) with a resolution of 400 cells, plotted over a reference solution with 9600 cells. This solution is shown at $t=0.038$, when the original shocks and rarefactions have interacted several times. The test was designed to capture a code's ability to maintain narrow features.}
\label{fig:blast_1D}
\end{figure}

We show plots of the density computed with both PPMP and PPMC in Figure~\ref{fig:blast_1D}. We plot a low resolution solution with 400 grid cells over a high resolution reference solution computed with 9600 cells. As can be seen in the figure, PPMP does an excellent job keeping the contact discontinuities at $x = 0.6$ and $x = 0.8$ contained within just two zones, as compared to the solution computed with PPMC in which the contacts are smeared over many cells. In addition, PPMC tends to more severely cut off the maximum at $x = 0.75$, while PPMP does a decent job of keeping the full height although the peak is slightly offset. Both reconstruction techniques do a good job reproducing the shocks and the rarefaction fan between $x = 0.65$ and $x = 0.7$.

\subsection{2D Hydrodynamics}

\subsubsection{Implosion Test}

\begin{figure}
\centering
\includegraphics[width=1.0\linewidth]{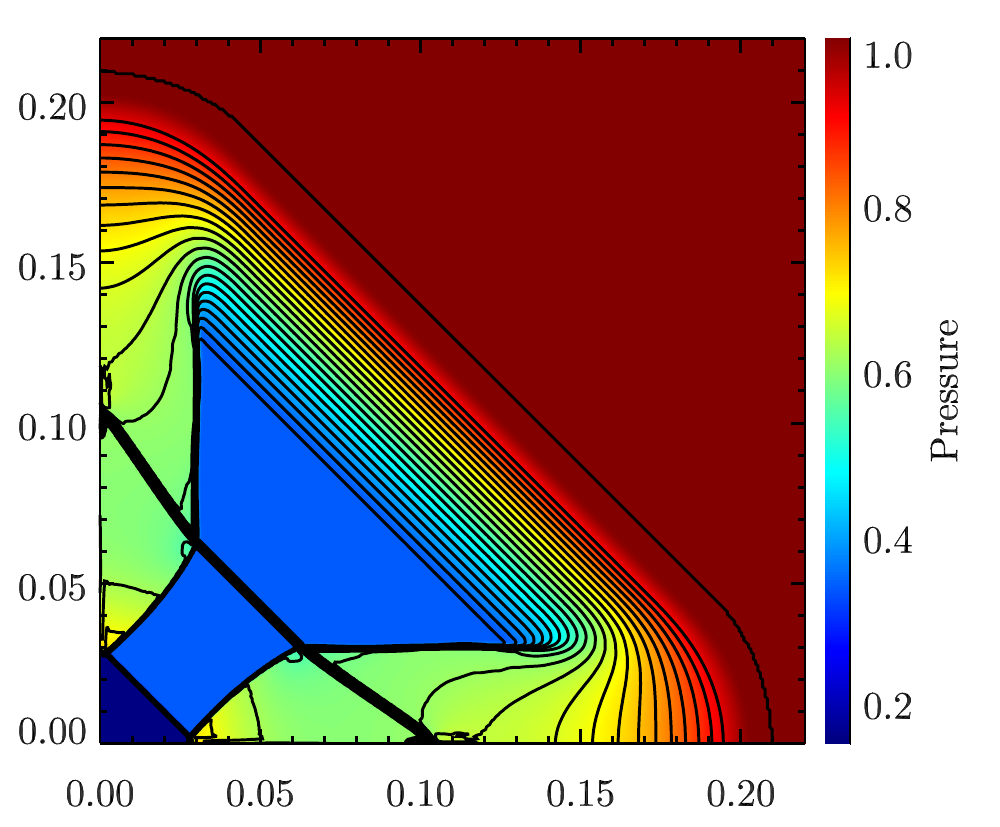}
\includegraphics[width=1.0\linewidth]{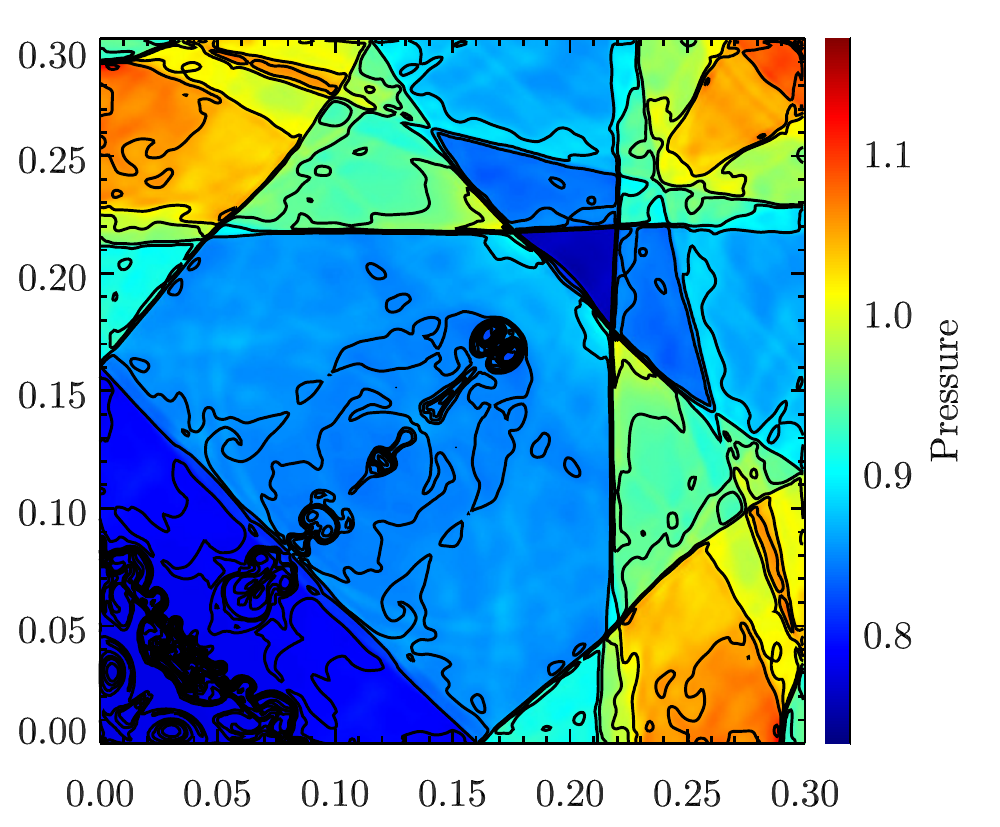}
\includegraphics[width=1.0\linewidth]{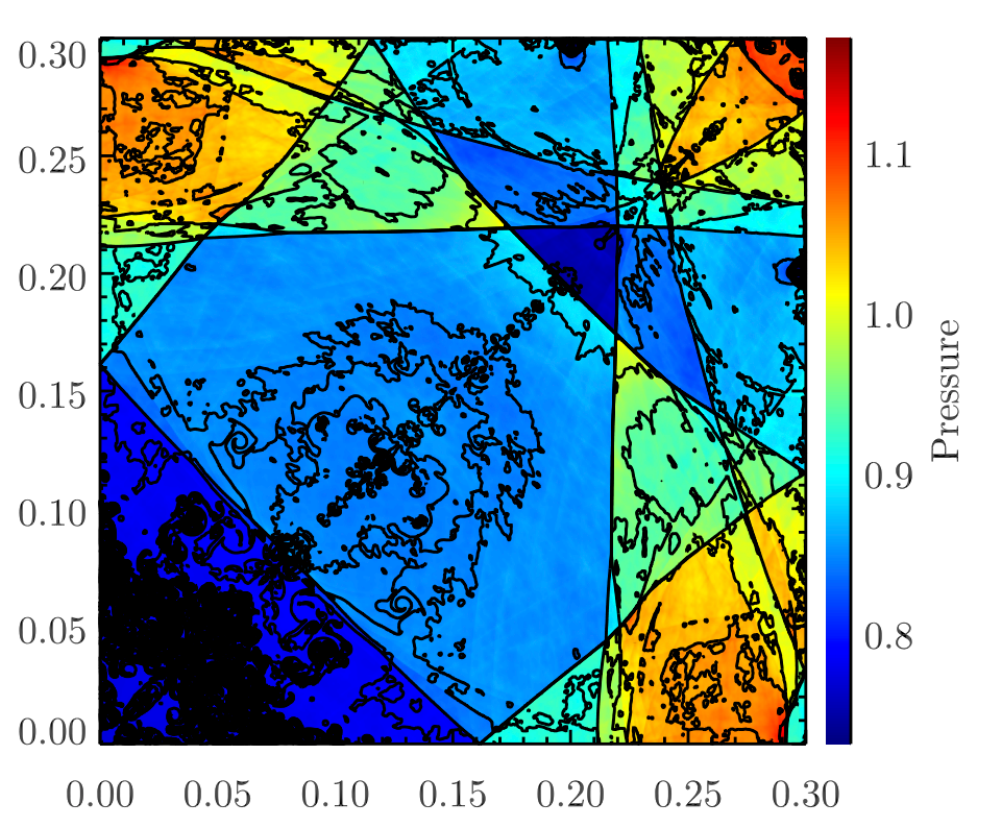}
\caption{Numerical solutions for an implosion test. 
Top: The $400\times400$ implosion test at $t = 0.045$; 31 density contours from 0.35 to 1.1 are overlaid on a color-scale pressure map, with only the inner region of the computational domain shown, $x = [0, 0.22]$ and $y = [0, 0.22].$ Middle: The same test at $t = 2.5$; 36 density contours from 0.125 to 1 are overlaid on a color-scale pressure map. Bottom: A $4096\times4096$ version of the implosion test at $t=2.5$. The same contour levels are drawn.}
\label{fig:implosion_2D}
\end{figure}

The {\it implosion test} is a 
converging shock problem first presented in \cite{Hui99a}. The version presented here is described in \cite{LW03}, and
begins with a square region of high density and pressure containing a diamond-shaped region of 
low density and pressure. These initial conditions evoke the traditional Sod shock tube problem extended to two dimensions, but inclined to the grid by 45 degrees rather than aligned as in the one dimensional case. 
As the test begins material moves inward rapidly toward the center, leading to an implosion. 
When run for a short amount of time, this test demonstrates 
the ability of a code to resolve contact discontinuities and other fluid features 
for a non-grid aligned shock tube. When run for enough time to evolve well past the 
initial shock tube solution, the test illustrates the symmetry (or lack thereof) of a code.

Figure \ref{fig:implosion_2D} shows the results of the implosion test run with PPMC and an exact solver at an early time $t = 0.045$ and a later time $t = 2.5$. 
The problem was run on a $400\times400$ 
grid with a domain $x = [0, 0.3]$, $y = [0, 0.3]$ and reflecting boundary conditions at every boundary, comprising
the upper right quadrant of the axisymmetric test described above. 
The initial density and pressure within the diamond-shaped region are $\rho = 0.125$ and $p = 0.14$, while outside the density and pressure are $\rho = 1.0$ and $p = 1.0$. Initial velocities are zero, as in the Sod test. 
A discontinuous interface is located along the diagonal running from $(0.15, 0)$ to $(0, 0.15)$. 
In the upper panel of Figure~\ref{fig:implosion_2D} a rarefaction fan can be seen expanding outward from this interface. 
As the upper panel shows, \cholla does an excellent job resolving the contact at early times, as can be seen 
along the diagonal from $(0, 0.1)$ to $(0.1, 0)$.

At the later time a jet has appeared in the solution. The production of the jet is a direct result 
of ability to preserve symmetry in the \cholla solution to numerical accuracy. \citet{LW03} demonstrated 
that codes that employ non-symmetry preserving methods like Strang splitting may
fail to produce the jet-like feature. The fact that this test is so sensitive to the symmetry of the problem makes it useful for diagnosing potential coding errors, but the test also demonstrates the extent 
to which a non-symmetric algorithm can impact the physical accuracy of the result. As this test shows, relatively large-scale features in the solution can be completely lost if a code fails to maintain a sufficient level of symmetric accuracy.

The bottom panel of Figure~\ref{fig:implosion_2D} shows the results of this
test recomputed at a much higher resolution of $4096\times4096$. 
The same large-scale features are apparent in the solution, but as expected
the small-scale density perturbations and shape of the jet have clearly not converged.
However, this high resolution test serves as further evidence of the ability of 
\cholla to preserve axisymmetry even in a very difficult problem. At this extreme 
resolution, the code must perform over $200,000$ time steps and more than
$3\times10^{12}$ cell updates to reach time $t=2.5$. 
At that point, the results are still exactly symmetric (to floating-point precision), 
demonstrating that symmetry preservation in \cholla is a robust feature of the code.

\subsubsection{Explosion Test}

\begin{figure}
\centering
\includegraphics[width=1.0\linewidth]{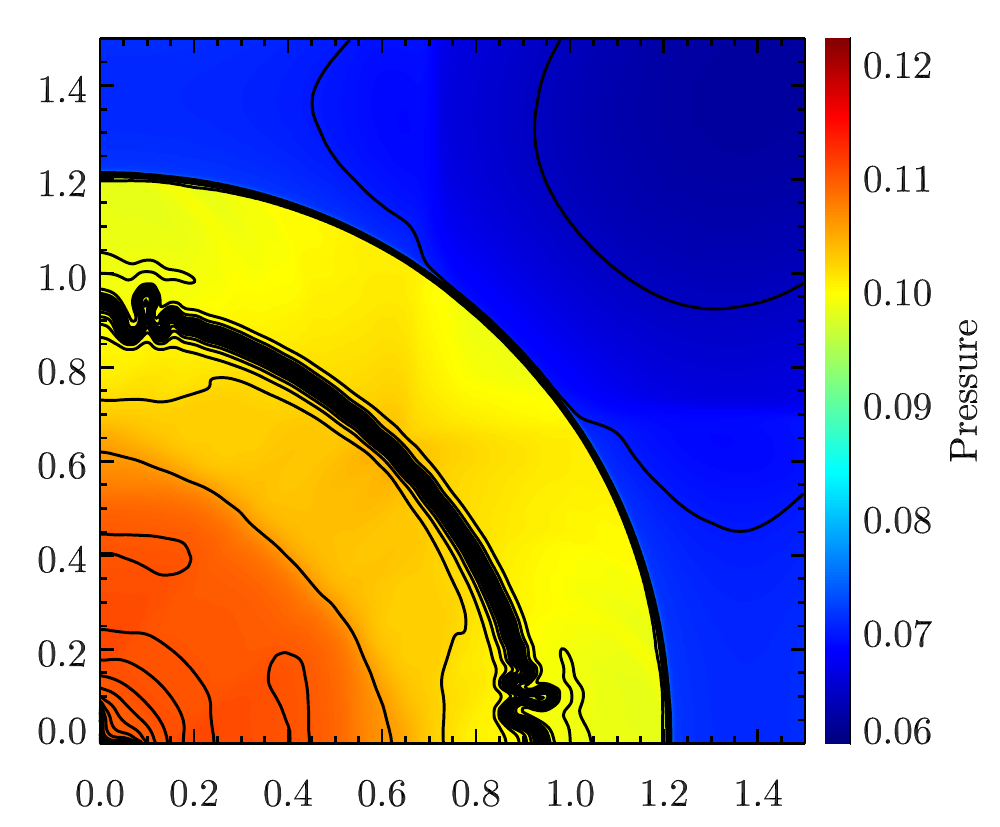}
\includegraphics[width=1.0\linewidth]{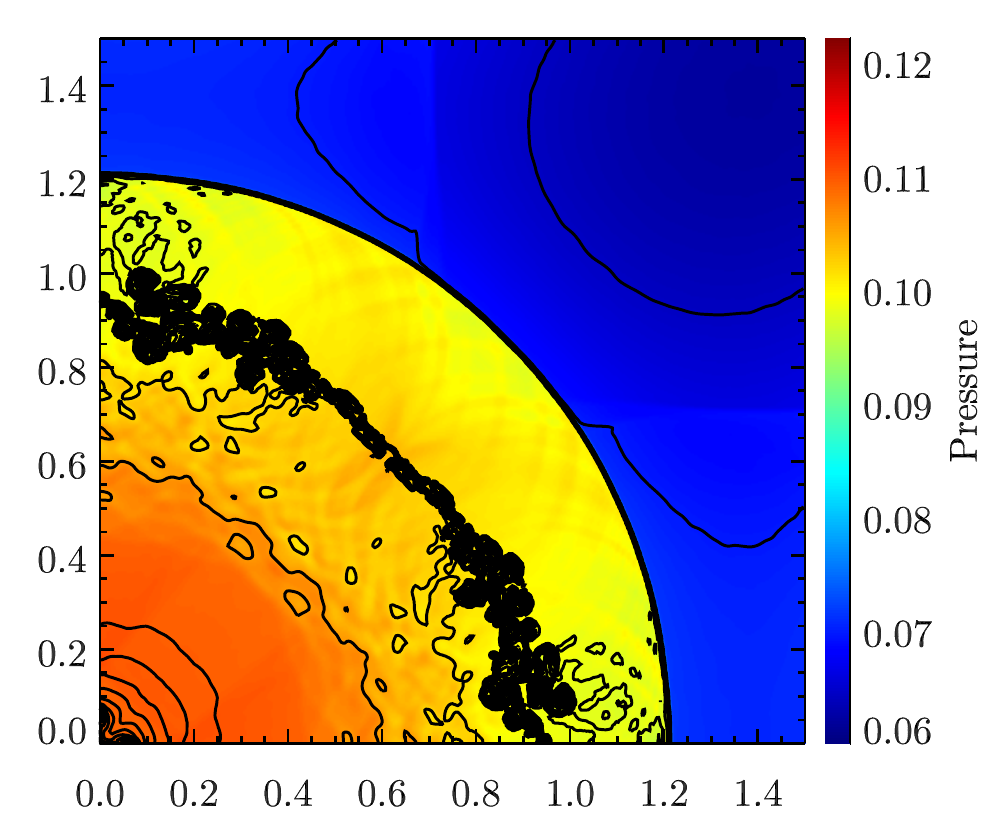}
\caption{Numerical solution to the 
2D explosion test at $t = 3.2$ using PLMP (top) and PPMP (bottom), both at a resolution of $400\times400$. We show a color map of the pressure overlaid by 27 density contours, from $0.08$ to $0.21$ with step $0.005$. The lower order reconstruction method is more diffusive for the contact discontinuity, but is less susceptible to the instability that causes the contact interface to be unstable.}
\label{fig:explosion_2D}
\end{figure}

The {\it explosion test}, also from \cite{LW03}, is designed to test the evolution of an unstable contact discontinuity and is highly sensitive to numerical noise in the initial conditions. This noise seeds an instability that grows as the solution evolves. The test starts 
with a domain $x = [0, 1.5]$, $y = [0, 1.5]$ that contains a circularly symmetric region of high density and pressure, with $\rho = 1$ and $p = 1$ inside a circle with radius $r =  0.4$. 
Reflecting inner boundaries and transmissive outer boundaries are used. 
Outside the circle the density and pressure are set to $\rho = 0.125$ and $p = 0.1$. The initial velocities are zero. Because the problem is sensitive to initial perturbations at the interface, 
the density and pressure for cells are area-weighted at the boundary. For each cell on the boundary of the circle
the percentage of the area inside the radial boundary is computed, and the initial cell data
weighted appropriately.

The test problem is performed on a grid of $400\times400$ cells, and Figure~\ref{fig:explosion_2D} 
shows the result of the calculation using PLMP and PPMP at $t = 3.2$. 
As expected, the higher order reconstruction method does a better job preserving the narrow structure 
of the contact, but is also more susceptible to structure along the interface as the instability develops. 
Thus, as the problem progresses, the lower order more diffusive method may result in a cleaner solution. We note that both methods preserve the exact symmetry of the problem, provided the initial conditions are symmetric.

\subsubsection{Kelvin-Helmholtz Instability}

\begin{figure}
\centering
\includegraphics[width=1.0\linewidth, height=0.50625\linewidth]{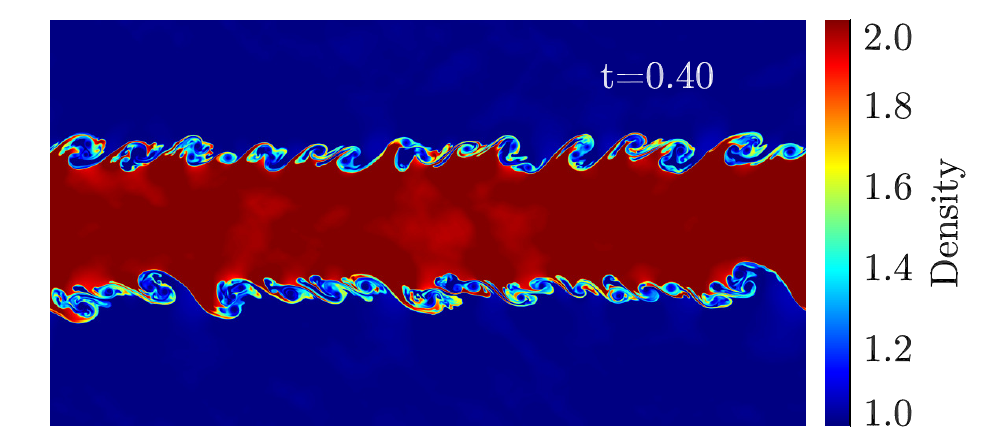}
\includegraphics[width=1.0\linewidth, height=0.5625\linewidth]{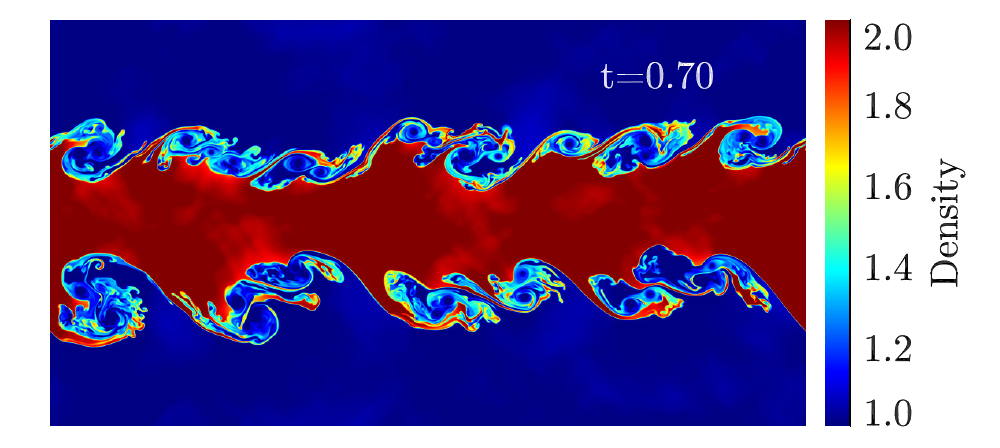}
\includegraphics[width=1.0\linewidth, height=0.5625\linewidth]{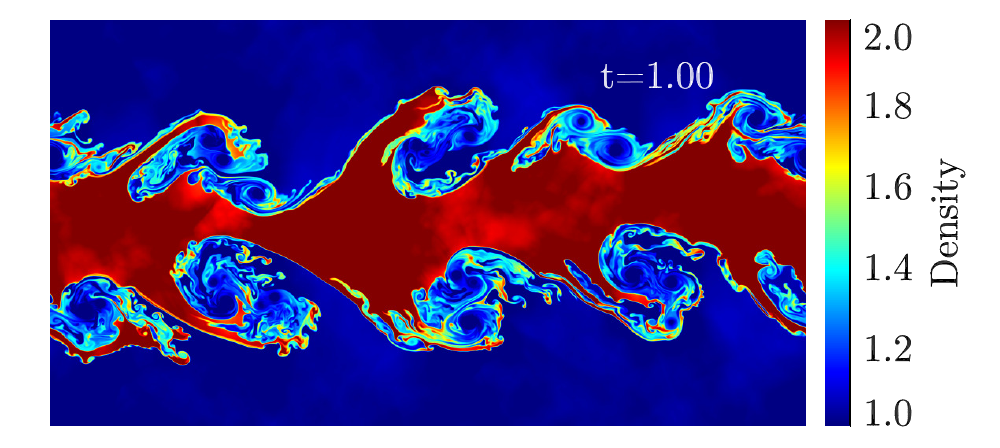}
\caption{Snapshots from a 2D Kelvin-Helmholtz instability test on a 1920 x 1080 grid at $t=0.4$, $0.7$, and $1.0$. The eddies at the discontinuous interface start out at the resolution scale of the simulation and grow in a predictable manner until they enter a non-linear regime.}
\label{fig:KH_2D}
\end{figure}

A {\it Kelvin-Helmholtz instability} test demonstrates the extent to which a hydrodynamics code resolves mixing caused by shear flows. In this test, two fluids at different densities flow past each other and characteristic eddies appear and grow at the interface between the fluids. The growth of the eddies in the linear regime can be analytically described \citep{Chandrasekhar61} and depends on properties of the fluid and the interface itself. 
At a discontinuous interface, small eddies will develop first at the grid scale of the simulation, and these will gradually grow and combine into larger eddies as shown in Figure \ref{fig:KH_2D}.

The exact nature of the instability depends sensitively on the resolution and initial conditions of the test  
\citep[e.g.,][]{robertson2010a}. The test shown in Figure \ref{fig:KH_2D} was run on a $1920\times1080$ grid, 
with a domain $x = [0, 1] $, $y = [0, 0.5625]$ in order to maintain square cells. The simulation initial conditions include a dense fluid with density $\rho = 2.0$ in the middle third of the box, surrounded by a less dense fluid with density $\rho = 1.0$ in the outer thirds. The denser fluid has a velocity $u = -0.5$, and the less dense fluid has a velocity $u = 0.5$; 
the $y$-velocities are initially $v=0$. The entire simulation volume is initialized in
pressure equilibrium with $p = 2.5$. A small-amplitude perturbation is added to the $x$- and $y$-velocities 
of every cell in the grid, in proportion to the $x$-position following $u = u + 0.01\mathrm{sin}(2 \pi x)$ 
and $v = v + 0.01 \mathrm{sin}(2 \pi x)$. 
The simulation is evolved to $t = 1.0$, by which time the growth of the eddies has entered the non-linear regime.

As the eddies grow, more mixing between the high density and low density material occurs. Resolving this mixing is an important task for a hydrodynamics code, as the amount of mixing can have a significant impact on broad features in the simulation outcome. The level of mixing tends to increase with resolution as well as with higher order reconstruction techniques. Therefore, Kelvin-Helmholtz instabilities highlight the importance of having an efficient high order reconstruction method and a fast code. As expected for a high resolution grid code with a high order reconstruction method, \cholla does an excellent job of resolving the shear mixing. 


\subsection{3D Hydrodynamics}

\subsubsection{Noh's Strong Shock}\label{sec:noh3D}

The {\it Noh strong shock test}, originally described in one dimension by \cite{Noh87}, demonstrates 
how well a code can track a strong, high mach number shock. This test is considered difficult to perform in either two or three dimensions, as many hydrodynamics codes cannot run the test accurately and some fail completely \citep{LW03}. The test starts with a constant density of $\rho_0 = 1.0$ throughout the grid, with zero pressure and constant velocity $|\bm{\mathrm{V}}| = 1.0$ toward the origin. For this test, the adiabatic index is set to $\gamma = \frac{5}{3}$. These initial conditions result in a formally infinite strength shock reflecting outward from the origin with spherical symmetry. 
\cholla cannot be run with zero pressure, so we set the initial pressure to a low number, $p_0 = 10^{-6}$, but we note that the results are relatively insensitive to the initial pressures below $p_0 \sim10^{-3}$. 

The Noh test is initialized in an octant on the domain $[0, 1]$ with reflecting inner boundaries. The outer boundaries are set according to the analytic solution for the density and energy, which in two or three dimensions is 
\begin{displaymath}
\rho(t) = \rho_0\left(1 + \frac{t}{r}\right)^{n-1},
\end{displaymath}
where $r$ is the radius in polar or spherical coordinates, and $n$ is the dimensionality of the problem. The momentum follows from the velocity and the solution for the density, and the total energy is set to
\begin{displaymath}
E(t) = \frac{p_0}{\gamma - 1} + 0.5\rho(t).
\end{displaymath}
We evolve the solution to $t = 2.0$, by which time the shock has propagated through more than half of the computational domain. The density immediately in front of the shock as well as the density of the post shock gas can also be calculated analytically. In the 3D case, the gas immediately before the shock has a density of $\rho = 16$, and the post-shock gas has a corresponding density of $\rho=64$.

Running the Noh test on a Cartesian grid creates strong, grid-aligned shocks that provoke a behavior in the numerical solution known as the carbuncle instability. The carbuncle instability arises as a result of oscillatory crossflow solutions to the Riemann problem near such shocks \citep{Quirk94}. This problem is addressed by implementing a form of the H correction, as described in \cite{Sanders98} and detailed in Appendix~\ref{app:h_correction}. By incorporating information about the fastest transverse wave speeds, the H correction adds dissipation to the 1D fluxes calculated by the Roe Riemann solver that reduces the carbuncle strength.

\begin{figure}
\centering
\includegraphics[width=1.0\linewidth]{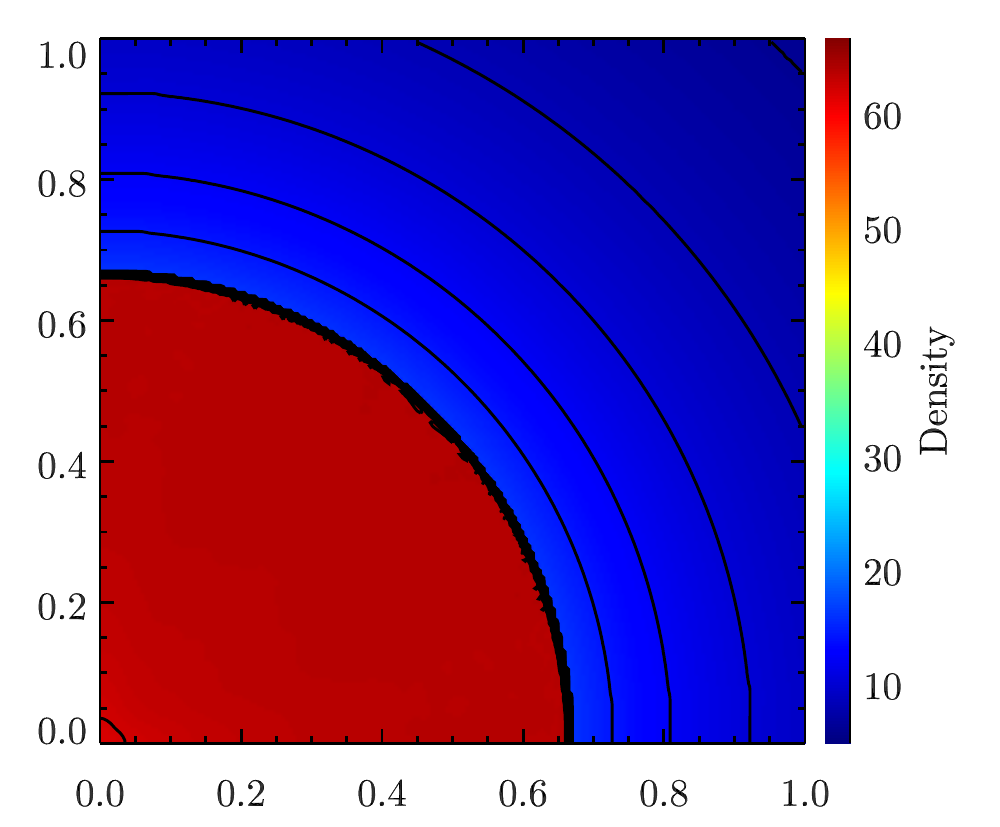}
\includegraphics[width=1.0\linewidth]{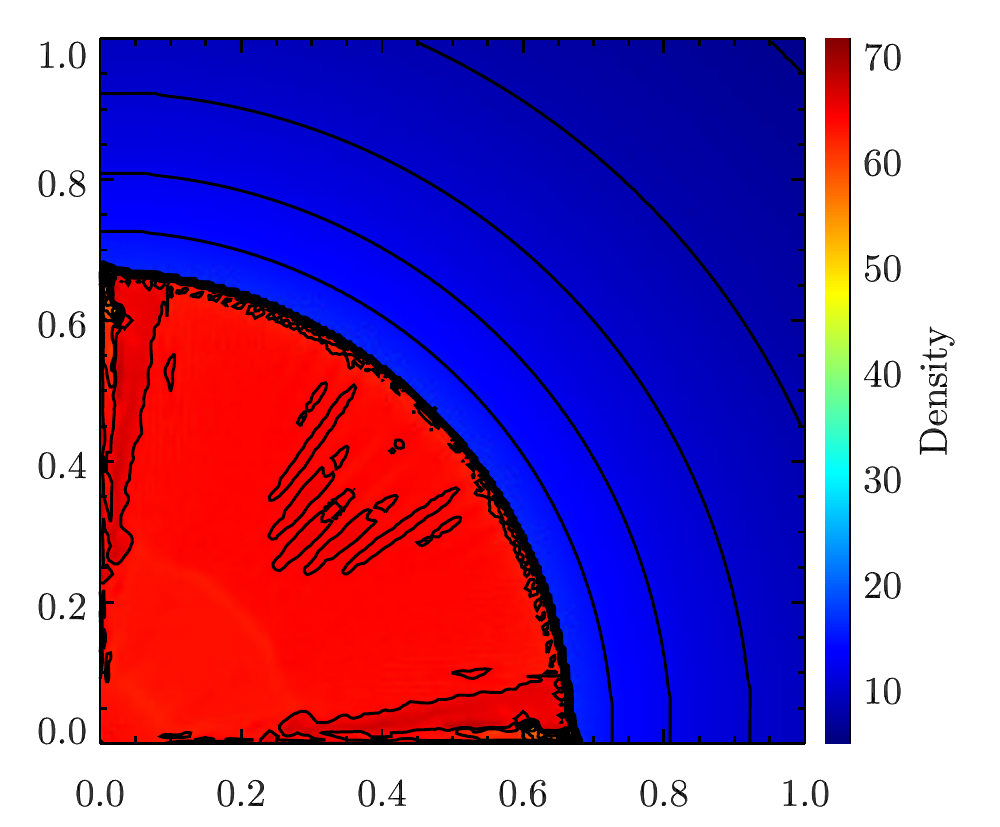}
\caption{Numerical solution of the Noh strong shock test at $t = 2.0$ on a $200^3$ grid. These figures show an $xy$ slice through the $z = 0$ plane. 31 density contours from $4$ to $64$ are overlaid on a color-scale density map. Top: With h correction. Bottom: Without h correction.}
\label{fig:noh_3D}
\end{figure}

The result of the Noh test using PPMC with and without the H correction can be seen in Figure~\ref{fig:noh_3D}. Without the H correction, the solution suffers from strong oscillatory behavior, particularly along the edges where the shock is aligned with the grid. In the version with the H correction applied, the unstable behavior along the axes is effectively absent. The region near the origin where the density dips down is a density error known as ``wall heating" and is not related to the carbuncle instability. Wall heating is the feature that the 1D Noh test was originally designed to demonstrate. The slight noise along the shock front is a result of the strength of the shock, and is similar to the minor oscillations seen in the 1D strong shock test. We note that implementing the H correction increases the stencil required for CTU. Because the current version of \cholla is designed to accommodate a maximum of four ghost cells, we currently implement the H correction only with PPMC or lower-order reconstruction methods.


\section{New Results for Astrophysical Phenomena: Shockwave-ISM Interactions}\label{sec:cloud}

We now showcase the power of \cholla in an astrophysical context by simulating the interaction of supernova blast waves with clouds in the interstellar medium (ISM). Advancing theoretical understanding of the conditions in the ISM and the effects of supernova feedback is an active area of research \citep[e.g.][]{Agertz13a, Kim14a, Martizzi14a}. Stellar feedback is thought to affect the evolution of galaxies on large scales by generating outflows and regulating star formation rates, but the scales on which this feedback couples to the ISM are unresolved in large cosmological simulations \citep{Martin99a, MacLow04a}. Using high resolution methods to constrain the physics of the ISM on sub-parsec scales is thus critical to improving the subgrid prescriptions applied in simulations of galaxy formation. In addition, simulating the ISM on smaller scales provides high-resolution numerical results that can be compared to observations of gas within our galaxy.

The superior shock-capturing abilities of grid-based hydrodynamic codes enables the to serve as an important tool for simulating the types of high-mach number shocks observed in star-forming regions of the ISM. In addition, simulating the interaction of these shocks with gas clouds benefits from high-order spatial reconstruction techniques that accurately trace the hydrodynamic instabilities that develop in ISM gas. Fast, physically accurate codes like \cholla are therefore well-suited for simulating problems like shockwave-ISM interactions. 

Theoretical work describing the interaction between shocks and gas clouds has a long history extending back at least to the calculations of \cite{McKee75a}. In order to treat the problem analytically, these authors presented early computations of a high mach number, planar shock hitting a spherical cloud. Using this simple setup, the speed of the shock within the cloud, $v_\mathrm{cs}$, can be calculated using only the density contrast between the cloud and the ambient medium, $\chi = n_{cl} / n_\mathrm{ism}$, and the speed of the shock in the ambient medium, $v_s$, as follows,
\begin{equation}
\label{eqn:shock_speed}
v_\mathrm{cs} = \chi^{-\frac{1}{2}} v_s.
\end{equation}
\cite{Klein94a} carried out a formative numerical study of the cloud-shock problem, in which they defined a characteristic timescale for the evolution of the cloud. This ``cloud crushing time", $t_\mathrm{cc}$, corresponds roughly to the internal shock crossing time, i.e. $t_\mathrm{cc} = r_\mathrm{cl} / v_\mathrm{cs}$. Using Equation~\ref{eqn:shock_speed}, the cloud crushing time can be related to the radius of the cloud, $r_\mathrm{cl}$, the density contrast $\chi$, and $v_s$, via
\begin{equation}
\label{eqn:t_cc}
t_{\mathrm{cc}} = r_{\mathrm{cl}}\chi^{\frac{1}{2}}/v_s.
\end{equation}

Using this characteristic timescale, various stages of the cloud's evolution can be described. The mixing time of dense cloud gas and the ambient medium holds interest for both quantifying the impact of supernovae on their immediate environments and for the survival time of dense gas in galactic outflows. The shocked cloud experiences various hydrodynamic instabilities that cause its destruction, most importantly the Kelvin-Helmholtz instability (KHI). The long wavelength modes of the KHI tend to break the cloud apart, while the shorter wavelength modes mix cloud material with the surrounding medium. Using 2D adiabatic simulations with density contrasts in the range $10 < \chi < 100$, \cite{Klein94a} demonstrated a spherical cloud is destroyed by large-scale instabilities in $t_\mathrm{dest}  \simeq 3.5 t_\mathrm{cc}$, where the destruction time, $t_{\mathrm{dest}}$, is defined as the time it takes for the mass in the core of the cloud to be reduced to a fraction $1 / e$ of the initial cloud mass. Meanwhile, small-scale instabilities efficiently mix cloud material with the ambient medium in a time of order $4-5 t_\mathrm{cc}$. These results were corroborated in 3D simulations by \cite{Xu95a}.

Subsequent work on the cloud-shock problem has examined how additional physics such as magnetic fields, radiative cooling, self-gravity, and thermal conduction affect the cloud's evolution \citep[e.g.][]{MacLow94a, Fragile05a, Orlando05a, Melioli06a, Shin08a}. These studies have produced many useful results, ranging from the stabilizing effects of magnetic fields in certain configurations to the structural properties of the cloud necessary for gravitational collapse. However, still missing from the literature is an attempt to connect the evolution of clouds with realistic density structures to the analytic theory derived for spheres. Almost all studies of the cloud-shock problem have investigated only spherical or elliptical over-densities, despite the approximately log-normal density distributions of ISM clouds \citep{Padoan02a}. One exception is the work of \cite{Nakamura06a}, which showed that clouds with a steeply tapering density profiles were destroyed and mixed more quickly than those with a shallow density gradient. Another is the study by \cite{Cooper09a}, which included a simulation of a cloud with a fractal density distribution. However, that work focused primarily on the long-term survival of radiatively cooling cloud fragments in a galaxy-scale hot wind and did not attempt to produce an analytic timescale for cloud destruction or mixing for the fractal cloud case. A numerical study evaluating the evolution of a cloud with a realistic density distribution in the context of the analytic timescales defined by \cite{Klein94a} remains to be performed.

In this section, we apply \cholla to a preliminary study of this problem. Our goal is to determine whether a realistic cloud with a given mean density is mixed with the ISM on a timescale comparable to a spherical cloud with the same mass and mean density. If not, we wish to adapt the analytical framework of \cite{Klein94a} to more realistic clouds to characterize their destruction process. To address these issues we carry out a series of high-resolution, 3D hydrodynamic simulations comparing clouds with spherical density distributions to clouds with more realistic density distributions created using a Mach $\sim 5$ turbulence simulation, and we devise a simple alteration to the \cite{Klein94a} formalism that enables and analytical description of the cloud evolution for both spherical and realistic density distributions.

\subsection{The Simulations}

\begin{figure*}[ht]
\centering
\includegraphics[scale=0.7]{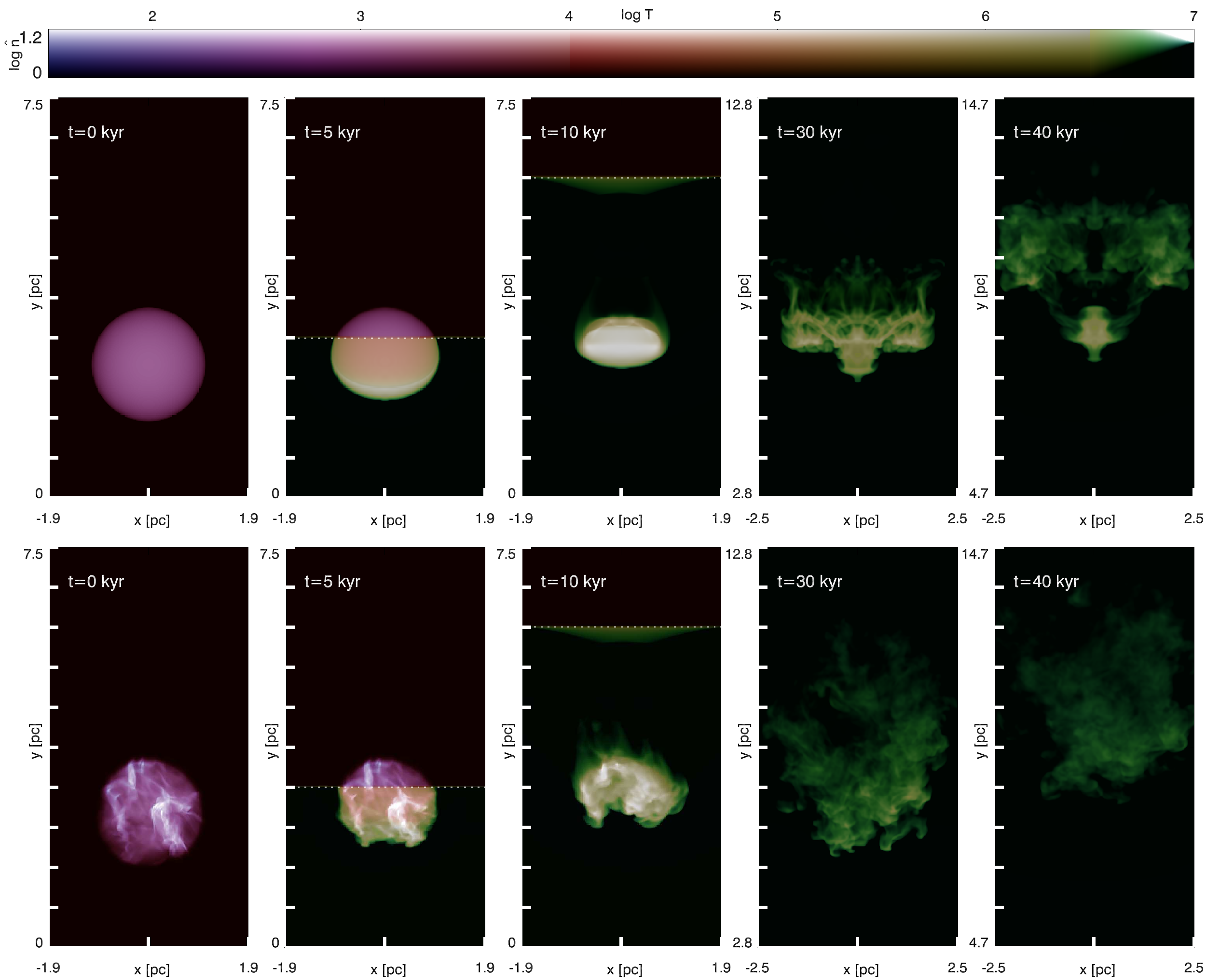}
\caption{Snapshots at $t=0$, $t = 5\, \mathrm{kyr} = 0.375\, t_\mathrm{cc}$, $t= 10\, \mathrm{kyr} = 1.0\, t_\mathrm{cc}$, $t= 30\, \mathrm{kyr} =3.5\, t_\mathrm{cc}$, and $t= 40\, \mathrm{kyr} = 4.75\, t_\mathrm{cc}$ from the intermediate-density cloud-shock simulations. Each snapshot displays an $x-z$ density projection of the part of the domain containing the cloud. Yellow and green regions indicate high temperature shocked gas, while high intensity indicates higher density material. The density scale represents the average of the projected density, $\hat{n}$. At $t = 0$ the spherical cloud is at rest in a $10\,\mathrm{pc} \times 10\,\mathrm{pc} \times 20\,\mathrm{pc}$ simulation box. The cloud is in pressure equilibrium with the ambient medium, which has a temperature of $10^4$ K. At $t = 0.375\, t_\mathrm{cc}$ a Mach 50 shock wave has propagated upward from the bottom of box and is sweeping over the cloud. The shock position of the shock wave is indicated by the white dotted line. At $t = 1.0\, t_\mathrm{cc}$ the internal shock wave has just crossed the spherical cloud, compressing it and increasing the mean density by a factor of 4. The realistic cloud has already begun to re-expand. At $t = 3.5\, t_\mathrm{cc}$ the cloud is being accelerated by the post-shock wind, and large-scale instabilities have disrupted the cloud. At $t = 4.75\, t_\mathrm{cc}$, 50\% of the cloud material has been ablated and mixed with the ISM.}
\label{fig:cloud}
\end{figure*}

We run three sets of simulations with low, medium, and high mean density contrasts, as listed in Table~\ref{tab:mixing_times}. Each simulation is run in a $512 \times 512 \times 1024$ box with side lengths $l = 10\,\mathrm{pc} \times 10\,\mathrm{pc} \times 20\,\mathrm{pc}$, corresponding to a resolution of $0.02\,\mathrm{pc}$. In each simulation, a cloud is placed with its center at $(0, 0, 2.5)$. In all simulations the cloud is initially at rest, and the temperature of the gas is set such that the cloud is in pressure equilibrium with the ambient medium. In the low density simulations, both the spherical cloud and the realistic cloud have a mean density that is 10 times the initial ambient density, $\chi = \hat{n}_\mathrm{cl} / n_\mathrm{ism} = 10$. In the intermediate-density simulations, $\chi = 20$, and in the high-density simulations, $\chi = 40$. The realistic cloud consists of a spherical region excised from a Mach $\sim 5$ turbulence simulation \citep{Robertson12} and has a log-normal density distribution that is truncated at densities below that of the ambient medium. Higher density regions are scaled such that the mean density of the realistic cloud matches that of the spherical cloud. The highest density regions in the realistic clouds are three orders of magnitude above the ambient density, and the lowest temperatures are of order $10$ K. The radius of the spherical cloud in all simulations is $R_{\mathrm{cl}} = 1.07$ pc, and is set such that the total mass in the spherical cloud matches that of the realistic cloud. The low-density clouds (both spherical and realistic) have an initial mass $m_\mathrm{cl, 0} \approx 0.13 \Msun$, the clouds in the intermediate simulation have an initial mass $m_\mathrm{cl, 0} \approx 0.25 \Msun$ while the high-density clouds have an initial mass $m_\mathrm{cl, 0} \approx 0.51 \Msun$. The initial cloud masses and mean densities include all material above the ambient density. Figure~\ref{fig:cloud} shows $x-z$ projections of the initial conditions and several later snapshots for the intermediate-density simulations.

We consider the interaction of small clouds with an old supernova blast wave, such that the radius of shock wave can be assumed to be infinite with respect to the cloud radius, and the shock can therefore be treated as planar. The simulation starts with a planar shock wave propagating upward through the box in the $+z$ direction through an ambient medium with an initial number density of hydrogen atoms $n_h = 0.1\,\mathrm{cm}^{-3}$, and initial temperature of $T = 10^4\,\mathrm{K}$. Using the shock jump conditions in the strong shock limit \citep{Zeldovich66}, the post-shock density, velocity, and pressure are given by
\begin{equation}
\begin{aligned}
n_\mathrm{psh} &= \frac{\gamma+1}{\gamma-1} n_\mathrm{ism}, \\
v_\mathrm{psh} &= \frac{2}{\gamma+1} v_s, \\
p_\mathrm{psh} &\approx \frac{2\gamma}{\gamma+1} M^2 p_\mathrm{ism},
\end{aligned}
\end{equation}
where $v_s = \mathcal{M} c_\mathrm{ism}$ is the shock speed, $\mathcal{M}$ is the Mach number of the shock, and $c_\mathrm{ism}$ is the sound speed in the interstellar medium. For the given initial ISM conditions and an adiabatic index $\gamma = \frac{5}{3}$, a Mach 50 shock travels at $v_s \approx 585\,\mathrm{km}\,\mathrm{s}^{-1}$. The post-shock density is $n_h = 0.4\,\mathrm{cm}^{-3}$, $v_\mathrm{psh} \approx 440\,\mathrm{km}\,\mathrm{s}^{-1}$, and the post-shock temperature is $T = 7.8\times10^6$ K. We set an inflowing $-z$ boundary with the post-shock quantities. All other boundaries are outflowing.

\subsection{Results}

Using Equation~\ref{eqn:t_cc} we can calculate cloud-crushing times for the spherical clouds. For the the intermediate-density simulation, $t_\mathrm{cc} = 8.00\,\mathrm{kyr}$, while for the low- and high-density simulations $t_\mathrm{cc} = 5.65\,\mathrm{kyr}$ and $t_\mathrm{cc} = 11.33\,\mathrm{kyr}$, respectively. In order to analyze the mixing of the clouds in the following analysis, we define the cloud mass as the sum of material with a density $n > 2 n_\mathrm{psh}$, where $n_\mathrm{psh}$ is the post-shock density of the ambient medium, which is $n_h = 0.8\, \mathrm{cm}^{-3}$ for these simulations. As mentioned previously, \cite{Klein94a} defined a destruction time corresponding to the cloud breaking into large fragments. While this timescale is useful for the case of a spherical cloud, the realistic cloud contains many separate regions of high density, with no single well-defined core. Therefore, we investigate instead the mixing time $t_\mathrm{mix}$, defined as the time at which the cloud mass (material with $n > 2 n_\mathrm{psh}$) is reduced to 50\% of the initial cloud mass. This definition leads to mixing times for the spherical clouds that agree to within a few percent of those quoted in \cite{Klein94a}, and an easy comparison between timescales in the spherical and realistic cloud simulations.

\begin{figure}
\centering
\includegraphics[width=1.0\linewidth]{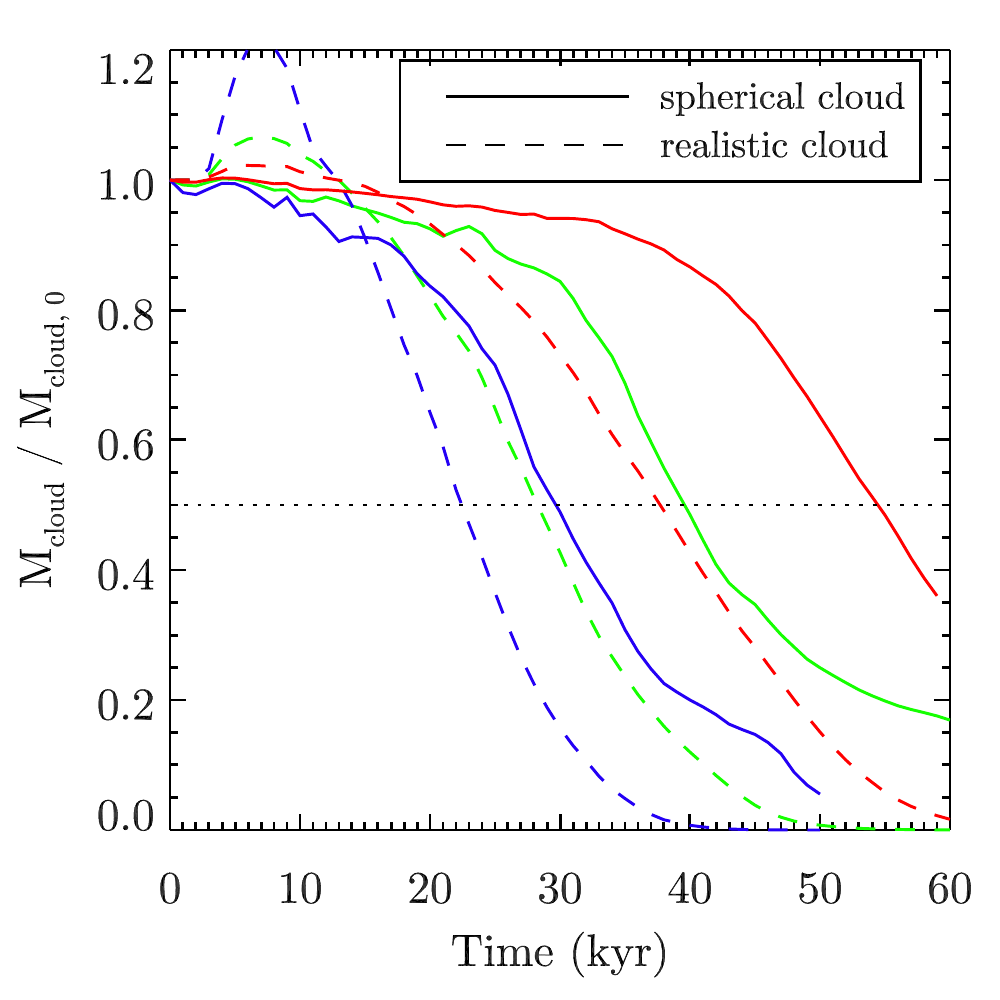}
\caption{The fraction of each cloud's mass as compared to its initial mass, plotted as a function of time. Low-density (blue), intermediate density (green), and high-density (red) cases are shown. The dotted horizontal line indicates the mixing time, $t_\mathrm{mix}$, when 50\% of the cloud material has fallen below the density threshold, $n = 2 n_\mathrm{psh}$. The shock wave hits at $\sim 2$ kyr.}
\label{fig:mass_fractions}
\end{figure}

The mass fractions of the spherical and realistic clouds as a function of time for all cases are shown in Figure~\ref{fig:mass_fractions}. Here we briefly discuss the overall evolution of the intermediate density cases. The shock wave hits at $\sim 2$ kyr, at which point the clouds start to compress. Owing to the definition of cloud mass, the compression of low-density cloud material causes the realistic cloud to initially gain mass. In contrast, the spherical cloud lacks internal low density material that can be compressed above the $n > 2 n_\mathrm{push}$ threshold to add to the cloud mass, and the effective cloud mass begins to slowly decrease. At 10 kyr the shock wave has passed through the spherical cloud, which begins to re-expand (see Figure~\ref{fig:cloud}, middle panels). The shock wave finishes its passage through the realistic cloud about 1 kyr earlier. In both simulations, the cloud is accelerated by the post-shock wind, and hydrodynamic instabilities begin to mix cloud material into the ambient medium. Figure~\ref{fig:cloud} shows the continued evolution of each cloud in snapshots at $t= 3.5\, t_\mathrm{cc}$, the typical cloud destruction time as defined by \cite{Klein94a}, and at $t = 4.75 t_\mathrm{cc}$, the measured mixing time for the intermediate density spherical cloud in our simulations.

As can be seen in Figure~\ref{fig:mass_fractions}, the realistic cloud is mixed significantly earlier than the spherical cloud. In the intermediate-density case, the spherical cloud is mixed in 38 kyrs, or $t = 4.75 t_\mathrm{cc}$, comparable to the results from previous studies \citep[e.g.][]{Klein94a, Nakamura06a}. Table~\ref{tab:mixing_times} shows that the mixing time as a function of $t_\mathrm{cc}$ does not vary substantially as a function of density for the spherical clouds. In contrast, the realistic cloud is mixed in only 27 kyrs. Using a cloud-crushing time estimated from the mean density, this mixing time corresponds to $t_\mathrm{mix} = 3.125 t_\mathrm{cc}$, a much shorter timescale than in the spherical cloud simulation.

\begin{deluxetable}{l l c c c}
\centering
\tablecolumns{5}
\tablecaption{Cloud Properties}
\tablehead{
\colhead{} & \colhead{Parameter} & \colhead{$\chi = 10$} & \colhead{$\chi = 20$} & \colhead{$\chi = 40$}}
\startdata
Spherical Cloud & $\hat{n}_\mathrm{cl}$ &  9.98 & 19.97 & 39.93 \\
 & $t_\mathrm{cc}$ (kyr) &  5.65 & 8.00 & 11.33 \\
   & $t_\mathrm{mix}$ (kyr) &  28 & 38 & 53 \\
 & $t_\mathrm{mix} / t_\mathrm{cc}$ &  4.96 & 4.75 & 4.68 \\
 \\
Realistic Cloud & $\hat{n}_\mathrm{cl}$ &  11.45 & 20.73 & 38.22 \\
 & $n_\mathrm{cl, med}$ &  6.11 & 10.68 & 18.95 \\
 & $t_\mathrm{cc, med}$ (kyr) &  4.42 & 5.84 & 7.79 \\
  & $t_\mathrm{mix}$ (kyr) &  21 & 27 & 36 \\
 & $t_\mathrm{mix} / t_\mathrm{cc, med}$ & 4.75 & 4.62 & 4.62
\enddata
\label{tab:mixing_times}
\tablecomments{Average parameters for both the spherical and realistic clouds for the low-density, intermediate-density, and high-density simulations are listed. Parameters include average cloud density, $\hat{n}_\mathrm{cl}$, median cloud density, $n_\mathrm{cl, med}$, cloud crushing times using either the average (spherical cloud) or median (realistic cloud) density, $t_\mathrm{cc}$ or $t_\mathrm{cc, med}$, mixing times in kyr, and mixing times as a function of cloud-crushing time.}
\end{deluxetable}

Evolution of the low- and high-density simulations is qualitatively similar (see Table~\ref{tab:mixing_times}). In each case, the realistic cloud is mixed much more quickly than the spherical cloud, despite having approximately the same initial mass and mean density. Over the density range explored in this work, the mixing time for the realistic clouds is $68 - 75\%$ that of a spherical cloud with the same mean density. Clearly, the definition of the cloud crushing time using the mean density does not fit the evolution of the realistic cloud very well. In our simulations, a realistic cloud is mixed on a timescale comparable to a spherical cloud with half its mass.

We can explain this new result in terms of the structural properties of the spherical and realistic clouds. While the mean density and mass of the realistic cloud matches that of the spherical cloud, $73\%$ of the volume in the realistic cloud is filled with gas below the mean density. As a result, the internal shock propagates more quickly through the cloud, expediting the introduction of velocity shear between high density cloud material and the post-shock wind. Rather than a single reflected bow shock, the realistic clouds experience many reflected shocks from each high-density region, as can be seen in the middle panel of Figure~\ref{fig:cloud}. The individual high-density regions fill a much smaller volume than the spherical cloud, effectively decreasing the cloud radius term that appears in Equation~\ref{eqn:t_cc}. The increased surface area for velocity shear for each high density region in the realistic cloud leads to material being more quickly ablated and mixed with the ambient medium. In contrast, the high-density core of the spherical cloud is protected by the outer material and survives beyond the mixing time.

The evolution of the realistic cloud can still be described within the analytic framework of \cite{Klein94a} if a more appropriate density is used in the definition of the cloud crushing time. Rather than using the average density, we analyze the mixing time for the realistic cloud using the median density, as the median better represents the density of the volume-fillling gas within the cloud. Median cloud densities and the corresponding cloud-crushing times are given in the bottom half of Table~\ref{tab:mixing_times}. Using these cloud-crushing times, the mixing time for the realistic cloud matches much better the results for the spherical case.

Further examinations of this problem are left for future work, but potentially interesting investigations include a more detailed parameter study incorporating clouds that are not initially at rest (i.e. have a realistic momentum distribution), and shocks with a finite radius of curvature (i.e. a nearby supernova). Nonetheless, our results clearly show that significant quantitative and qualitative differences exist between the destruction and mixing of an idealized, spherical cloud and a cloud with a log-normal density distribution, even when each has a similar mean density. Further, we newly show that these differences can be understood by applying the \cite{Klein94a} formalism adapted to use the median cloud density.


\section{Conclusions}\label{sec:conclusion}

In this work we have presented \chollans, a new, massively-parallel, three-dimensional hydrodynamics code 
optimized for Graphics Processor Units (GPUs). \cholla uses the unsplit Corner Transport Upwind algorithm \citep{Colella90, GS08}, multiple Riemann solvers, and a variety of reconstruction methods to model numerical solutions to the Euler equations on a static mesh. 

In writing the code, we have maintained a modular structure that allows for the implementation of different
hydrodynamical schemes. 
\cholla features five methods for interface reconstruction, including the first-order piecewise constant method, 
two second-order linear reconstruction methods, and two third-order methods based on the original piecewise parabolic method 
developed by \cite{CW84}.  
There are multiple Riemann solvers, including the 
exact solver from \cite{Toro09} and a linear solver based on the method of \cite{Roe81}. 
Incorporating multiple reconstruction and Riemann solver methods 
provides
the ability to test results for a dependence on the particular numerical techniques used, and
the strengths and weaknesses of the different methods are discussed. 
\cholla also implements an optional diffusive correction called the 
H correction \citep{Sanders98} to suppress instabilities along grid-aligned shocks. The H correction adds additional diffusion to the Roe fluxes based on the fastest transverse velocities. 
The Appendices of this paper detail all of the equations used in the code, and supplement the discussion of
each method presented in the main text.

The strategies employed in designing \cholla to run natively on GPUs are extensively detailed. 
The necessity of transferring data to and from the GPU with every time step requires a specific memory layout to improve efficiency. Once information has been transferred to the GPU, the CTU integration algorithm can be effectively divided into kernel functions that execute on the device, similar to functions in a traditional CPU code. Each of these kernels is self-contained and contributes to the modularity of the code. Because the GPU has limited global memory storage, the simulation volume must often be subdivided to optimize performance and compensate for GPU memory constraints. 
The strategy we employ to execute this ``subgrid splitting" efficiently is also presented.

The architectural differences between CPUs and massively-parallel GPUs require the illucidation of
several key concepts in GPU programming.
When a device kernel is called, the GPU launches a large set of individual computational elements called threads. A single kernel call can launch millions of threads. In \chollans, each thread is assigned the work of computing the updated hydrodynamic conserved variables for a single real cell in the simulation volume. The streaming multiprocessors on the GPUs 
handle the work of assigning threads to GPU cores, which means that \cholla will be easily transportable to newer generations of GPU hardware or other coprocessors. 
Thousands of cores operating simultaneously on each device results in thousands of simultaneous cell updates, making the hydrodynamics solver in \cholla very fast. GPUs are designed to optimize for throughput as opposed to latency. 
As demonstrated in the GPU-based time step calculation presented in Section~\ref{sec:timestep},
adding additional floating point calculations to GPU kernels is relatively inexpensive.
This feature can be exploited to incorporate more physics on the GPU, such as cooling, at relatively little computational cost.

The scalability of \cholla and its performance on a range of hydrodynamics problems was also documented. Using the Message Passing Interface (MPI) library \citep{MPIForum94}, \cholla can be run across many GPUs in parallel. 
The code incorporates both slab-based and block-based domain decompositions and exhibits excellent strong and weak scaling in 
block-decomposed tests using at least 64 GPUs. We present the results of a suite of canonical hydrodynamics tests in one, two, 
and three dimensions. The state-of-the art hydrodynamics algorithms employed enables \cholla to perform accurately on a wide variety of tests, and the GPU architecture makes the computations fast without sacrificing physical accuracy. 
Since a single GPU can compute a simulation with nontrivial resolution, using \cholla with a 
cluster presents the option of running many problems to explore large parameter spaces rapidly. 
Further, the excellent weak scaling of \cholla suggests that very large problem sizes can be tackled
on large GPU-enabled supercomputers.

We demonstrate the power of \cholla for astrophysics by addressing the classic numerical problem of a shock hitting a small gas cloud in the interstellar medium. Calculations of the evolution of such clouds require high numerical resolution and excellent shock-capturing abilities \citep{Klein94a}, and \cholla was designed to address such astrophysical problems. Comparing the well-studied ideal case of spherical overdensities to realistic clouds with log-normal density distributions created by supersonic turbulence, we present new results showing that realistic clouds are destroyed more quickly than spherical clouds of the same mass and mean density. In our simulations, realistic clouds are mixed with the ambient medium on timescales comparable to spherical clouds with half their mean density. We posit that the faster destruction time is a result of the lower median density of the gas in the realistic cloud. The shock propagates more quickly through the realistic clouds, allowing the hydrodynamic instabilities responsible for disrupting the cloud and mixing the gas to develop sooner than in the spherical case. We further show that the \cite{Klein94a} formalism can be used to describe the destruction of our realistic cloud simulations provided that the median density is used in place of the average cloud density. These results demonstrate the successful application of \cholla in the performance of high-resolution, physically-accurate astrophysical simulations. We plan to use similar calculations in the future to connect small-scale stellar feedback to galaxy evolution on larger scales.

Lastly, we note that in contrast to many MPI-enabled codes, we have designed \cholla to perform almost all hydrodynamics calculations on the GPU without transferring information to the CPU during the course of a single timestep, leaving the computational power of the CPU to address other tasks (see e.g. Figure~\ref{fig:cholla_diagram}). By performing calculations on the CPU and GPU simultaneously, additional physics could be modeled on the CPU during the hydrodynamical computation on the GPU. 
Logical extensions to \cholla include using Fourier transforms to solve gravity or drive turbulence. The addition of a magnetohydrodynamics module on the GPU is also an attractive possibility, as \cholla uses an unsplit integration algorithm that is optimized for MHD \citep{GS08}.

\acknowledgments

We are grateful to Jordan Stone for creating the acronym \chollans. This material is based upon work supported by the National Science Foundation Graduate Research Fellowship under Grant No.~DGE-1143953, as well as a National Science Foundation Grant No.~1228509.


\appendix

\section{Reconstruction Methods}\label{app:reconstruction}

To calculate the input states for the Riemann solvers used in \chollans, 
appropriate values of the conserved variables at each interface must be reconstructed
using the cell-averaged quantities. Each Riemann problem requires input states at 
either side of a cell interface, referred to as $\bm{U}^*_{L}$ and $\bm{U}^*_{R}$ in 
the context of the CTU algorithm presented in Section~\ref{sec:CTU}. 
While previously $L$ and $R$ indicated the left and right of the interface,
in this Appendix a cell-centered labeling is used to 
document the procedure for computing the input states at the left and right boundaries of a single cell.

The input states calculated at the left and right sides of the cell will be 
labeled $\bm{U}^*_L$ and $\bm{U}^*_R$ in the conserved variables, 
or $\bm{W}^*_L$ and $\bm{W}^*_R$ in the primitive variables. As described in Table~\ref{tab:notation}, the asterisk indicates the time-evolved input state. 
The  boundary values reconstructed before time evolution will be labeled $\bm{W}_L$ and $\bm{W}_R$.  
For each method described below, only the steps involved in the reconstruction for the $x$-interfaces are shown. 
The $y$- and $z$-reconstructions proceed in the same manner but with an appropriate change of stencil. 
The notation will drop the $i$, $j$, and $k$ subscripts unless they are needed for clarification. 


\subsection{PLMP}

The simplest practical reconstruction method implemented in \cholla is PLMP, 
a piecewise linear reconstruction with slope limiting applied in the primitive variables. 
The stencil to calculate the boundary values for cell $i$ contains cells $i-1$ to the left and $i+1$ to the right. 
The first step in the method converts the cell-averaged values of the conserved variables 
into the primitive variables, $\bm{w} = [\rho, u, v, w, p]^{T}$. 
The cell-averaged primitive
values are then used to reconstruct boundary 
values in the primitive variables, $\bm{W}_L$ and $\bm{W}_R$ at the left and right sides of cell $i$ using a local, piece-wise linear reconstruction \citep{Toro09}:
\begin{equation}
\begin{aligned}
\bm{W}_L &= \bm{w}_{i} - \frac{1}{2} \delta\bm{w}_i, \\
\bm{W}_R &= \bm{w}_{i} + \frac{1}{2} \delta\bm{w}_i,
\end{aligned}
\end{equation}
where $\delta\bm{w}_i$ is a vector containing the slope of each primitive variable across cell $i$. 
To compute $\delta\bm{w}_i$, 
we first calculate the left, right, and centered differences in the primitive variables across each of the cell interfaces:
\begin{equation}\label{eqn:diffs}
\delta \bm{w}_L = \bm{w}_{i} - \bm{w}_{i-1}, \quad \delta \bm{w}_R = \bm{w}_{i+1} - \bm{w}_{i}, \quad
\delta \bm{w}_C = 0.5 (\bm{w}_{i+1} - \bm{w}_{i-1}).
\end{equation}
A monotonized central-difference limiter \citep{vanLeer77} is then used to compute $\delta\bm{w}_i$:
\begin{equation}\label{eqn:vanLeer_limiter}
\delta\bm{w}_i =
\begin{cases}
\mathrm{sgn}(\delta\bm{w}_C)\mathrm{min}(|\delta\bm{w}_C|, 2|\delta\bm{w}_L|, 2|\delta\bm{w}_R|), & \delta \bm{w}_L \delta \bm{w}_R > 0 \\
0 ,& \mathrm{otherwise},
\end{cases}
\end{equation}
where $\mathrm{sgn}$ is defined as
\begin{equation}
\mathrm{sgn}(x) =
\begin{cases}
-1, & x < 0 \\
1, & \mathrm{otherwise}.
\end{cases}
\end{equation}
Each of the primitive variables is treated independently in the limiting process, so the vector of
primitive variable slopes (for extrapolations to the left cell face) can be simply written as
\begin{equation}
\delta \bm{w}_L = \{ \delta \rho_L, \delta u_L, \delta v_L, \delta w_L, \delta p_L\}^\mathrm{T}.
\end{equation}
The primitive variable slopes for extrapolating to the right cell face can be similarly defined.

The last step in computing input states for the Riemann problem is to evolve the reconstructed boundary values by half a 
time step $\Delta t / 2$. The time evolution is modeled using the conserved form of the Euler equations, 
and $\bm{W}_L$ and $\bm{W}_R$ are therefore converted back into conserved variables, 
$\bm{U}_L$ and $\bm{U}_R$ and used to calculate the associated fluxes via Equation~\ref{eqn:x_flux}. 
These fluxes are used to evolve the reconstructed boundary values and obtain input states 
appropriate for the Riemann problem:
\begin{equation}
\bm{U}^*_L = \bm{U}_L + 0.5 \frac{\Delta t}{\Delta x}[\bm{F}(\bm{U}_L) - \bm{F}(\bm{U}_R)]
\label{eqn:plmp_evolution_L}
\end{equation}
\begin{equation}
\bm{U}^*_R = \bm{U}_R + 0.5 \frac{\Delta t}{\Delta x}[\bm{F}(\bm{U}_L) - \bm{F}(\bm{U}_R)].
\label{eqn:plmp_evolution_R}
\end{equation}


\subsection{PLMC}

The second reconstruction method, PLMC, is also based on a piecewise linear reconstruction but with the slope limiting
computed using the characteristic variables. The stencil again contains one cell to the left and right of cell $i$. 
After converting the cell-averaged quantities from conserved to primitive variables, an eigenvector decomposition of the 
Euler equations is performed using the characteristic variables as described in Section~\ref{sec:hydrodynamics}. 
First, the eigenvalues of the linear system of equations for cell $i$ are calculated. For adiabatic hydrodynamics, 
the eigenvalues correspond to the three wave speeds,
\begin{equation}\label{eqn:plmc_eigenvalues}
\lambda^{m} = u_i - a_i, \quad
\lambda^{0} = u_i, \quad
\lambda^{p} = u_i + a_i,
\end{equation}
where $a_i$ is the average sound speed in cell $i$. The quantities
$\lambda^m$ and $\lambda^p$ are speeds of the acoustic waves and 
$\lambda^0$ is the speed of the contact wave. The corresponding
eigenvalue for any advected scalar quantity (such as the transverse velocities in multidimensional problems) is simply the speed of the fluid in the normal direction $u_i$. 

The left ($\delta \bm{w}_L$), right ($\delta \bm{w}_R$), and centered ($\delta \bm{w}_C$) differences in the primitive variables shown in Equation~\ref{eqn:diffs} are then calculated. These differences are projected 
onto the characteristic variables, $\delta \bm{\xi}$, using the left eigenvectors given in Appendix A of \cite{Stone08}. Rather than reproduce the the expressions for each eigenvector, equations describing the final
projections are shown since they
are actually used in the GPU kernel. The projection of the left difference is
\begin{equation}
\delta\bm{\xi}_L = 
\begin{bmatrix} -0.5 (\rho_i \delta u_{L} / a_i + \delta p_{L} / a_i^2) \\ 
			\delta \rho_{L} - \delta p_{L} / a_i^2 \\ 
			\delta v_{L} \\
			\delta w_{L} \\
			0.5 (\rho_i \delta u_{L} / a_i + \delta p_{L} / a_i^2) 
\end{bmatrix},
\end{equation}
where $\delta \rho_L$, $\delta u_L$, $\delta v_L$, $\delta w_L$, and $\delta p_L$ are the components of the primitive variable difference vector $\delta \bm{w}_L$. The projections for the right and central differences 
are calculated in the same manner, yielding $\delta \bm{\xi}_R$ and $\delta \bm{\xi}_C$.

The characteristic differences are then monotonized using the \citet{vanLeer77} limiter, computed as
\begin{equation}
\delta \bm{\xi} = 
\begin{cases}
\mathrm{sgn}(\delta \bm{\xi}_C) \mathrm{min}( |\delta \bm{\xi}_C|, 2 |\delta \bm{\xi}_L|, 2 |\delta \bm{\xi}_R|), & \delta \bm{\xi}_L \delta \bm{\xi}_R > 0 \\
0 ,& \mathrm{otherwise}.
\end{cases}
\end{equation}
We project the monotonized differences in the characteristic variables back onto the primitive variables, 
providing slopes in each variable that are analogous to the limited slopes described in PLMP:
\begin{equation}\label{eqn:plmc_lim_slopes}
\delta\bm{w}_i =
\begin{bmatrix} \delta \xi_0 + \delta \xi_1 + \delta \xi_4 \\ 
			(a_i / \rho_i)( -\delta \xi_0 + \delta \xi_4) \\ 
			\delta \xi_2 \\
			\delta \xi_3 \\
			a_i^2 (\delta \xi_0 + \delta \xi_4)
\end{bmatrix}.
\end{equation}
Here, the numeric subscripts to refer to the components of the vector $\bm{\xi}$.

As in PLMP, these slopes are subsequently used
to create a linear interpolation for reconstructing boundary values of the primitive variables:
\begin{equation}\label{eqn:plm_interpolation}
\begin{aligned}
\bm{W}_{L,A} &= \bm{w}_{i} - \frac{1}{2} \delta\bm{w}_i, \\
\bm{W}_{R,A} &= \bm{w}_{i} + \frac{1}{2} \delta\bm{w}_i.
\end{aligned}
\end{equation}
The primitive variable boundary values 
are further monotonized to ensure that they are numerically bounded by the neighboring cell values: 
\begin{equation}
\begin{aligned}
\bm{W}_{L,B} &= \mathrm{max}[\mathrm{min}(\bm{w}_i, \bm{w}_{i-1}), \bm{W}_{L,A}] \\
\bm{W}_L &= \mathrm{min}[\mathrm{max}(\bm{w}_i, \bm{w}_{i-1}), \bm{W}_{L,B}] \\
\bm{W}_{R,B} &= \mathrm{max}[\mathrm{min}(\bm{w}_i, \bm{w}_{i+1}), \bm{W}_{R,A}] \\
\bm{W}_R &= \mathrm{min}[\mathrm{max}(\bm{w}_i, \bm{w}_{i+1}), \bm{W}_{R,B}],
\end{aligned}
\end{equation}
enabling a slope vector to be computed from these adjusted boundary values as
\begin{equation}
\delta\bm{w}_i = \bm{W}_R - \bm{W}_L.
\end{equation}

The reconstructed boundary values must be evolved in time to calculate appropriate input states for the Riemann problem. Instead of simply evolving the reconstructed values using associated fluxes as in Equations~\ref{eqn:plmp_evolution_L} and \ref{eqn:plmp_evolution_R}, the characteristic tracing method 
of \cite{CW84} is employed. To obtain a first approximation for the input states, an 
integration under the linear interpolation 
function is performed using the minimum wave speed to define the domain of dependence for the left side of the cell, $\lambda^m$, and the maximum wave speed for the right side of the cell, $\lambda^p$:
\begin{equation}
\begin{aligned}
\bm{\tilde{W}}^*_L &= \bm{W}_L - 0.5 \frac{\Delta t}{\Delta x}\mathrm{min}(\lambda^m, 0) \delta\bm{w}_i  \\
\bm{\tilde{W}}^*_R &= \bm{W}_R - 0.5 \frac{\Delta t}{\Delta x}\mathrm{max}(\lambda^p, 0) \delta\bm{w}_i
\end{aligned}
\end{equation}
The input states are then corrected by accounting for the portion 
of each wave that does not reach the interface over a time $\Delta t / 2$ as a result of the presence of the other waves. 
Correction terms are only needed for characteristics propagating toward each interface. 
The eigenvector projection and correction for each element of $\bm{W}^*$ is shown below, 
tracking the correction terms in the vectors $\bm{s}_L$ and $\bm{s}_R$. For the left side of cell $i$,
\begin{equation}
\bm{W}^*_L = \bm{\tilde{W}}^*_L  + 0.5\frac{\Delta t}{\Delta x} \bm{s}_L
\end{equation}
with
\begin{equation}
\bm{s}_L = 
\begin{bmatrix} (\lambda^m - \lambda^0)(\delta\rho_i - \delta p_i / a_i^2) \\ 
			0 \\ 
			(\lambda^m - \lambda^0) \delta v_i \\
			(\lambda^m - \lambda^0) \delta w_i \\
			0
\end{bmatrix}
+
\begin{bmatrix}  0.5 (\lambda^m - \lambda^p)(\rho_i \delta u_i / a_i + \delta p_i / a_i^2) \\ 
			0.5 (\lambda^m- \lambda^p)[\delta u_i + \delta p_i/(a_i\rho_i)] \\ 
			0 \\
			0 \\
			0.5 (\lambda^m - \lambda^p)(\rho_i \delta u_i a_i + \delta p_i)
\end{bmatrix}.
\end{equation}
The first term is associated with the contact wave and is added only if $\lambda^0 < 0$. The second term is associated with the right acoustic wave and is added only if $\lambda^p < 0$. If both $\lambda^0$ and $\lambda^p$ are greater than 0, there is no need for a correction because those waves cannot affect the left interface of the cell.
For the right side of cell $i$,
\begin{equation}
\bm{W}^*_R = \bm{\tilde{W}}^*_R  + 0.5\frac{\Delta t}{\Delta x} \bm{s}_R
\end{equation}
with
\begin{equation}
\bm{s}_R = 
\begin{bmatrix} 0.5(\lambda^p - \lambda^m)(-\rho_i \delta u_i / a_i + \delta p_i / a_i^2) \\ 
			0.5(\lambda^p - \lambda^m)[\delta u_i - \delta p_i / (a_i \rho_i)] \\ 
			0 \\
			0  \\
			0.5(\lambda^p - \lambda^m)(-\rho_i \delta u_i a_i + \delta p_i)
\end{bmatrix}
+
\begin{bmatrix} (\lambda^p - \lambda^0)(\delta\rho_i - \delta p_i / a_i^2) \\ 
			0 \\ 
			(\lambda^p - \lambda^0) \delta v_i \\
			(\lambda^p - \lambda^0) \delta w_i \\
			0
\end{bmatrix}.
\end{equation}
Here the first term is associated with the left acoustic wave and is added only if $\lambda^m > 0$, while the second term is associated with the contact wave and is added only if $\lambda^0 > 0$. As with the left interface, the corrections only apply if the waves are moving toward the interface. Once the corrections have been made, $\bm{W}^*_{L}$ and $\bm{W}^*_{R}$ can be used as inputs to the Riemann problem.


\subsection{PPMC}

\cholla also includes implementations of third-order spatial reconstruction techniques, including
two varieties of the piecewise parabolic method developed by \cite{CW84}. 
The piecewise parabolic method with slope limiting applied in the characteristic variables (PPMC) is presented
first as it shares several steps with PLMC. 
Our implementation of this method closely follows that outlined in \cite{Stone08}.

The first step in PPMC is to calculate monotonized slopes for cells $i-1$, $i$, and $i+1$. These slopes are labeled 
$\delta\bm{w}_{i-1}$, $\delta\bm{w}_{i}$, and $\delta\bm{w}_{i+1}$. The limited slopes for each cell are calculated in a 
manner identical to that described in PLMC, following Equations \ref{eqn:plmc_eigenvalues} - \ref{eqn:plmc_lim_slopes}. 
Since slopes must be calculated for all three cells, the stencil for PPMC contains two cells to the left and right of 
cell $i$. Once the limited slope vectors for all three cells have been calculated, the algorithm proceeds as follows.

Using the monotonized linear slopes, a parabolic interpolation is computed and used to 
calculate the reconstructed boundary values:
\begin{equation}\label{eqn:parabolic_interpolation}
\begin{aligned}
\bm{W}_{L,A} &= \frac{1}{2} (\bm{w}_{i} + \bm{w}_{i-1}) - \frac{1}{6} (\delta\bm{w}_i - \delta\bm{w}_{i-1}), \\
\bm{W}_{R,A} &= \frac{1}{2} (\bm{w}_{i+1} + \bm{w}_{i}) - \frac{1}{6} (\delta\bm{w}_{i+1} - \delta\bm{w}_{i}).
\end{aligned}
\end{equation}
Monotonicity constraints are applied to the reconstructed boundary values to 
ensure that they lie between the average values in neighboring cells. 
If the cell contains a local minimum or maximum, both interface values are set equal to the cell average:
\begin{equation}\label{eqn:max_min}
\bm{W}_{L,A} = \bm{W}_{R,A} = \bm{w}_i, \quad \mathrm{if} \
(\bm{W}_{R,A} - \bm{w}_i)(\bm{w}_i - \bm{W}_{L,A}) \leq 0
\end{equation}
If the parabolic interpolation violates monotonicity as a result of a steep gradient, the interface values are modified
as
\begin{equation}\label{eqn:steep_gradient}
\begin{aligned}
\bm{W}_{L,A} &= 3\bm{w}_i - 2\bm{W}_{R,A}, \quad \mathrm{if} \
(\bm{W}_{R,A} - \bm{W}_{L,A})[\bm{W}_{L,A} - (3\bm{w}_i - 2\bm{W}_{R,A})] < 0 \\
\bm{W}_{R,A} &= 3\bm{w}_i - 2\bm{W}_{L,A}, \quad \mathrm{if} \
(\bm{W}_{R,A} - \bm{W}_{L,A})[(3\bm{w}_i - 2\bm{W}_{L,A}) - \bm{W}_{R,A}] < 0
\end{aligned}
\end{equation}
These reconstructed boundary values are further adjusted using the minmod limiter operation:
\begin{equation}
\begin{aligned}
\bm{W}_{L,B} &= \mathrm{max}[\mathrm{min}(\bm{w}_i, \bm{w}_{i-1}), \bm{W}_{L,A}] \\
\bm{W}_L &= \mathrm{min}[\mathrm{max}(\bm{w}_i, \bm{w}_{i-1}), \bm{W}_{L,B}] \\
\bm{W}_{R,B} &= \mathrm{max}[\mathrm{min}(\bm{w}_i, \bm{w}_{i+1}), \bm{W}_{R,A}] \\
\bm{W}_R &= \mathrm{min}[\mathrm{max}(\bm{w}_i, \bm{w}_{i+1}), \bm{W}_{R,B}]
\end{aligned}
\end{equation}

At this point, a monotonized parabolic interpolation can be reconstructed. New slopes 
are computed that account for the adjusted boundary values:
\begin{equation}
\delta\bm{w}_i = \bm{W}_R - \bm{W}_L.
\end{equation}
These slopes are used
to compute the time-evolved left and right boundary values by integrating under a parabolic interpolation function:
\begin{equation}\label{eqn:integration_PPMC}
\begin{aligned}
\bm{\tilde{W}}^*_L &= \bm{W}_L - \frac{1}{2} \alpha^m \left[\delta\bm{w}_i + \bm{w}_6 (1 + \frac{2}{3}\alpha^m)\right], \\
\bm{\tilde{W}}^*_R &= \bm{W}_R - \frac{1}{2} \beta^p \left[\delta\bm{w}_i -  \bm{w}_6 (1 - \frac{2}{3} \beta^p)\right],
\end{aligned}
\end{equation}
where
\begin{equation}
\bm{w}_6 = 6\bm{w}_i - 3(\bm{W}_L - \bm{W}_R).
\end{equation}
Here we have borrowed from the notation of \cite{CW84} to define $\alpha^m$ and $\beta^p$, unit-free 
variables associated with the characteristic speeds
\begin{equation}
\begin{aligned}
\alpha^m = \frac{\Delta t}{\Delta x}\mathrm{min}(\lambda^m, 0), &\quad \beta^m = \frac{\Delta t}{\Delta x}\mathrm{max}(\lambda^m, 0), \\
\alpha^0 = \frac{\Delta t}{\Delta x}\mathrm{min}(\lambda^0, 0), &\quad \beta^0 = \frac{\Delta t}{\Delta x}\mathrm{max}(\lambda^0, 0), \\
\alpha^p= \frac{\Delta t}{\Delta x}\mathrm{min}(\lambda^p, 0), &\quad \beta^p = \frac{\Delta t}{\Delta x}\mathrm{max}(\lambda^p, 0).
\end{aligned}
\end{equation}
As in PLMC, the minimum characteristic speed, $\lambda^m$, is used to define the domain of dependence for the left 
interface, and the maximum characteristic speed, $\lambda^p$, is used for the right interface. The primitive variable
time-evolved boundary values $\bm{\tilde{W}}^*_L$ and $\bm{\tilde{W}}^*_R$ are first approximations to the input states for the Riemann problem.

The input states must now be corrected by accounting for the other characteristics propagating 
toward the interface. 
At the left side of the cell, we compute
\begin{equation}
\bm{s}_L = 
\begin{bmatrix} \bm{E}_0 - \bm{E}_4 / a_i^2 \\ 
			0 \\ 
			\bm{E}_2 \\
			\bm{E}_3 \\
			0
\end{bmatrix}
+
\begin{bmatrix}  0.5 (\rho_i \bm{B}_1 / a_i + \bm{B}_4 / a_i^2) \\ 
			0.5 \left[\bm{B}_1 + \bm{B}_4/(a_i\rho_i)\right] \\ 
			0 \\
			0 \\
			0.5 (\rho_i \bm{B}_1 a_i + \bm{B}_4)
\end{bmatrix},
\end{equation}
where the first term is added only if $\lambda^0 < 0$, and the second only if $\lambda^p < 0$ - otherwise there is no correction. In the above,
\begin{equation}
\bm{E} = \frac{1}{2}\alpha^m \left[\delta\bm{w}_i + \bm{w}_6 (1 + \frac{2}{3}\alpha^m) \right] - \frac{1}{2}\alpha^0 \left[\delta\bm{w}_i + \bm{w}_6 (1 + \frac{2}{3}\alpha^0) \right],
\end{equation}
is associated with the contact wave, and
\begin{equation}
\bm{B} = \frac{1}{2}\alpha^m \left[\delta\bm{w}_i + \bm{w}_6 (1 + \frac{2}{3}\alpha^p) \right] - \frac{1}{2}\alpha^p \left[\delta\bm{w}_i + \bm{w}_6 (1 + \frac{2}{3}\alpha^p) \right].
\end{equation}
is associated with the right-most acoustic wave. Subscripts denote the elements of $\bm{E}$ and $\bm{B}$.
Similarly,  for the right side of the cell,
\begin{equation}
\bm{s}_R = 
\begin{bmatrix} 0.5(-\rho_i \bm{C}_1 / a_i + \bm{C}_4 / a_i^2) \\ 
			0.5\left[\bm{C}_1 - \bm{C}_4 / (a_i \rho_i)\right] \\ 
			0 \\
			0  \\
			0.5(-\rho_i \bm{C}_1 a_i + \bm{C}_4)
\end{bmatrix}
+
\begin{bmatrix} (\bm{D}_0 - \bm{D}_4 / a_i^2) \\ 
			0 \\ 
			\bm{D}_2 \\
			\bm{D}_3 \\
			0
\end{bmatrix}.
\end{equation}
The first term accounts for the correction owing to the left-most acoustic wave and is added only if $\lambda^m > 0$, and the second term accounts for the correction from the contact wave and is added only if $\lambda^0 > 0$. In this case,
\begin{equation}
\begin{aligned}
\bm{C} = \frac{1}{2}\beta^p \left[\delta\bm{w}_i - \bm{w}_6 (1 - \frac{2}{3}\beta^p) \right] &- \frac{1}{2}\beta^m \left[\delta\bm{w}_i - \bm{w}_6 (1 - \frac{2}{3}\beta^m) \right], \\
\bm{D} = \frac{1}{2}\beta^p \left[\delta\bm{w}_i - \bm{w}_6 (1 - \frac{2}{3}\beta^p) \right] &- \frac{1}{2}\beta^0 \left[\delta\bm{w}_i + \bm{w}_6 (1 + \frac{2}{3}\beta^0) \right].
\end{aligned}
\end{equation}
With these correction terms input states for the Riemann problem can be calculated as
\begin{equation}
\bm{W}^*_L = \bm{\tilde{W}}^*_L + \bm{s}_L \quad \mathrm{and} \quad \bm{W}^*_R = \bm{\tilde{W}}^*_R + \bm{s}_R.
\end{equation}


\subsection{PPMP}

The \cholla implementation of the piecewise parabolic method computed in the primitive variables closely follows the 
original description in \cite{CW84}, with some additional notation adapted from \cite{Fryxell00}. 
For convenience, the following description assumes a uniform cell size.
PPMP is the most complicated of the reconstruction methods implemented 
in \chollans, and the algorithm follows this brief outline:
\begin{enumerate}
\item Reconstruct boundary values using parabolic interpolation.
\item Steepen contact discontinuities, if necessary.
\item Flatten shocks, if necessary.
\item Ensure that the parabolic distribution is monotonic.
\item Integrate under the parabolic interpolation to determine input states for the Riemann problem.
\item Use characteristic tracing to correct the input states.
\end{enumerate}

\subsubsection{Parabolic Interpolation}

The first step in PPMP is to reconstruct the boundary values using a parabolic interpolation with limited slopes. The interpolation is identical to that shown in Equation~\ref{eqn:parabolic_interpolation}:
\begin{equation}
\begin{aligned}
\bm{W}_L &= \frac{1}{2}(\bm{w}_{i} + \bm{w}_{i-1}) - \frac{1}{6} (\delta\bm{w}_i - \delta\bm{w}_{i-1}), \\
\bm{W}_R &= \frac{1}{2}(\bm{w}_{i+1} + \bm{w}_{i}) - \frac{1}{6} (\delta\bm{w}_{i+1} - \delta\bm{w}_{i}).
\end{aligned}
\end{equation}
However, in PPMP the slopes are limited in the primitive variables using the \citet{vanLeer77} limiter.
The slopes $\delta\bm{w}_{i-1}$, $\delta\bm{w}_i$, and $\delta\bm{w}_{i+1}$ are all calculated
following Equations~\ref{eqn:diffs} and \ref{eqn:vanLeer_limiter}.

\subsubsection{Contact Discontinuity Steepening}

Once the interface values have been reconstructed, contact discontinuity steepening is applied 
to the interface values for the density, $\bm{W}_L[0]=\rho_L$ and $\bm{W}_R[0]=\rho_R$.
Whether steepening is applied depends on a number of necessary criteria. 
First, to avoid steepening density jumps owing to numerical noise, steepening is only applied if the density difference
between cells exceeds a minimum relative size:
\begin{equation}
\frac{|\rho_{i+1} - \rho_{i-1}|}{\mathrm{min}(\rho_{i+1}, \rho_{i-1})} > 0.01.
\label{eqn:density_jump}
\end{equation}
If the density jump is large enough, we further require that the pressure jump across the cell be sufficiently small:
\begin{equation}
\frac{|p_{i+1} - p_{i-1}|}{\mathrm{min}(p_{i+1}, p_{i-1})} < 0.1\gamma \frac{|\rho_{i+1} - \rho_{i-1}|}{\mathrm{min}(\rho_{i+1}, \rho_{i-1})}.
\label{eqn:pressure_jump}
\end{equation}
Next, the second derivative of the density distribution across cells $i-1$ and $i+1$ is estimated as
\begin{equation}
\delta^2\rho_i = \frac{\rho_{i+1}-2\rho_i+\rho_{i-1}}{6 \Delta x^2}.
\end{equation}
The product of the second derivatives then determines whether the local density profile on either side of the
cell $i$ has the same concavity by requiring
\begin{equation}
\delta^2\rho_{i-1}\delta^2\rho_{i+1} > 0.
\label{eqn:density_profile}
\end{equation}
Assuming the three conditions listed in Equations~\ref{eqn:density_jump}, \ref{eqn:pressure_jump}, and \ref{eqn:density_profile} are satisfied, a dimensionless quantity involving the first and third derivatives of density is calculated as
\begin{equation}
\tilde{\eta}_i = -\frac{(\delta^2\rho_{i+1} - \delta^2\rho_{i-1})\Delta x^2}{\rho_{i+1} - \rho_{i-1}}.
\end{equation}
This quantity is used to compute the {\it steepening coefficient} $\eta_i$ from 
the parameters determined heuristically in \cite{CW84}:
\begin{equation}
\eta_i = \mathrm{max}\left[0, \mathrm{min}(20\tilde{\eta}_i - 1, 1)\right].
\end{equation}
The steepening coefficient $\eta_i$ and the monotonized slopes $\delta \bm{w}_{i-1}[0] = \delta\rho_{i-1}$ and $\delta \bm{w}_{i+1}[0] = \delta\rho_{i+1}$ are then used 
to steepen the left and right interface density values, providing
\begin{equation}
\begin{aligned}
\rho_L &= \rho_L(1-\eta_i) + (\rho_{i-1} + 0.5\delta\rho_{i-1})\eta_i \\
\rho_R &= \rho_R(1-\eta_i) + (\rho_{i-1} - 0.5\delta\rho_{i+1})\eta_i
\end{aligned}
\end{equation}

\subsubsection{Shock Flattening}

Because of their self-steepening property, shocks in PPMP can become under-resolved, i.e. narrow enough to be contained within a single cell. Tests have demonstrated that shocks contained within a single cell tend to lead to severe oscillations near the shock front, while those spread over two or more cells do not pose a problem \citep{CW84, Fryxell00}. 
A solution is to flatten numerically the interpolation near problematic shocks,
reverting to a first-order reconstruction in circumstances where the shock is empirically deemed too narrow.
 To determine whether a shock needs flattening, a shock steepness parameter $S$ is calculated 
 to compare the pressure gradient across two and four cells:
\begin{equation}
S_i = \frac{p_{i+1} - p_{i-1}}{p_{i+2} - p_{i-2}}
\end{equation}
The steepness parameter is used to construct a dimensionless coefficient
\begin{equation}
\tilde{F}_i = \mathrm{max}(0, \mathrm{min}[1, 10(S_i - 0.75)])
\end{equation}
that may cover the range $\tilde{F}_i=[0,1]$.
This formulation is designed to ensure that only shocks contained within fewer than two cells 
are steepened. 
Further, we set $\tilde{F}_i=0$ if the relative pressure jump is not large and the shock is not steep, when
\begin{equation}
\frac{|p_{i+1} - p_{i-1}|}{\mathrm{min}(p_{i+1}, p_{i-1})} < \frac{1}{3}
\end{equation}
or if the velocity gradient is positive (indicating that the fluid is not being compressed in the direction along which we are reconstructing boundary values), when
\begin{equation}
u_{i+1} - u_{i-1} > 0.
\end{equation}
The same rules are applied to the dimensionless parameters $\tilde{F}_{i-1}$ and $\tilde{F}_{i+1}$. This procedure means that PPMP requires a stencil with three cells on both sides of cell $i$. Here we are calculating shocks in the 
$x$-direction; in the $y$- and $z$-directions, the components 
$v$ and $w$ are used to test the velocity gradient. The final flattening coefficient for cell $i$ is set as
\begin{equation}
F_i = 
\begin{cases}
\mathrm{max}(\tilde{F}_i, \tilde{F}_{i+1}), \quad &\mathrm{if} \ p_{i+1} - p_{i-1} < 0, \\
\mathrm{max}(\tilde{F}_i, \tilde{F}_{i-1}), \quad &\mathrm{otherwise}.
\end{cases}
\end{equation}
We use this value of $F_i$ to modify the interface values. Unlike in discontinuity steeping, for shock
flattening every primitive variable is modified:
\begin{equation}
\begin{aligned}
\bm{W}_L &= F_i \bm{w}_i + (1 - F_i) \bm{W}_L \\
\bm{W}_R &= F_i \bm{w}_i + (1 - F_i) \bm{W}_R
\end{aligned}
\end{equation}
If $F_i = 0$, the expression has no effect on the interface values.
If $F_i = 1$ the zone average values 
are used for the interface values, effectively
replacing the original limited slope for cell $i$ with a flatter slope.

\subsubsection{Monotonicity}

The next step in the reconstruction is to ensure that the parabolic distribution of each of the variables is monotonic by checking for local maxima and minima and modifying steep gradients, as in Equations~\ref{eqn:max_min} and \ref{eqn:steep_gradient}:
\begin{equation}
\begin{aligned}
\bm{W}_L = \bm{W}_R = \bm{w}_i, \quad &\mathrm{if} \
(\bm{W}_R - \bm{w}_i)(\bm{w}_i - \bm{W}_L) <= 0 \\
\bm{W}_L = 3\bm{w}_i - 2\bm{W}_R, \quad &\mathrm{if} \
(\bm{W}_R - \bm{W}_L)[\bm{W}_L - (3\bm{w}_i - 2\bm{W}_R)] < 0 \\
\bm{W}_L = 3\bm{w}_i - 2\bm{W}_L, \quad &\mathrm{if} \
(\bm{W}_R - \bm{W}_L)[(3\bm{w}_i - 2\bm{W}_L) - \bm{W}_R] < 0.
\end{aligned}
\end{equation}

\subsubsection{Calculation of the Input States}

By this stage, reconstruction of the boundary values has been completed and the input states for the Riemann problem
can be calculated.
Once again, the characteristic tracing method of \cite{CW84} is used. 
First, the speeds of the three characteristics are defined as in PLMC and PPMC:
\begin{equation}
\lambda^{m} = u_i - a_i, \quad
\lambda^{0} = u_i, \quad
\lambda^{p} = u_i + a_i.
\end{equation}
Again, $a_i$ is the sound speed in cell $i$ calculated using average values of the density and pressure. Because we have adjusted the boundary values from the original parabolic interpolation, we must adjust the values of the slopes across the cell so that the parabolic interpolation retains the correct average value:
\begin{equation}
\delta\bm{w}_i = \bm{W}_R - \bm{W}_L, \quad \bm{w}_6 = 6\left[\bm{w}_i - 0.5(\bm{W}_L + \bm{W}_R)\right].
\end{equation}
We now define $\alpha$ and $\beta$, two variables that are associated with the characteristic wave speeds approaching the left and right interfaces,
\begin{equation}
\begin{aligned}
\alpha^m = \frac{\Delta t}{\Delta x}\mathrm{min}(\lambda^m, 0), &\quad \beta^m = \frac{\Delta t}{\Delta x}\mathrm{max}(\lambda^m, 0), \\
\alpha^0 = \frac{\Delta t}{\Delta x}\mathrm{min}(\lambda^0, 0), &\quad \beta^0 = \frac{\Delta t}{\Delta x}\mathrm{max}(\lambda^0, 0), \\
\alpha^p= \frac{\Delta t}{\Delta x}\mathrm{min}(\lambda^p, 0), &\quad \beta^p = \frac{\Delta t}{\Delta x}\mathrm{max}(\lambda^p, 0).
\end{aligned}
\end{equation}
We use these variables to calculate a time-evolved boundary value associated with each characteristic:
\begin{equation}\label{eqn:integration_PPMP}
\begin{aligned}
\bm{W}^m_L &= \bm{W}_L - \frac{1}{2} \alpha^m \left[\delta\bm{w}_i + \bm{w}_6(1 + \frac{2}{3} \alpha^m)\right] \\
\bm{W}^0_L &= \bm{W}_L - \frac{1}{2} \alpha^0 \left[\delta\bm{w}_i + \bm{w}_6(1 + \frac{2}{3} \alpha^0)\right] \\
\bm{W}^p_L &= \bm{W}_L - \frac{1}{2} \alpha^p \left[\delta\bm{w}_i + \bm{w}_6(1 + \frac{2}{3} \alpha^p)\right] \\
\bm{W}^m_R &= \bm{W}_R - \frac{1}{2} \beta^m \left[\delta\bm{w}_i - \bm{w}_6(1 - \frac{2}{3} \beta^m)\right] \\
\bm{W}^0_R &= \bm{W}_R - \frac{1}{2} \beta^0 \left[\delta\bm{w}_i - \bm{w}_6(1 - \frac{2}{3} \beta^0)\right] \\
\bm{W}^p_R &= \bm{W}_R - \frac{1}{2} \beta^p \left[\delta\bm{w}_i - \bm{w}_6(1 - \frac{2}{3} \beta^p)\right]
\end{aligned}
\end{equation}
For example, $\bm{W}^m_L$ is the time-evolved boundary value at the left interface of the cell obtained by integrating under the characteristic associated with the left acoustic wave. If the fluid flow is supersonic toward the right, the left acoustic wave is not approaching the interface, and $\bm{W}^m_L$ is simply equal to the reconstructed boundary value. The same integration appeared in Equation~\ref{eqn:integration_PPMC}, though in this case we explicitly integrate under every characteristic and not just the characteristic approaching the interface with the greatest speed. 
For the density, normal velocity, and pressure, we refer to value calculated using the characteristic approaching the interface at the greatest speed as the ``reference state", e.g. $\bm{W}^m_L$ for the left interface and $\bm{W}^p_R$ for the right. As with PPMC, this reference state is our first guess at the input state for the Riemann problem. For the transverse velocities, we use the states associated with the advection speed, $\bm{W}^0_L$ and $\bm{W}^0_R$.

These reference states are only first-order accurate approximations and the input states can be further corrected
to account for the presence of other characteristics approaching the interface. 
Following the notation in \cite{CW84}, the description of the algorithm continues in terms of the primitive variables. 
The sound speeds for the reference states on the left and right of the cell are computed
\begin{equation}
a_L = \sqrt{\frac{\gamma p_L^m}{\rho_L^m}} \quad \mathrm{and} \quad a_R = \sqrt{\frac{\gamma p_R^p}{\rho_R^p}},
\end{equation}
along with correction terms that are added to the reference state,
\begin{equation}
\begin{aligned}
\chi^p_L &= -\frac{1}{2\rho^m_L a_L}\left(u^m_L - u^p_L - \frac{p^m_L + p_L^p}{\rho^m_L a_L}\right), \\
\chi^m_R &= \frac{1}{2\rho^p_s a_R}\left(u^p_R - u^m_R - \frac{p^p_R - p_R^m}{\rho^p_R a_R}\right), \\
\chi^0_L &= \frac{p^m_L - p_L^0}{(\rho^m_L a_L)^2} + \frac{1}{\rho^m_L} - \frac{1}{\rho_L^0}, \\
\chi^0_R &= \frac{p^p_R - p_R^0}{(\rho^p_R a_R)^2} + \frac{1}{\rho^p_R} - \frac{1}{\rho_R^0}.
\end{aligned}
\end{equation}
In the event that the characteristic is not traveling toward the interface these correction terms are set to zero,
\begin{equation}
\begin{aligned}
\chi^0_L &= 0 \quad \mathrm{if} \lambda^0 > 0, \\
\chi^p_L &= 0 \quad \mathrm{if} \lambda^p > 0, \\
\chi^m_R &= 0 \quad \mathrm{if} \lambda^m < 0, \\
\chi^0_R &= 0 \quad \mathrm{if} \lambda^0 < 0.
\end{aligned}
\end{equation}
The correction terms are then used with the reference state 
integration to calculate the final input states for the Riemann problem on each side of the cell:
\begin{equation}
\begin{aligned}
\rho_L = \left(\frac{1}{\rho^m_L} - \chi^0_L - \chi^p_L \right)^{-1}, &\quad \rho_{R,i} = \left(\frac{1}{\rho^p_R} - \chi^m_R - \chi^0_R \right)^{-1}, \\
u_L = u^m_L + \rho^m_L a_L \chi^p_L, &\quad u_R = u^p_R - \rho^p_R a_R \chi^m_R, \\
p_L = p^m_L + (\rho^m_L a_L)^2 \chi^p_L, &\quad p_R = p^p_R + (\rho^p_R a_R)^2 \chi^m_R, \\
v_L = v^0_L, &\quad v_R = v^0_R \\
w_L = w^0_L, &\quad w_R = w^0_R.
\end{aligned}
\end{equation}


\section{Riemann Solvers}\label{app:riemann_solvers}

\subsection{The Exact Solver}

The exact solver used in \cholla follows the solver presented in 
Chapter 4 of \citet{Toro09}, adapted for implementation in CUDA C. 
The algorithm to solve the Riemann problem is presented below, using an $x$-interface
as an example. 
In the following section the CTU notation from Section~\ref{sec:CTU} is used, where states are
labeled at the left and right of the interface.

First, the input states at the left and right of the interface are converted to the primitive 
variables, $\bm{W^*}_L$ and $\bm{W^*}_R$. (Between GPU kernels like the interface reconstruction and the Riemann problem, calculated values are stored in the conserved form.) 
These vectors are used to compute the corresponding sound 
speed on either side of the interface
\begin{equation}
a_L = \sqrt{\frac{\gamma p_L}{\rho_L}} \quad \mathrm{and} \quad a_R = \sqrt{\frac{\gamma p_R}{\rho_R}}.
\end{equation}
To determine the Riemann solution, 
the exact pressure  $p^m$ and velocity $u^m$ in the intermediate state must be computed
(see Figure~\ref{fig:riemann_problem}).
We use the \citet{Toro09} primitive variable Riemann solver to provide an initial 
approximation to the intermediate state pressure, given by
\begin{equation}
\tilde{p} = 0.5(p_L + p_R) + 0.125(u_L - u_R)(\rho_L + \rho_R)(a_L + a_R).
\end{equation}
Because $\tilde{p}$ is an approximation and the solution for $p$ cannot be negative, we set $\tilde{p} = 0$ if 
the calculated pressure is below zero. A two-shock Riemann solver is then used
to calculate a more accurate estimate, 
\begin{equation}
p_0 = \frac{g_L p_L + g_R p_R - (u_R-u_L)}{g_L + g_R},
\end{equation}
with
\begin{equation}
g_k = \sqrt{\frac{A_k}{\tilde{p} + B_k}}, \quad A_k = \frac{2}{(\gamma + 1)\rho_k}, \quad B_k = \frac{\gamma -1}{\gamma + 1}p_k, \quad \mathrm{where} \ k = L, \ R.
\end{equation}
To maintain positivity, a pressure floor of $p_0\ge10^{-6}$ is enforced in this estimate. 
The pressure $p_0$ is then used as a starting point in a Newton-Raphson iteration 
to compute the exact solution for the pressure in the intermediate region. We define 
the pressure functions $f_L$ and $f_R$,
\begin{equation}\label{eqn:pressure_functions}
f_k = 
\begin{cases}
\frac{2a_k}{\gamma -1}\left[\left(\frac{p_0}{p_k}\right)^{\frac{\gamma-1}{2\gamma}} -1 \right] &\quad \mathrm{if} \ p_0 \leq p_k \ \mathrm{(rarefaction)}, \\
(p_0 - p_k)\left(\frac{A_k}{p_0 + B_k}\right)^{\frac{1}{2}} &\quad \mathrm{if} \ p_0 > p_k \ \mathrm{(shock)},
\end{cases}
\end{equation}
and their first derivatives $f'_L$ and $f'_R$,
\begin{equation}
f'_k = 
\begin{cases}
\frac{1}{\rho_k a_k} \left(\frac{p_0}{p_k} \right)^{-\frac{\gamma + 1}{2\gamma}} &\quad \mathrm{if} \ p_0 \leq p_k \ \mathrm{(rarefaction)}, \\
\left(\frac{A_k}{B_k + p_0} \right)^{\frac{1}{2}} \left[1 - \frac{p_0 - p_k}{2(B_k + p_0)} \right] &\quad \mathrm{if} \ p_0 > p_k \ \mathrm{(shock)}.
\end{cases}
\end{equation}
Again, $k = L$ or $R$, and $A_k$ and $B_k$ are as defined above. These quantities are used 
to calculate the pressure in the intermediate state, $p^m$,
\begin{equation}
p^m = p_0 - \frac{f_L + f_R + u_R - u_L}{f'_L + f'_R}.
\end{equation}
We then compare the newly computed pressure, $p^m$, to the previously computed pressure,
\begin{equation}\label{eqn:change}
\Delta = 2 \frac{|p^m - p_0|}{|p^m + p_0|}.
\end{equation}
If $\Delta$ is greater than a relative tolerance (e.g., $10^{-6}$),
the values of $f_k$, $f'_k$, and $p^m$ are recomputed using the 
updated value of  $p^m$ in place of $p_0$ in Equations~\ref{eqn:pressure_functions} - \ref{eqn:change}. 
When $\Delta$ is less than the tolerance, the procedure halts. Having calculated a suitably accurate pressure $p^m$, 
the pressure can then be used to compute the velocity in the intermediate state as
\begin{equation}
u^m = 0.5(u_L + u_R + f_R - f_L).
\end{equation}

Once the values of $p^m$ and $u^m$ in the intermediate state have been computed, 
the values for each of the primitive variables can be calculated at the cell interface. 
To do this, a number of conditions are tested to determine where in the Riemann solution 
the interface lies. For pure hydrodynamics there are ten possible outcomes.

If $u^m \geq 0$, the contact discontinuity is to the right of the cell interface. We then check to see if there is a rarefaction wave on the left, i.e. if $p^m \leq p_L$. If so, there are three possible solutions.
\begin{itemize}
\item If $u_L - a_L \geq 0$ the interface is in the left data state, and the solution is simply the input data on the left:
\begin{equation}
\rho = \rho_L, \quad u = u_L, \quad p = p_L.
\end{equation}
\item If $u^m - a_L \left(\frac{p^m}{p_L} \right)^{\frac{\gamma-1}{2\gamma}} < 0$ the interface is in the intermediate data state to the right of the fan, but left of the contact:
\begin{equation}
\rho = \rho_L \left(\frac{p^m}{p_L} \right)^{\frac{1}{\gamma}}, \quad u = u^m, \quad p = p^m.
\end{equation}
\item Otherwise, the interface is within the rarefaction fan:
\begin{equation}
\rho = \rho_L \left(\frac{a}{a_L} \right)^{\frac{2}{\gamma-1}}, \quad u = a, \quad p = p_L \left(\frac{a}{a_L} \right)^{\frac{2 \gamma}{\gamma-1}}, \quad \mathrm{where} \ a = \frac{2}{\gamma +1}\left(a_L + \frac{\gamma-1}{2}u_L\right).
\end{equation}
\end{itemize}
If there is a shock to the left, rather than a rarefaction, i.e. if $p^m > p_L$, the shock speed is calculated as
\begin{equation}
s_L = u_L - a_L \sqrt{\frac{(\gamma + 1)}{2\gamma} \frac{p^m}{p_L} + \frac{\gamma - 1}{2\gamma}}.
\end{equation}
\begin{itemize}
\item If $s_L \geq 0$ the interface samples the left data state:
\begin{equation}
\rho = \rho_L, \quad u = u_L, \quad p = p_L.
\end{equation}
\item Otherwise, the interface samples the intermediate data state to the left of the contact:
\begin{equation}
\rho = \rho_L \left(\frac{p^m}{p_L} +\frac{\gamma-1}{\gamma+1} \right) / \left( \frac{p^m}{p_L}\frac{\gamma-1}{\gamma+1} + 1 \right), \quad u = u^m, \quad p = p^m.
\end{equation}
\end{itemize}

If instead $u^m < 0$, the contact discontinuity is to the left of the cell interface and there is a
similar set of five possible outcomes. If there is a rarefaction wave on the 
right of the Riemann solution, i.e. if $p^m \leq p_R$ there are three possibilities:
\begin{itemize}
\item If $u_R + a_R \leq 0$ the interface samples the right data state:
\begin{equation}
\rho = \rho_R, \quad u = u_R, \quad p = p_R.
\end{equation}
\item If $u^m + a_R \left(\frac{p^m}{p_R} \right)^{\frac{\gamma-1}{2\gamma}} \geq 0$ the interface samples
the intermediate state to the left of the fan, but right of the contact:
\begin{equation}
\rho = \rho_R \left(\frac{p^m}{p_R} \right)^{\frac{1}{\gamma}}, \quad u = u^m, \quad p = p^m.
\end{equation}
\item Otherwise, the interface samples the rarefaction fan:
\begin{equation}
\rho = \rho_R \left(\frac{a}{a_R} \right)^{\frac{2}{\gamma-1}}, \quad u = -a, \quad p = p_R \left(\frac{a}{a_R} \right)^{\frac{2 \gamma}{\gamma-1}}, \quad \mathrm{where} \ a = \frac{2}{\gamma +1}\left(a_R + \frac{\gamma-1}{2}u_R\right).
\end{equation}
\end{itemize}
If $p^m > p_R$ there is a shock to the right rather than a rarefaction, the shock speed is calculated as
\begin{equation}
s_LR= u_R + a_R \sqrt{\frac{(\gamma + 1)}{2\gamma} \frac{p^m}{p_R} + \frac{\gamma - 1}{2\gamma}}.
\end{equation}
\begin{itemize}
\item If $s_R \leq 0$ the interface samples the right data state:
\begin{equation}
\rho = \rho_R, \quad u = u_R, \quad p = p_R.
\end{equation}
\item Otherwise, the interface samples the intermediate data state to the right of the contact:
\begin{equation}
\rho = \rho_R \left(\frac{p^m}{p_R} +\frac{\gamma-1}{\gamma+1} \right) / \left( \frac{p^m}{p_R}\frac{\gamma-1}{\gamma+1} + 1 \right), \quad u = u^m, \quad p = p^m.
\end{equation}
\end{itemize}

After determining where in the Riemann solution the interface samples, the evolved primitive variables 
$\rho$, $u$, and $p$ at the cell interface are determined. The fluxes of the conserved variables
can then be calculated following Equation~\ref{eqn:x_flux}:
\begin{equation}
\bm{F} = 
\begin{bmatrix}
		\rho u \\
		\rho u^{2} + p \\
		\rho u v_k \\
		\rho u w_k \\
		(E + p) u
\end{bmatrix},
\end{equation}
where $k$ = $L$ or $R$; $L$ if $u \geq 0$, and $R$ if $u < 0$. These conditions
reflect the fact that transverse velocities are simply advected with the flow. 
Equation~\ref{eqn:y_flux} is used to compute similar
fluxes in the $y$-direction, and Equation~\ref{eqn:z_flux} is used for fluxes in the $z$-direction.

\subsection{The Roe Solver}

Rather than using an expensive iterative procedure to calculate the exact solution to the Riemann problem, the 
\citet{Roe81} Riemann 
solver calculates an exact solution to a linearized version of the Euler equations. Below the procedure for 
calculating the Roe fluxes at an $x$-interface is detailed. The $y$- and $z$-interface calculations are identical modulo
an appropriate change of variables. Should the approximate Roe solver fail, we include a failsafe based on the method
of \citet{Stone08} where HLLE fluxes \citep{Harten83, Einfeldt88} are substituted. The Roe and HLLE solvers are described
below.

The Roe solver starts with the calculated input vectors of primitive variables at the left and right of an 
interface, $\bm{W^*}_L$ and $\bm{W^*}_R$. Fluxes corresponding to the left and right state, $\bm{F}_L$ and $\bm{F}_R$, are
calculated following Equation~\ref{eqn:x_flux}. The Roe average state, $\bm{\tilde{u}} = (\tilde{\rho}, \tilde{u}, \tilde{v}, \tilde{w}, \tilde{H})^T$, is then computed:
\begin{equation}
\begin{split}
\tilde{\rho} &= \sqrt{\rho_L}\sqrt{\rho_R} \\
\tilde{u} &= (\sqrt{\rho_L}u_L + \sqrt{\rho_R}u_R) / (\sqrt{\rho_L} + \sqrt{\rho_R}) \\
\tilde{v} &= (\sqrt{\rho_L}v_L + \sqrt{\rho_R}v_R) / (\sqrt{\rho_L} + \sqrt{\rho_R}) \\
\tilde{w} &= (\sqrt{\rho_L}w_L + \sqrt{\rho_R}w_R) / (\sqrt{\rho_L} + \sqrt{\rho_R}) \\
\tilde{H} &= (\sqrt{\rho_L}H_L + \sqrt{\rho_R}H_R) / (\sqrt{\rho_L} + \sqrt{\rho_R}),
\end{split}
\end{equation}
with the enthalpy, $H = (E + p) / \rho$, used instead of the pressure. The average sound speed
\begin{equation}
\tilde{a} = \sqrt{(\gamma - 1)(\tilde{H} - 0.5\bm{\tilde{\mathrm{V}}}^2)}, \quad \mathrm{where} \ \bm{\tilde{\mathrm{V}}}^2 = \tilde{u}\tilde{u} + \tilde{v}\tilde{v} + \tilde{w}\tilde{w}
\end{equation}
is also needed.
These Roe average states are used to calculate the eigenvalues of the Roe Jacobian,
 \begin{equation}
 \lambda^m = \tilde{u} - \tilde{a}, \quad \lambda^0 = \tilde{u}, \quad \mathrm{and} \quad \lambda^p = \tilde{u} + \tilde{a}.
 \end{equation}
If the flow is supersonic ($\lambda^m \geq 0$ to the right, or $\lambda^p \leq 0$ to the left), the solver
returns the appropriate left or right state fluxes $\bm{F}_L$ or $\bm{F}_R$.
If flow is subsonic, the calculation of the Roe fluxes proceeds.

Differences in the conserved variables between the left and right states are computed:
\begin{equation}
\begin{split}
\delta\rho &= \rho_R - \rho_L \\
\delta m_x &= m_{x, R} - m_{x, L} \\
\delta m_y &= m_{y, R} - m_{y, L} \\
\delta m_z &= m_{z, R} - m_{z, L} \\
\delta E &= E_R - E_L.
\end{split}
\end{equation}
These differences are projected onto the characteristics by multiplying by the left eigenvector 
associated with each eigenvalue. The resulting characteristics are the wave strengths, $\bm{\xi}$, 
from Equation~\ref{eqn:Roe_wave_strengths}:
\begin{equation}
\begin{split}
\xi_0 &= \delta \rho N_a (0.5 \gamma' \bm{\tilde{\mathrm{V}}}^2 + \tilde{u}\tilde{a}) - \delta m_x N_a (\gamma' \tilde{u}+ \tilde{a}) - \delta m_y N_a \gamma' \tilde{v} - \delta m_z N_a \gamma' \tilde{w} + \delta E N_a \gamma' \\
\xi_1 &= -\delta \rho \tilde{v} + \delta m_y \\
\xi_2 &= -\delta \rho \tilde{w} + \delta m_z \\
\xi_3 &= \delta \rho(1 - N_a \gamma' \bm{\tilde{\mathrm{V}}}^2) + \delta m_x \gamma' \tilde{u} / \tilde{a}^2 + \delta m_y \gamma' \tilde{v}/ \tilde{a}^2 + \delta m_z \gamma' \tilde{w}/ \tilde{a}^2 - \delta E \gamma' / \tilde{a}^2 \\
\xi_4 &= \delta \rho N_a(0.5 \gamma' \bm{\tilde{\mathrm{V}}}^2 - \tilde{u}\tilde{a}) - \delta m_x N_a (\gamma' \tilde{u} - \tilde{a}) - \delta m_y N_a \gamma' \tilde{v} - \delta m_z N_a \gamma' \tilde{w} + \delta E N_a \gamma'
\end{split}
\end{equation}
where $N_a = 1 / (2 \tilde{a}^2)$ and $\gamma' = \gamma -1$. 
Numeric subscripts denote the elements of $\bm{\xi}$. 
Each characteristic variable is multiplied by its associated eigenvalue, yielding a vector of coefficients, $\bm{C}$:
\begin{equation}
\label{eqn:Roe_coefficients}
C_0 = \xi_0 |\lambda^m|, \quad C_1 = \xi_1 |\lambda^0|, \quad C_2 = \xi_2 |\lambda^0|, \quad C_3 = \xi_3 |\lambda^0|, \quad C_4 = \xi_4 |\lambda^p|
\end{equation}
The product of these coefficients with the right eigenvectors are then summed, 
keeping track of the summation in the vector $\bm{s}$:
\begin{equation}
\begin{split}
s_0 &= C_0 + C_3 + C_4 \\
s_1 &= C_0 (\tilde{u} - \tilde{a}) + C_3 \tilde{u} + C_4 (\tilde{u} + \tilde{a}) \\
s_2 &= C_0 \tilde{v} + C_1 + C_3 \tilde{v} + C_4 \tilde{v} \\
s_3 &= C_0 \tilde{w} + C_2 + C_3 \tilde{2} + C_4 \tilde{w} \\
s_4 &= C_0 (\tilde{H} - \tilde{u}\tilde{a}) +  C_1 \tilde{v} + C_2 \tilde{w} + 0.5 C_3 \bm{\tilde{\mathrm{V}}}^2 + C_4 (\tilde{H} + \tilde{u}\tilde{a})
\end{split}
\end{equation}
By this stage, all information needed to compute the Roe fluxes has been obtained.  The Roe fluxes are then
computed as
\begin{equation}
\bm{F}_{Roe} = \frac{1}{2} \left( \bm{F}_L + \bm{F}_R - \bm{s} \right)
\label{eqn:roe_fluxes}
\end{equation}

Before returning these fluxes, however, the intermediate states are examined for possible negative
densities or pressures.
The intermediate states, labeled  $\bm{U}^m_L$ and $\bm{U}^m_R$  in analogy with the exact Riemann solver,
are calculated by projecting the characteristic variables onto the conserved variables. 
Each characteristic variable is multiplied by its associated right eigenvector and the results summed, in turn, 
to the left state. 
The left intermediate state, $\bm{U}^m_L$, is calculated as
\begin{equation}
\begin{split}
\rho^m_L &= \rho_L + \xi_0 \\
\rho u^m_L &= \rho u_L + \xi_0 (\tilde{u} - \tilde{a}) \\
\rho v^m_L &= \rho v_L + \xi_0 \tilde{v} \\
\rho w^m_L &= \rho w_L + \xi_0 \tilde{w} \\
E^m_L &= E_L + \xi_0 (\tilde{H} - \tilde{u}\tilde{a}).
\end{split}
\end{equation}
If $\lambda^0 > \lambda^m$, we check for negative density and pressure. We then move on to the right intermediate state, $\bm{U}^m_R$:
\begin{equation}
\begin{split}
\rho^m_R &= \rho^m_L + \xi_3 \\
\rho u^m_R &= \rho u^m_L + \xi_3 \tilde{u} \\
\rho v^m_R &= \rho v^m_L + \xi_1 + \xi_3 \tilde{v} \\
\rho w^m_R &= \rho w^m_L + \xi_2 + \xi_3 \tilde{w} \\
E^m_R &= E^m_L + \xi_1 \tilde{v} + \xi_2 \tilde{w} + 0.5 \xi_3 \bm{\tilde{\mathrm{V}}}^2.
\end{split}
\end{equation}
If $\lambda^p > \lambda^0$ then negative densities and pressures are possible and, if present, the Roe fluxes
are replaced with HLLE fluxes. Otherwise the Roe fluxes calculated in Equation~\ref{eqn:roe_fluxes} are returned.

Little additional work must be done to calculate the HLLE fluxes since many of the required quantities are
computed for the Roe fluxes.  The HLLE solver constructs a single average state between the right and left 
input states, neglecting the contact wave. Ignoring the contact wave means that density discontinuities diffuse quickly, but 
the HLLE solver has the advantage of always producing positive intermediate densities and pressures, as shown in \cite{Einfeldt91}. 
The HLLE flux algorithm starts with the computation of the sound speed for the left and right input 
states, $a_L$ and $a_R$. 
Bounding speeds are then calculated, 
defined by the minimum and maximum Roe eigenvalues and the left and right acoustic waves:
\begin{equation}
b^m = \mathrm{min}[\mathrm{min}(\lambda^m, u_L - a_L), 0], \quad
b^p = \mathrm{max}[\mathrm{max}(\lambda^p, u_R + a_R), 0]
\end{equation}
These speeds are used to compute left and right fluxes:
\begin{equation}
\bm{F}_L = \bm{F}(\bm{W}_L) - b^m \bm{U}_L \quad \mathrm{and} \quad \bm{F}_R = \bm{F}(\bm{W}_R) - b^p \bm{U}_R,
\end{equation}
where $\bm{F}(\bm{W})$ is calculated using Equation~\ref{eqn:x_flux}, and $\bm{U}$ refers to the conserved variables on either side of the interface, i.e. the input states for the Riemann problem. These intermediate fluxes are
then employed to compute the HLLE fluxes, given by
\begin{equation}
\bm{F}_{\mathrm{HLLE}} = \frac{1}{2}\left[(\bm{F}_L + \bm{F}_R) + (\bm{F}_L - \bm{F}_R) \left(\frac{b^p + b^m}{b^p - b^m} \right)\right].
\end{equation}
If the Roe solver fails during the transverse flux or conserved variable update steps of the CTU algorithm,
these HLLE fluxes may be used.


\section{The H Correction}\label{app:h_correction}

As mentioned in the main text, strictly upwind multidimensional integration algorithms like CTU are susceptible to a 
particular type of numerical pathology, 
referred to as the carbuncle instability owing to its appearance \citep{Quirk94}. The carbuncle instability arises in problems containing strong, grid-aligned shocks as a result of insufficient dissipation when inserting a one dimensional Riemann flux into a multidimensional problem. In the scenario when a planar shock is present perpendicular to the direction of the fluid flow, crossflow oscillations arise that grow and lead to severe inaccuracies in the numerical model. To correct this phenomenon, \cholla implements a solution devised by \cite{Sanders98} that accounts for 
the perpendicular velocities when calculating the Riemann flux normal to an interface. The correction thereby adds 
sufficient dissipation to mitigate the carbuncle instability. This method for suppressing the carbuncle instability 
is called  {\it H correction} in reference to the shape of the required stencil.

In \chollans, H Correction is implemented in conjunction
with the Roe Riemann solver. The correction is only applied to the fluxes calculated during the second set of Riemann problems, between the transverse flux update and the final conserved variable update of the CTU algorithm. 
To apply the H correction, 
at every interface a quantity $\eta$ is computed that
depends on the velocity normal to the interface and the sound speed. 
For the $x$-interface at $(i+\frac{1}{2}, j)$ in a two dimensional simulation, this quantity can be computed as
\begin{equation}
\eta_{(i+\frac{1}{2}, j)} = \frac{1}{2}|(u_R + a_R) - (u_L + a_L)|,
\end{equation}
where $u_R$ and $u_L$ are the normal velocity at the interface calculated using the right and left input states, and $a_L$ and $a_R$ are the corresponding sound speeds. For $y$- and $z$-interfaces, the velocity components $v$ and $w$ would be
used. Once a value of $\eta$ has been calculated for each interface, a further quantity $\eta^H$ is computed
for each interface by finding the maximum value of $\eta$ for each of the surrounding interfaces:
\begin{equation}
\eta^H_{(i+\frac{1}{2}, j)} = \mathrm{max}[\eta_{(i-1, j+\frac{1}{2})}, \eta_{(i-1, j-\frac{1}{2}}, \eta_{(i+\frac{1}{2}, j)}, \eta_{(i+1, j+\frac{1}{2})}, \eta_{(i+1, j-\frac{1}{2})}].
\end{equation}
For $x$-interfaces,
information about the maximum wavespeeds at the adjacent $y$-interfaces that could affect the Riemann solution are thereby
incorporated. In three dimensions, the calculation of $\eta^H$ uses an appropriate nine-point stencil. The $\eta^H$ quantity is used in conjunction with the Roe Riemann solver when calculating the CTU fluxes for an interface ($\bm{F}^{n+\frac{1}{2}}$, $\bm{G}^{n+\frac{1}{2}}$, and $\bm{H}^{n+\frac{1}{2}}$) by replacing the eigenvalues $\lambda^\alpha$ in Equation~\ref{eqn:Roe_coefficients} with
\begin{equation}
|\tilde{\lambda}^\alpha| = \mathrm{max}(|\lambda^\alpha|, \eta^H)
\end{equation}
This substitution serves to add crossflow dissipation to the one-dimensional Roe fluxes and mitigates 
the carbuncle instability, as seen in Figure~\ref{fig:noh_3D}. 
Using the H correction comes at an additional cost, as its stencil
requires adding an extra ghost cell on each side of the simulation in every dimension.


\bibliography{code_paper}

\end{document}